\documentclass[12pt]{iopart}

\usepackage{axodraw}

\usepackage{graphicx}
\usepackage{rotating}
\usepackage{color}

\begin{document}
\begin{flushright}
%January 25, 2009 \\
%March 25, 2009 \\
%{D\O\ Note xxx}
\end{flushright}

\title{STATUS OF HIGGS BOSON SEARCHES AT THE TEVATRON}

\author{Andr\'e Sopczak}

\address{Lancaster University}
\ead{andre.sopczak@cern.ch}

\begin{center}
{\bf\small ABSTRACT}
\end{center}
{\large 
Over the last years the Tevatron Run-II has extended several
limits on Higgs boson masses and coupling 
which were pioneered during the LEP accelerator 
operation between 1989 and 2000.
Higgs boson searches will also be at the forefront of research at the LHC.
This review concisely discusses the experimental constraints 
set by the CDF and D\O\ collaborations in winter 2008/2009 at the beginning of the LHC era.
Model-independent and model-dependent limits on Higgs boson masses and couplings
have been set and interpretations are discussed both in the Standard Model and 
in extended models. Recently, for the first time the Tevatron excludes a SM Higgs boson mass range
(160-170~GeV) beyond the LEP limit at 95\% CL.
The experimental sensitivities are estimated for the completion of the Tevatron programme. 
}

\vspace*{3cm}
\begin{center}
{\em Contribution to WONP'09 conference\\
     Havana, Cuba, February 9-12}
\end{center}

\maketitle

\thispagestyle{empty}
\tableofcontents
\setcounter{page}{0}

%AS
\setlength{\textheight}{250mm}

\newcommand{\bb}{b\bar b}
\newcommand{\nn}{\nu\bar \nu}
\newcommand{\tautau}{\tau^+\tau^-}
\newcommand{\ee}{\mbox{$\mathrm{e}^{+}\mathrm{e}^{-}$}}

\newcommand{\Zo} {{\mathrm {Z}}}
\newcommand{\db}    {{d_{\rm B}}}%
\newcommand{\dgz}  {{\Delta g_1^{\Zo}}}%
\newcommand{\dkg}   {{\Delta \kappa_\gamma}}%

\newcommand{\pb}   {\mbox{$\rm pb^{-1}$}}
\newcommand{\fb}   {\mbox{$\rm fb^{-1}$}}

\clearpage
\section{Introduction}
\begin{figure}[h!]
\vspace*{-0.4cm}
\includegraphics[width=0.32\textwidth,height=5cm]{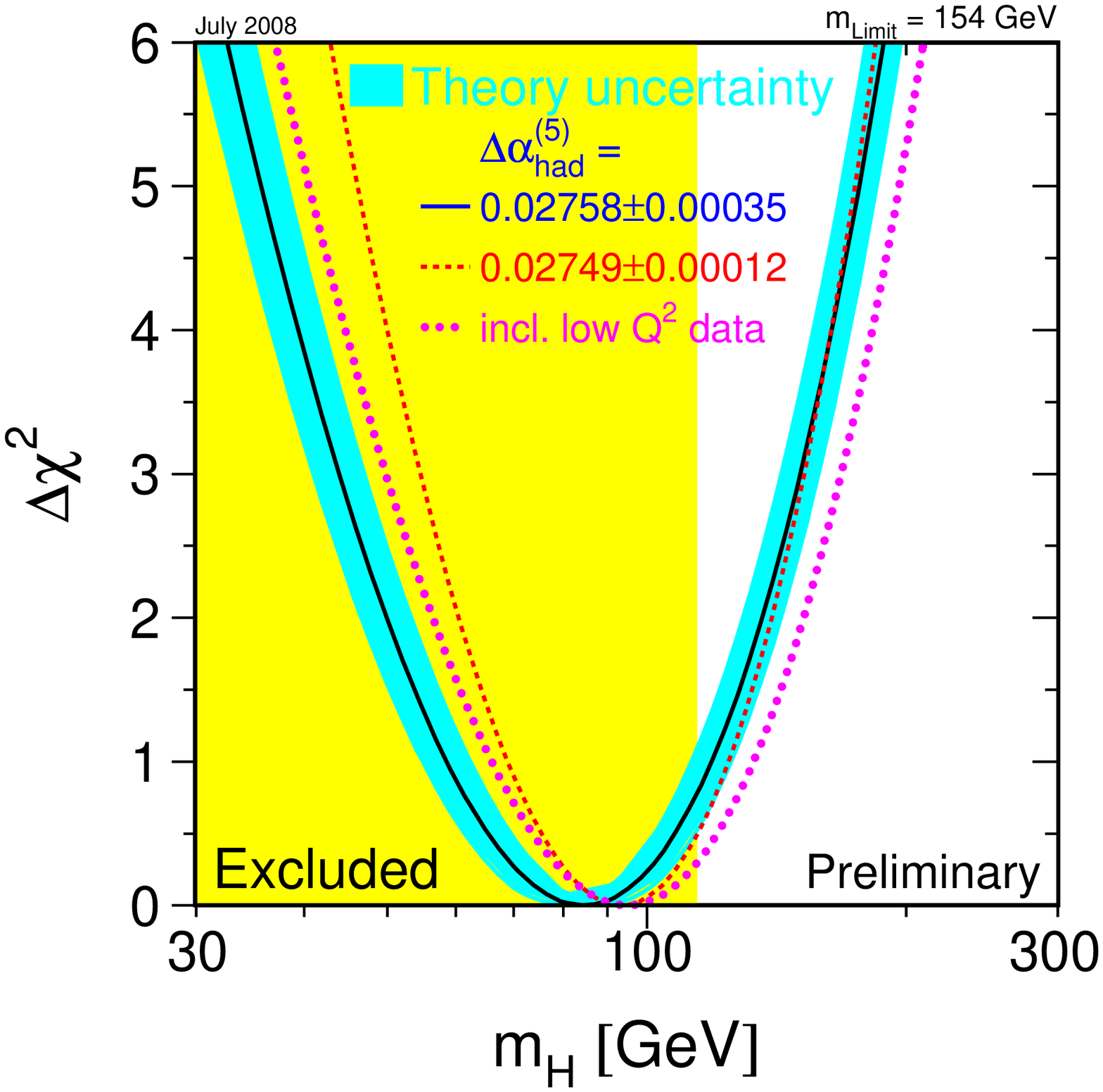}\hfill
\includegraphics[width=0.32\textwidth,height=5cm]{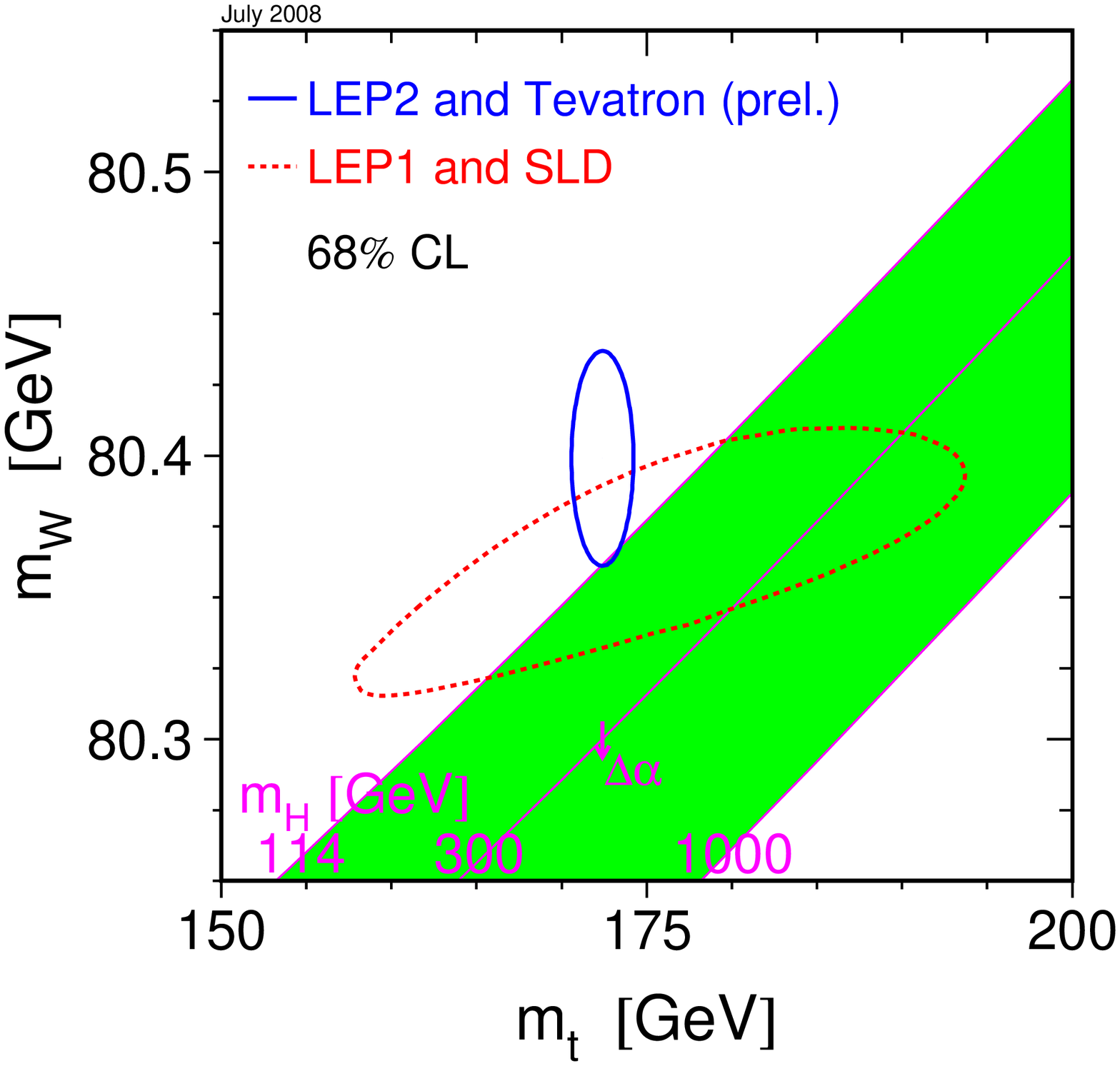}\hfill
\includegraphics[width=0.32\textwidth,height=5cm]{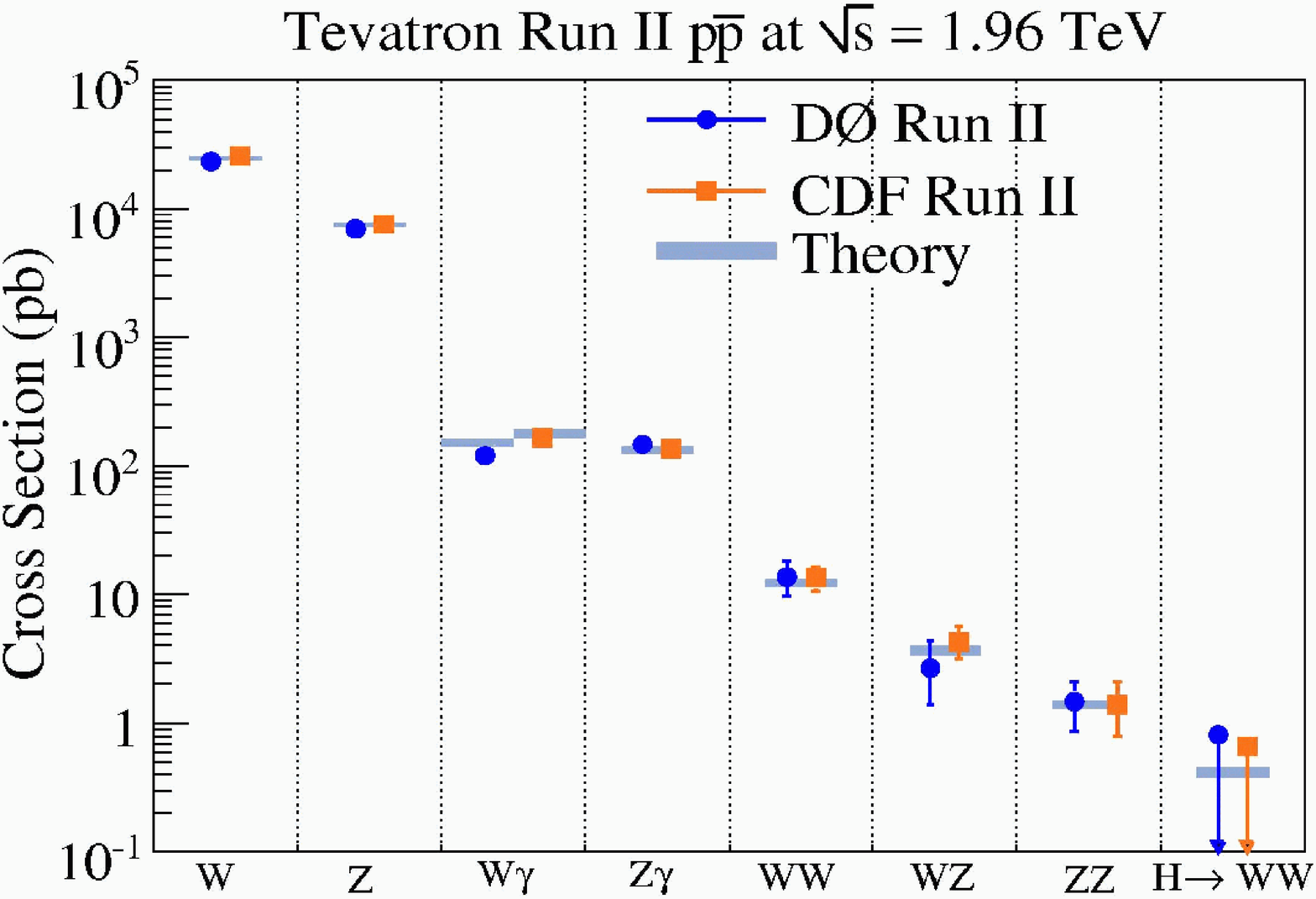}
\vspace*{-0.4cm}
\caption{
Left: Higgs boson mass prediction in the SM framework.
The upper SM Higgs boson mass limit at 95\% CL is 154~GeV.
Center: smaller ellipse including LEP-2 and Tevatron data
       (solid line) prefers a region 
       outside the SM Higgs boson mass band 
       ($m_{\rm H} = 114$ to 1000 GeV).
       The combined results from LEP-1 and SLD only 
       are shown separately (dashed line).
Right: overview of SM measurements by CDF and D\O, and indication of the
expected cross-section for a 160~GeV SM Higgs boson.
}
\label{fig:ew}
\end{figure}

The search for new particles is at the forefront of High Energy Physics. The discovery of a Higgs 
boson would shed light on electroweak symmetry breaking and the generation of mass in the Universe.
Many searches for new particles were performed at LEP and stringent limits on Higgs bosons 
in the Standard Model (SM) and beyond were set. 
These limits are summarized in Table~\ref{tab:lep} (from~\cite{lep05}) including model-independent LEP
limits and benchmark results in the Minimal Supersymmetric extension of the 
SM (MSSM)~\cite{susy05}.
In addition to the limits from direct searches, some indication on the Higgs boson mass exist from
precision electro-weak measurements,
as shown in Fig.~\ref{fig:ew} (left and center plots from~\cite{lep-ew}).
Up to about 4.2~fb$^{-1}$ of data have been analyzed so far (winter 2008/9) by each Tevatron experiment,
while about 5~fb$^{-1}$ data have been recorded (about 6~fb$^{-1}$ delivered). This report is structured 
similarly as Ref.~\cite{as06} 
to allow to compare more directly the experimental progress 
from summer 2005~\cite{as06}.

\begin{figure}[h!]
\begin{center}
\vspace*{-0.45cm}
\includegraphics[width=0.49\textwidth,height=5.2cm]{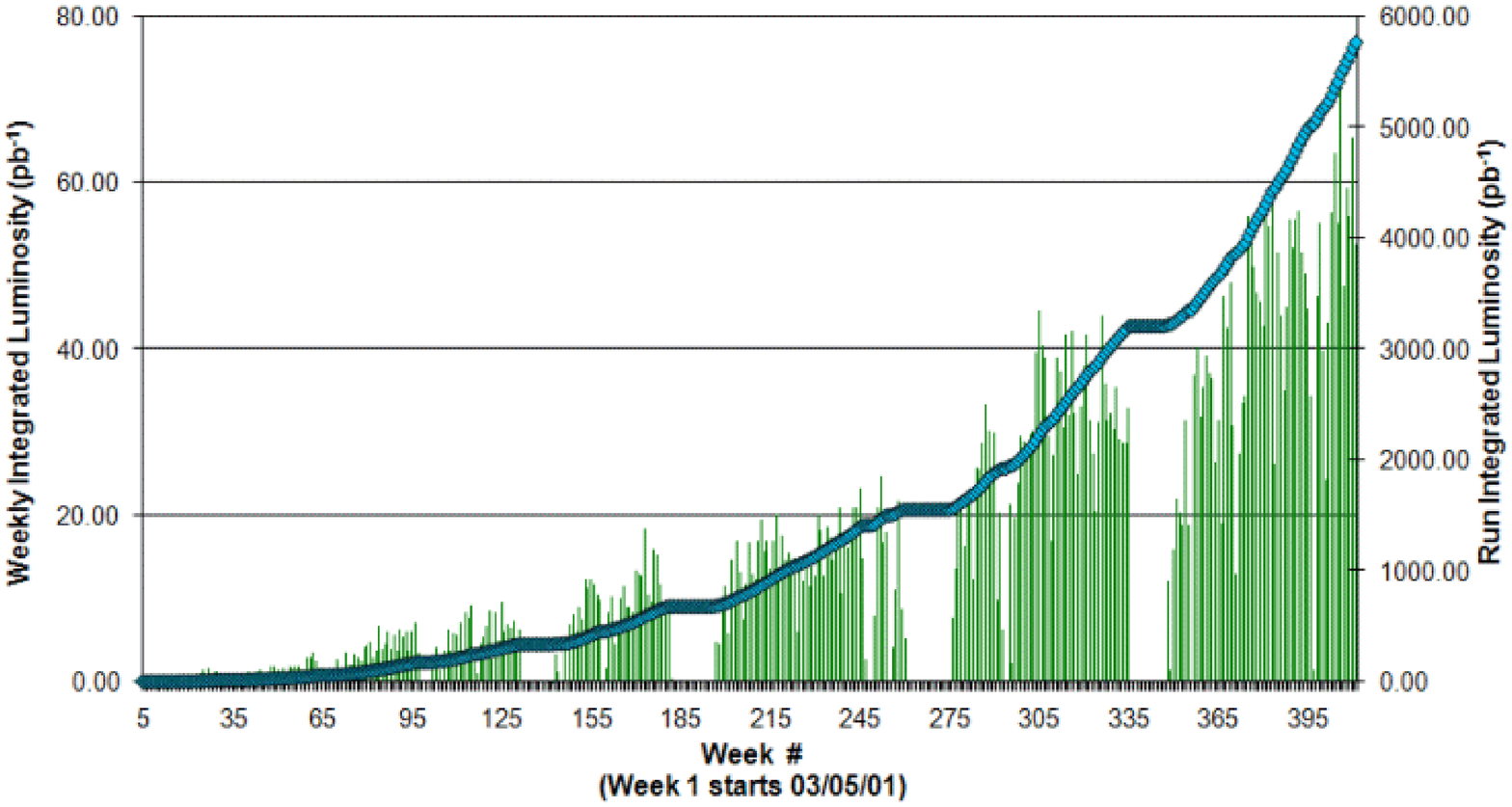} \hfill
\includegraphics[width=0.49\textwidth,height=5.2cm]{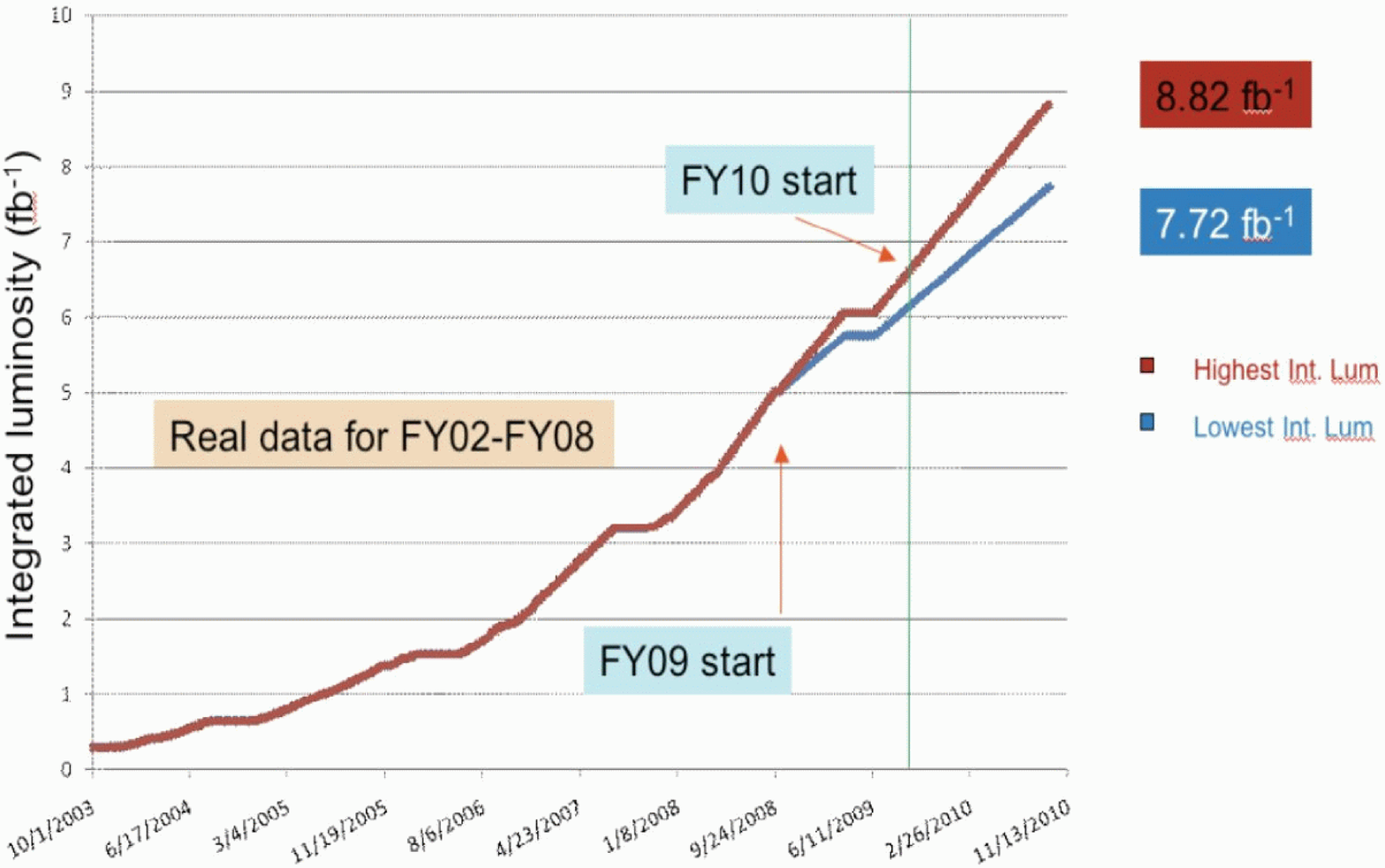}
\vspace*{-0.2cm}
\caption{Left: integrated delivered Tevatron luminosity.
         Right: expectation for 2010 data-taking.
}
\label{fig:lumi}
\vspace*{-0.7cm}
\end{center}
\end{figure}

Both CDF and D\O\ have measured with precision various SM processes
as illustrated in Fig.~\ref{fig:ew} (right plot from~\cite{cdf_d0-sm}).
The figure includes also recent ZZ measurements~\cite{cdf_zz,d0_zz}.
Figure~\ref{fig:lumi} (from~\cite{lumi}) shows the delivered luminosity and its expectations.

\section{Production and Decay}
\vspace*{-1mm}

\begin{figure}[bp]
\begin{minipage}{0.58\textwidth}
\begin{turn}{270}
\includegraphics[height=\textwidth,width=6cm]{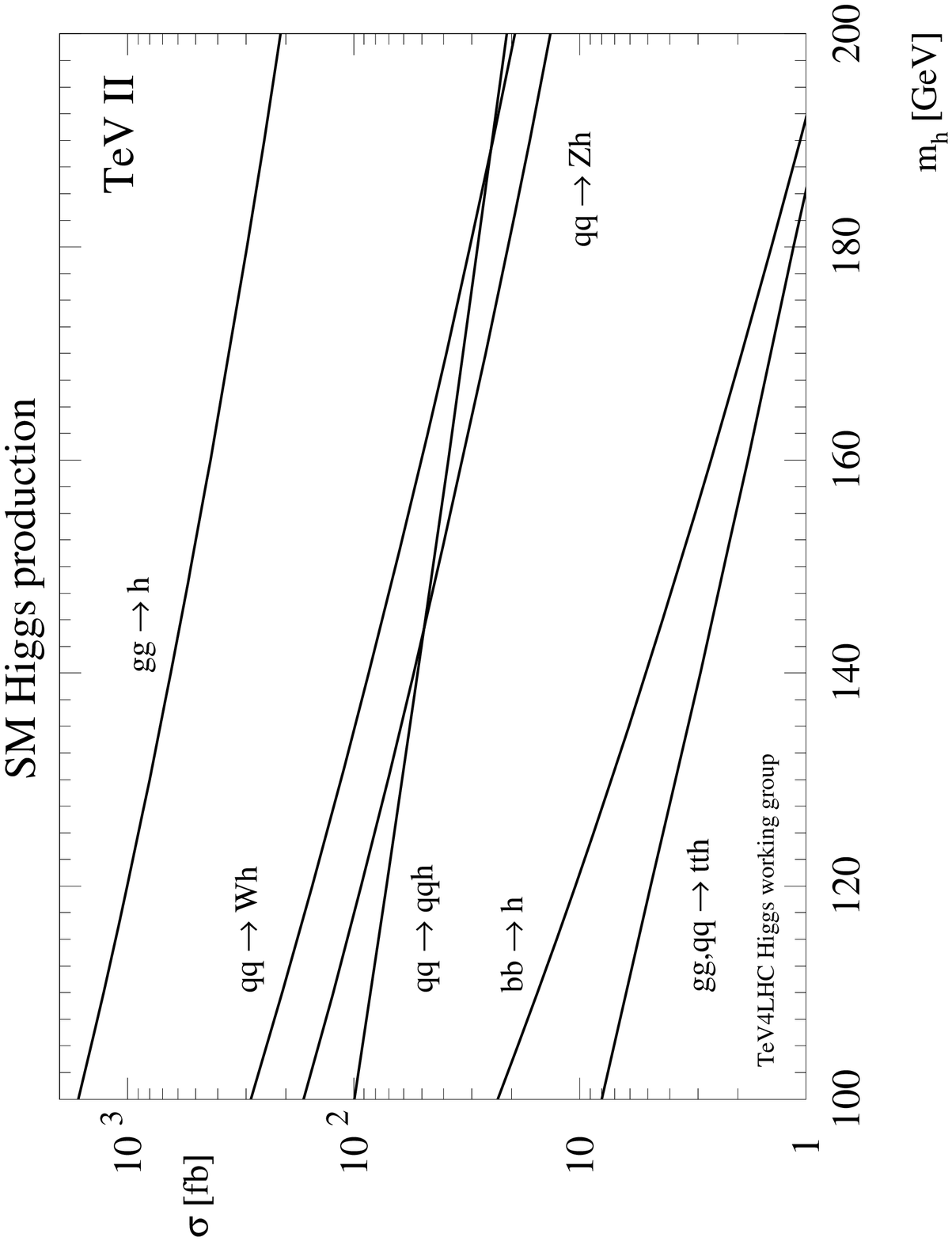} \hfill
\end{turn}\hfill
\end{minipage}
\begin{minipage}{0.40\textwidth}
\includegraphics[width=\textwidth,height=6cm]{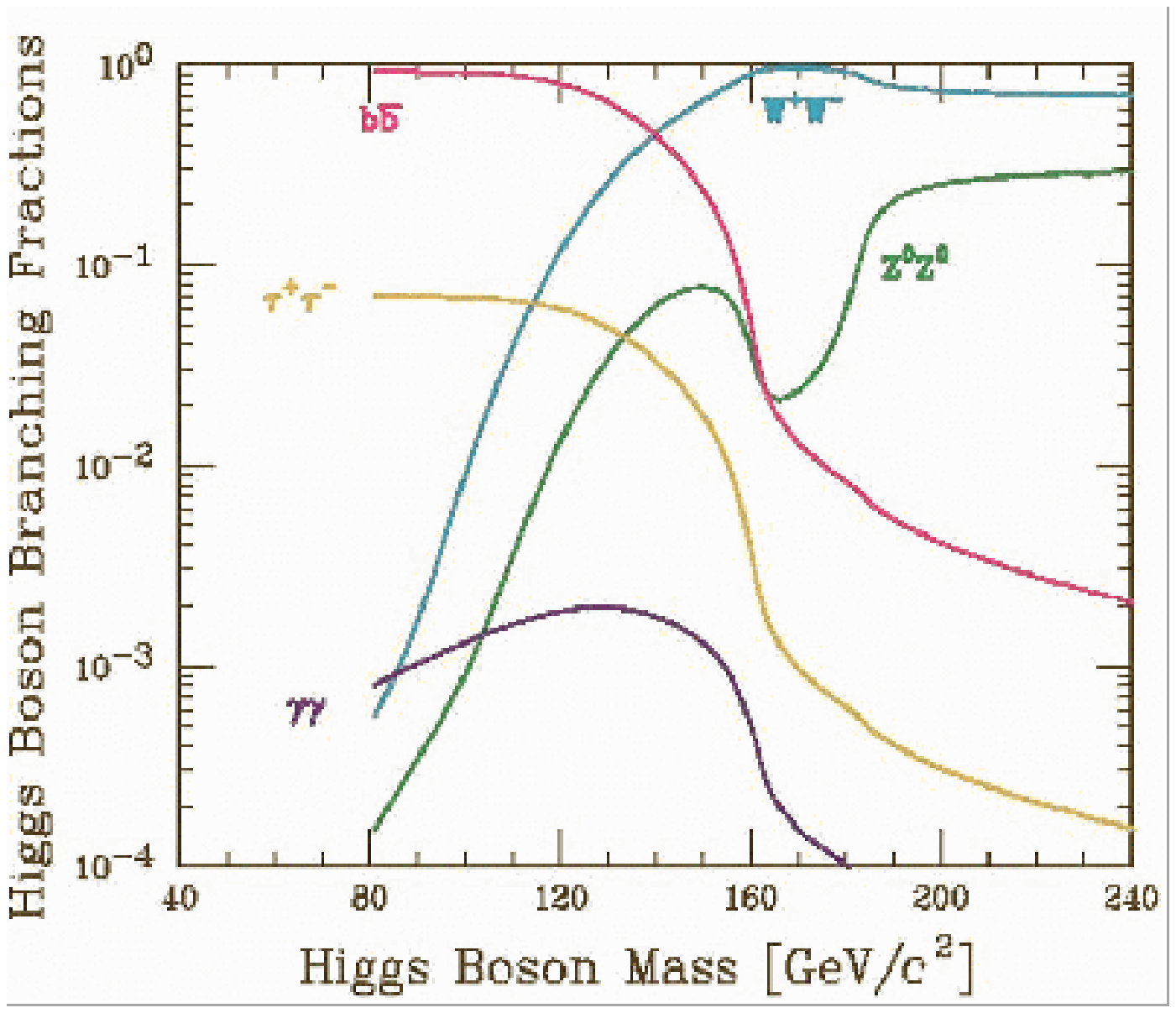}
\end{minipage}
\vspace*{-4mm}
\caption{
Left: expected SM Higgs boson production cross-sections at the Tevatron (1.96~TeV).
Right: expected Higgs boson decay branching ratios for a SM Higgs boson masses.
       At the Tevatron, $\rm \bb$ and WW decays are dominant, in addition the $\tautau$ decay
       mode has a significant contribution.
}
\label{fig:tev-xsec}
\vspace*{-0.3cm}
\end{figure}

The expected cross-section and branching ratios are shown 
in Fig.~\ref{fig:tev-xsec} (from~\cite{xsec} and~\cite{br}) as a function of the Higgs boson
mass.
It is interesting to note that corresponding to the current collected data sample
of about 5~\fb\ about 5000 SM Higgs bosons of 120 GeV could have already 
been produced in $\rm p\bar p$ collisions at each experiment.
For a SM Higgs boson mass below about 200~GeV the decay width is below 1~GeV which is much
below the detector resolution. 

\begin{table}[hp]
\vspace*{-0.1cm}
\renewcommand{\arraystretch}{0.65} % enlarge line spacing
\caption{Summary of Higgs boson mass limits at 95\% CL.
`LEP' indicates a combination of the results from ALEPH, DELPHI, L3 and OPAL.
If results from the experiments are not (yet) combined, examples
which represent the different search areas from individual experiments
are given. 
Details are given in Ref.~\cite{lep05}.
\label{tab:lep} }
\vspace*{-0.2cm}
\begin{center}
{
\begin{tabular}{c|c|r}
Search                      & experiment & limit \\\hline 
Standard Model              &   LEP  
   & $m^{\rm SM}_{\rm H} > 114.4$ GeV \\ 
Reduced rate and SM decay &       
  & $\xi^2>0.05:$ $ m_{\rm H} > 85$ GeV \\
& & $\xi^2>0.3:$ $ m_{\rm H} > 110$ GeV \\
Reduced rate and $\rm b\bar b$ decay  &
  & $\xi^2>0.04:$ $ m_{\rm H} > 80$ GeV \\
& & $\xi^2>0.25:$ $ m_{\rm H} >110$ GeV \\ 
Reduced rate and $\tau^+\tau^-$ decay & 
  & $\xi^2>0.2:$ $ m_{\rm H} > 113$ GeV \\ 
\hspace*{-4mm} Reduced rate and hadronic decay\hspace*{-2mm} &
  & $\xi^2=1:$   $m_{\rm H} >112.9$ GeV\\ 
& & $\xi^2>0.3:$ $ m_{\rm H} > 97$ GeV \\ 
&ALEPH& $\xi^2>0.04:$ $m_{\rm H} \approx 90$ GeV \\ 
Anomalous couplings & L3 & $d,~\db,~\dgz,~\dkg$ exclusions \\ \hline
MSSM (no scalar top mixing) & LEP 
  & almost entirely excluded\\ 
General MSSM scan & DELPHI &  $m_{\rm h} > 87$ GeV, $m_{\rm A} >90$ GeV\\ 
Larger top-quark mass     & LEP & strongly reduced $\tan\beta$ limits \\ \hline
MSSM with CP-violating phases  & LEP    &  strongly reduced mass limits  \\ \hline
Visible/invisible Higgs decays & DELPHI & $m_{\rm H} >111.8$ GeV\\ 
Majoron model (max. mixing) &  & $m_{\rm H,S} >112.1$ GeV\\ \hline
Two-doublet Higgs model   & DELPHI
  & $\rm hA\to b\bar b b\bar b:$
    $m_{\rm h}+m_{\rm A} >  150$ GeV\\
(for $\sigma_{\rm max}$) & 
  & $\tau^+\tau^-\tau^+\tau^-:$
    $m_{\rm h}+m_{\rm A} >  160$ GeV\\
& & $\rm (AA)A\to 6b:$ $m_{\rm h}+m_{\rm A} >  150$ GeV\\
& & $\rm (AA)Z\to 4b~Z:$ $m_{\rm h} >  90$ GeV\\
& & $\rm hA\to q\bar q q\bar q:$ 
      $m_{\rm h}+m_{\rm A} >  110$ GeV\\
Two-doublet model scan & OPAL
  & $\tan\beta > 1:$ $ m_{\rm h} \approx m_{\rm A} > 85$ GeV \\\hline 
Yukawa process & DELPHI & $C > 40:$ $m_{\rm h,A} > 40$ GeV \\\hline 
Singly-charged Higgs bosons & LEP 
  & $m_{\rm H^\pm} > 78.6$ GeV \\
$\rm W^\pm A$ decay mode & DELPHI& $m_{\rm H^\pm} > 76.7$ GeV \\ \hline
Doubly-charged Higgs bosons &  DELPHI/OPAL 
  & 
$m_{\rm H^{++}} > 99$ GeV \\
$\ee\to\ee$ &L3 &$h_{\rm ee} > 0.5:$ $m_{\rm H^{++}} > 700$ GeV \\ \hline
Fermiophobic $\rm H\to WW, ZZ, \gamma\gamma$ & L3 
  &  $m_{\rm H} > 108.3$ GeV \\
$\rm H\to \gamma\gamma$ &LEP &  $ m_{\rm H} > 109.7$ GeV \\ \hline
Uniform and stealthy scenarios & OPAL & depending on model parameters
\end{tabular}
}
\end{center}
\vspace*{-1.1cm}
\end{table}

\clearpage
\section{b-Quark Tagging}
\vspace*{-0.25cm}

\begin{figure}[hp]
\vspace*{1cm}
\includegraphics[width=0.32\textwidth,height=5cm]{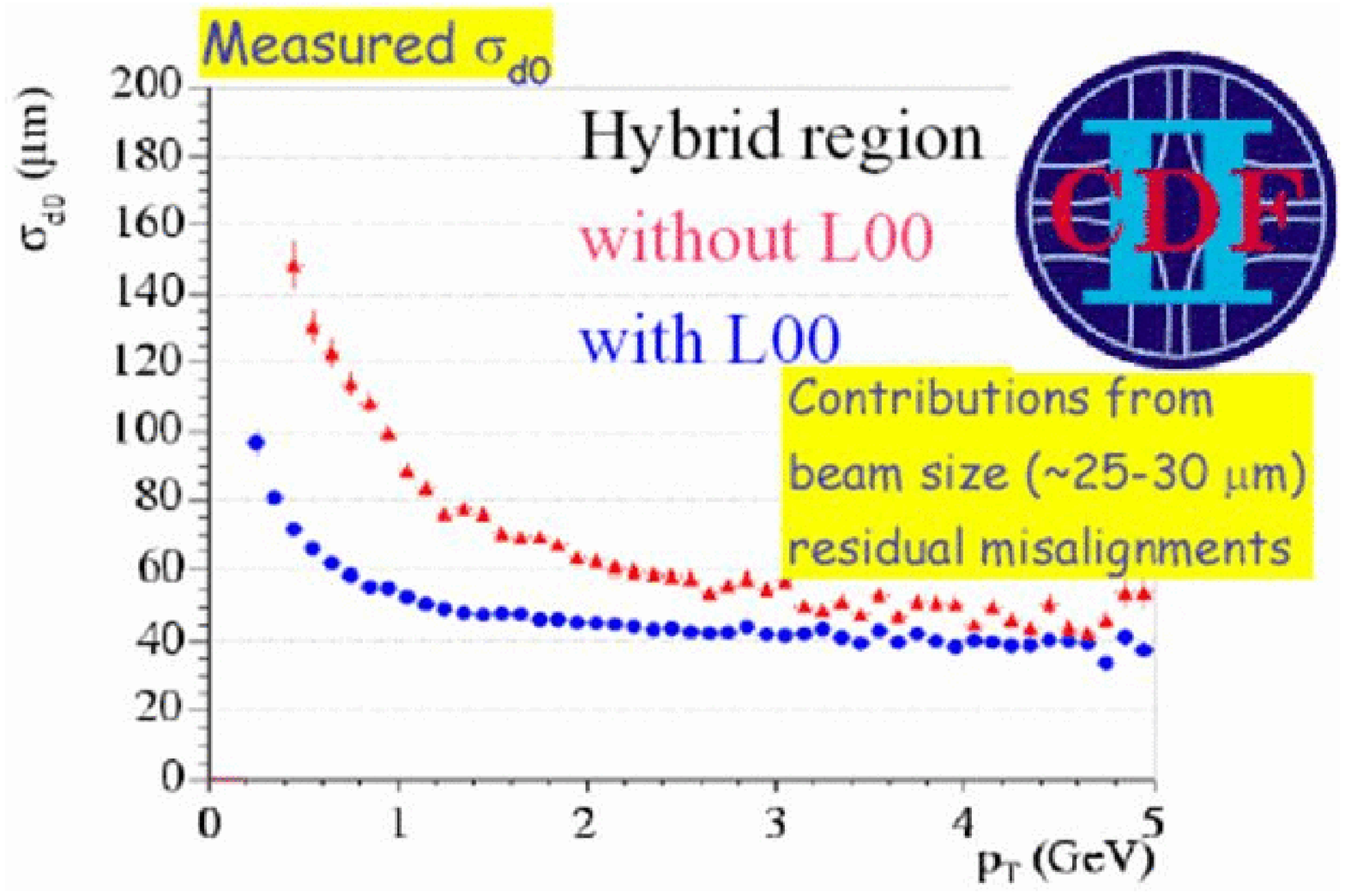}\hfill
\includegraphics[width=0.32\textwidth,height=5cm]{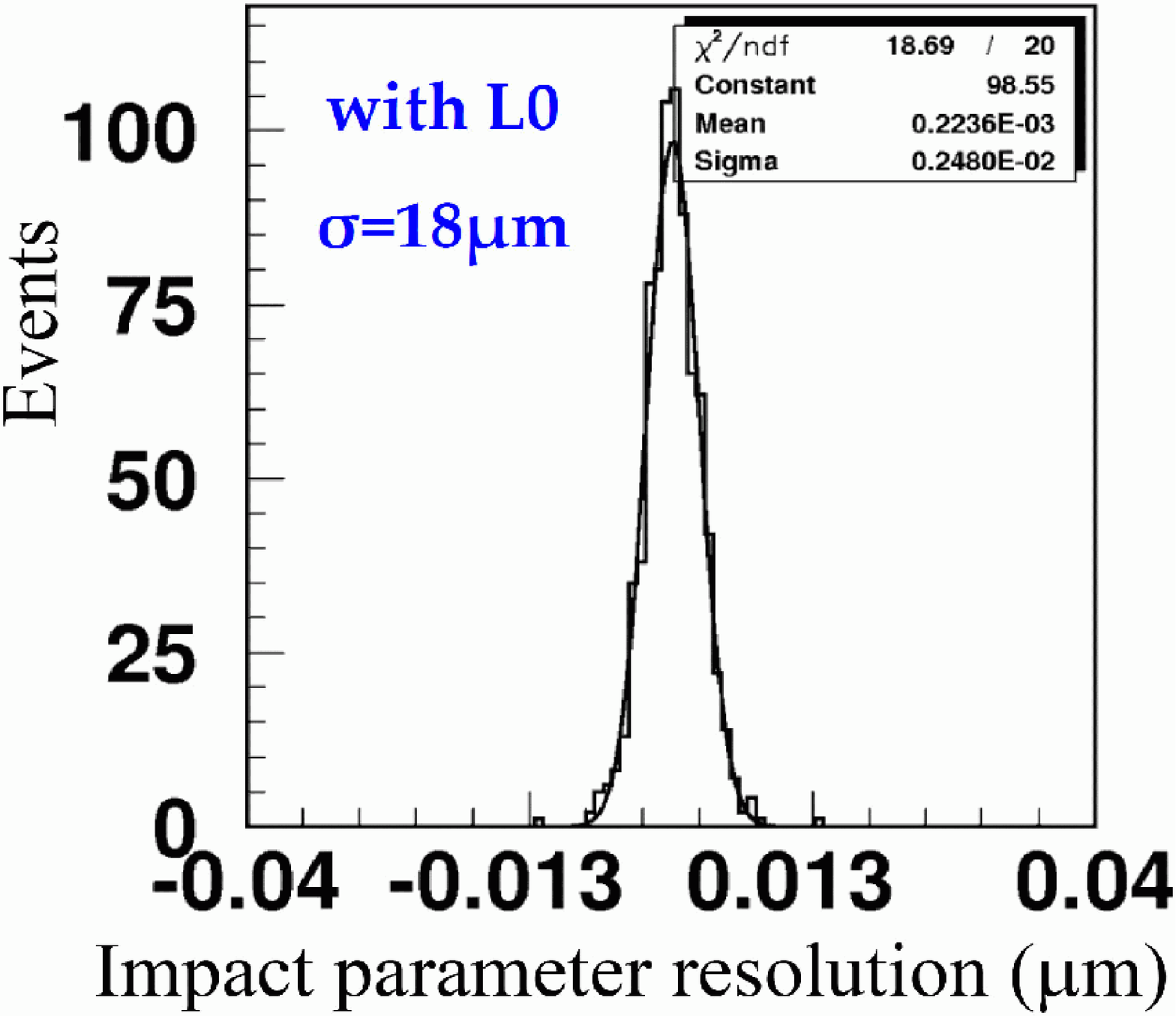} \hfill
\includegraphics[width=0.32\textwidth,height=5cm]{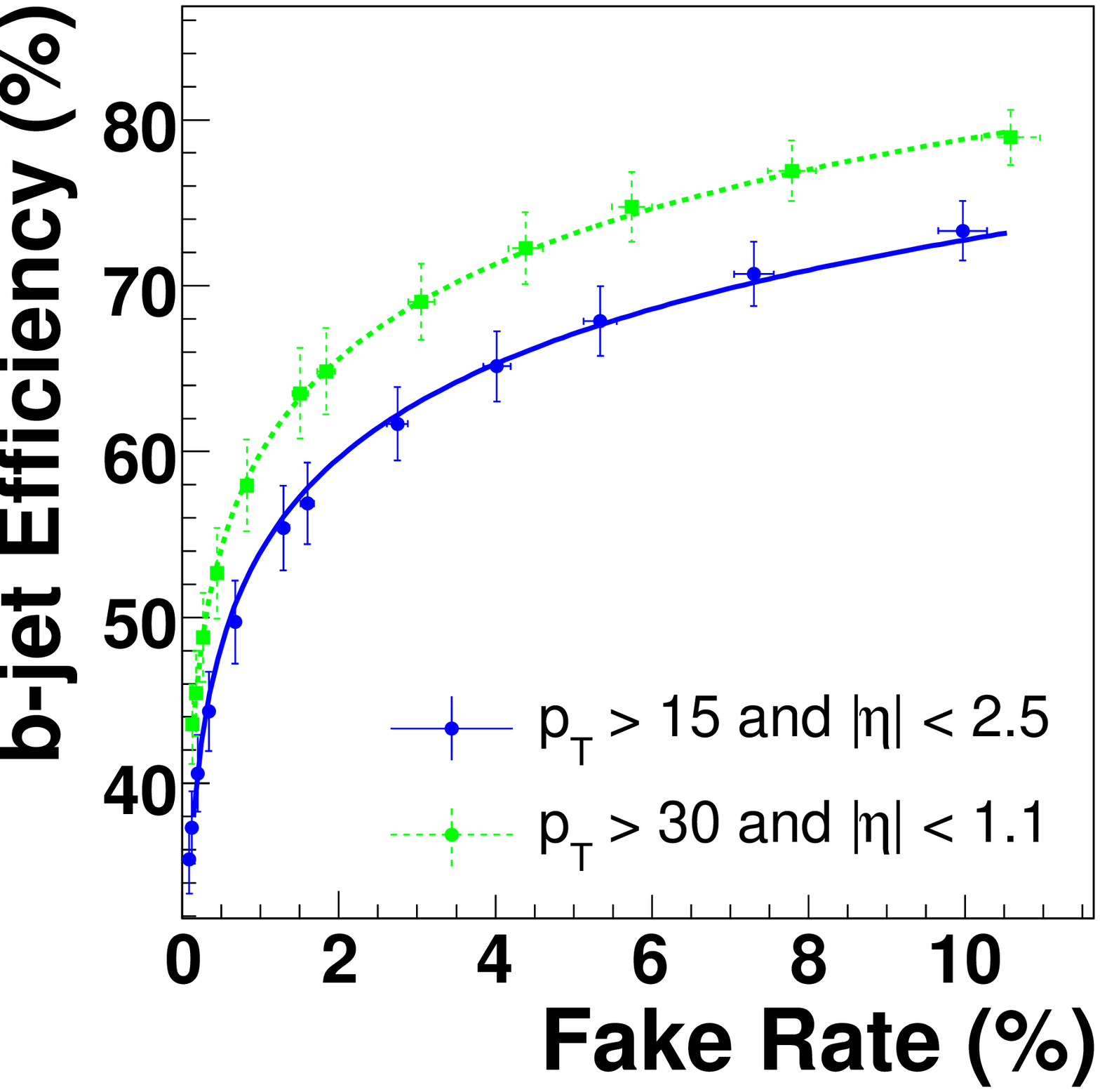}
\vspace*{-0.2cm}
\caption{
Left: CDF impact parameter resolution as a function of $p_T$ for 
tracks traversing passive material in vertex detector, with (blue dots) and 
without (red triangles) use of L00 hits.
Center: D\O\ impact parameter resolution after the installation of a new
vertex detector layer (L0), which improved the resolution by 40\%.
Right: D\O\ b-quark tagging performance for $\rm Z\to b\bar b$ and $\rm Z\to q\bar q$ events.
The error bars include statistical and systematic uncertainties.
}
\label{fig:vertex}
\vspace*{-1cm}
\end{figure}

\begin{figure}[htbp]
\vspace*{1cm}
\begin{minipage}{0.49\textwidth}
\includegraphics[width=\textwidth,height=6cm]{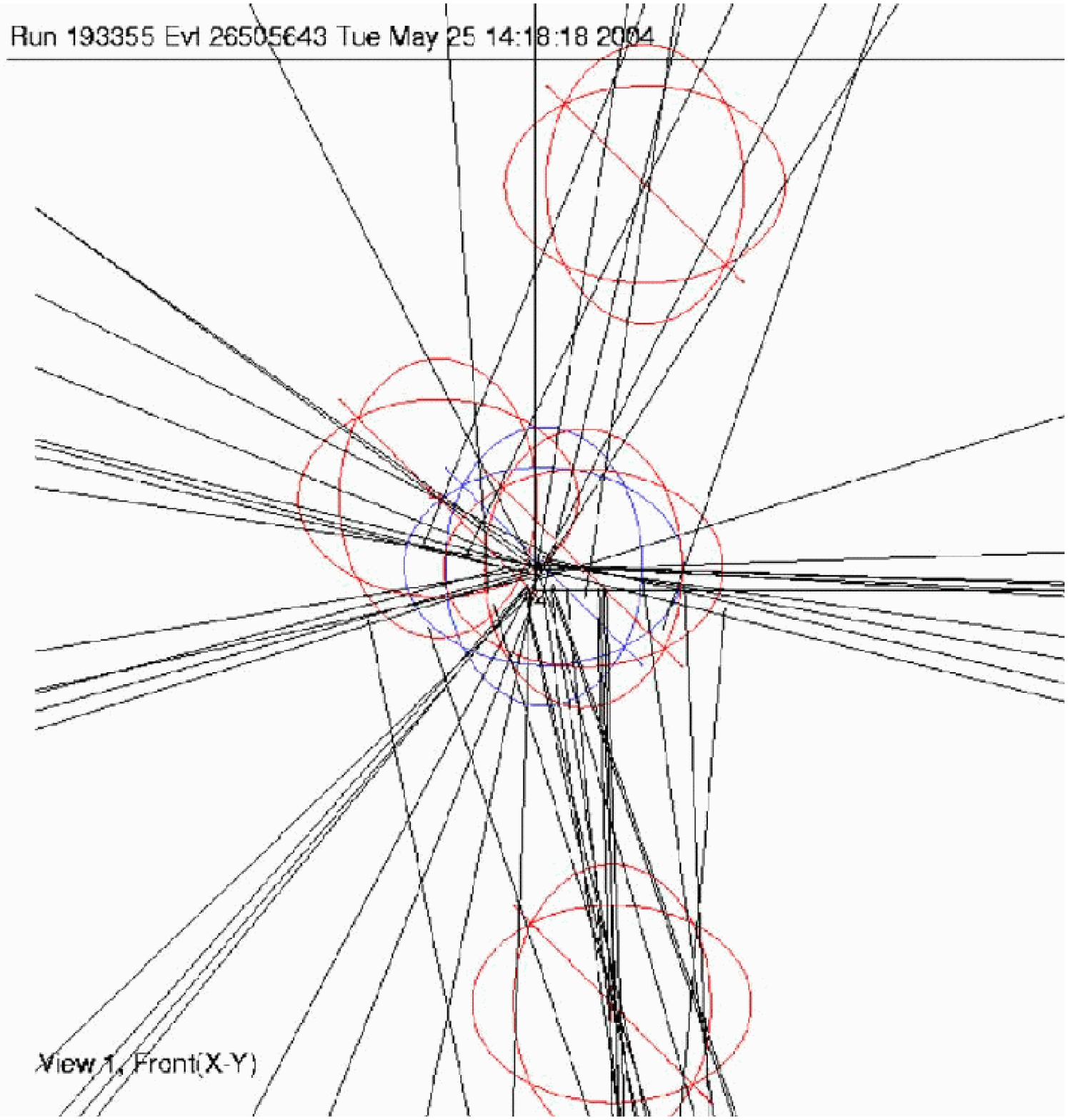} 
\end{minipage}\hfill
\begin{minipage}{0.49\textwidth}
\includegraphics[width=\textwidth,height=6cm]{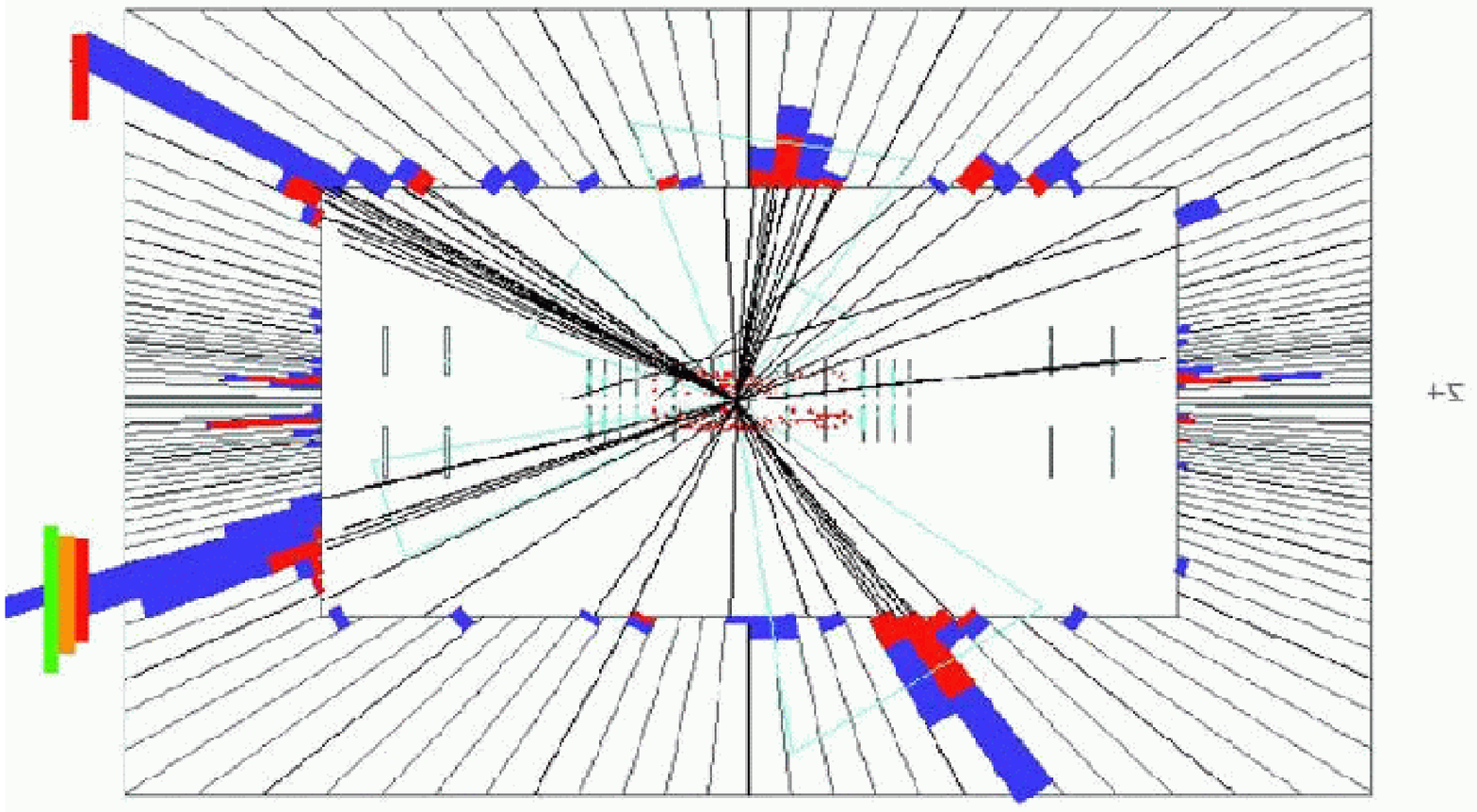} 
\end{minipage}
\caption{D\O\ example of b-tagged event. 
Left: reconstructed tracks near the
interaction point. 
Right: jets clearly visible in the calorimeter.
}
\label{fig:d0-b-event}
\end{figure}

The b-tagging capabilities are most important for the low-mass
Higgs boson searches and 
a critical parameter is the impact parameter resolution of the vertex detector. 
The improvement of the impact parameter resolution with a sensitive layer very 
close to the interaction point is illustrated in Fig.~\ref{fig:vertex} 
(left plot from~\cite{cdf-l00} and center plot from~\cite{d0-l0}).
In CDF this layer is called L00 and in D\O\ it is called L0.
These innermost layers contribute significantly to the b-tagging performance.
Figure~\ref{fig:vertex} (right plot from~\cite{d0-btag}) shows also 
the D\O\ b-quark tagging performance including L0.
An example of a quadruply b-tagged event is shown in 
Fig.~\ref{fig:d0-b-event} (from~\cite{d0-b-event}).

\begin{figure}[tp]
\vspace*{-0.2cm}
\includegraphics[width=0.32\textwidth,height=6cm]{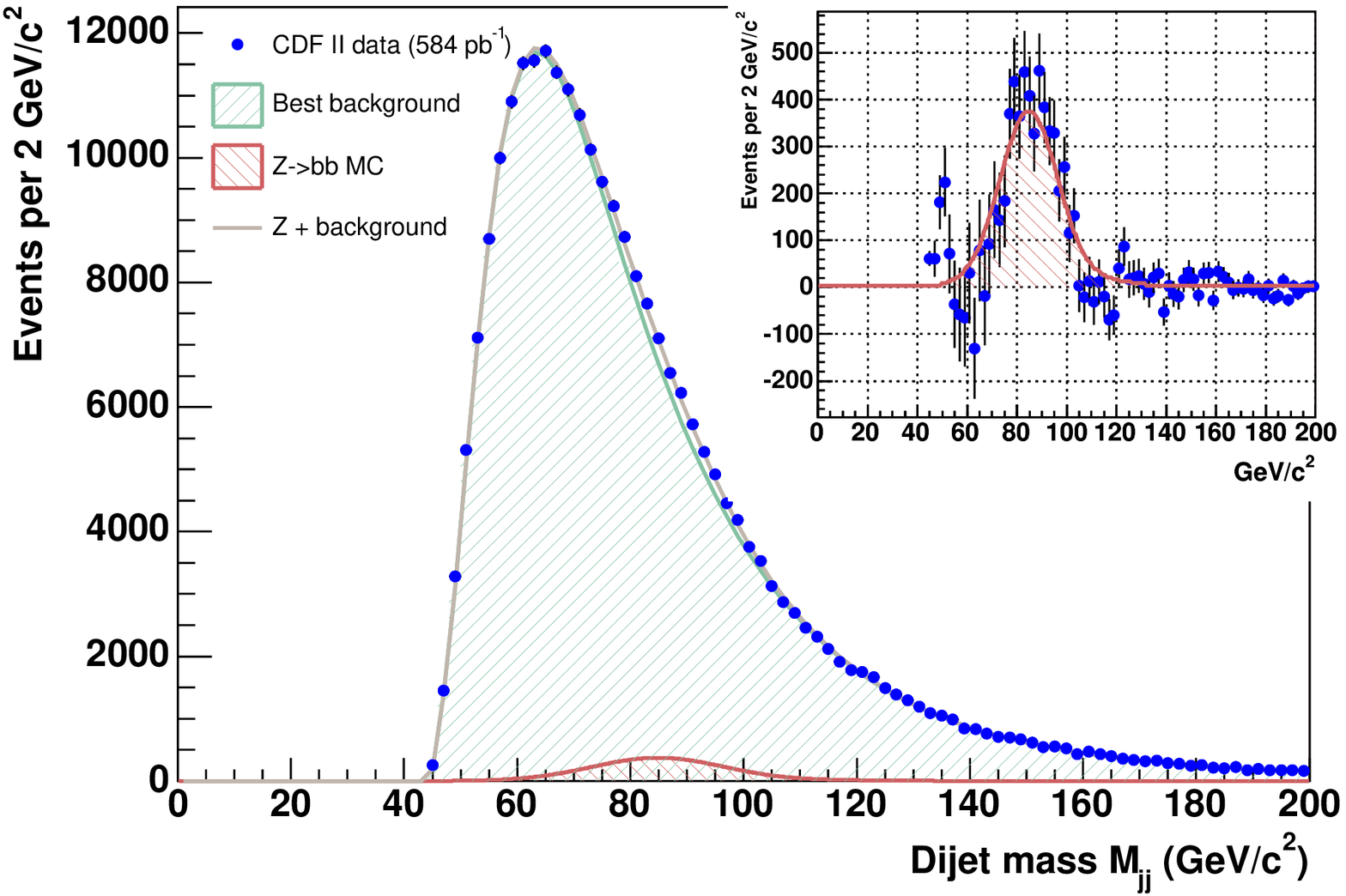} \hfill
\includegraphics[width=0.32\textwidth,height=6cm]{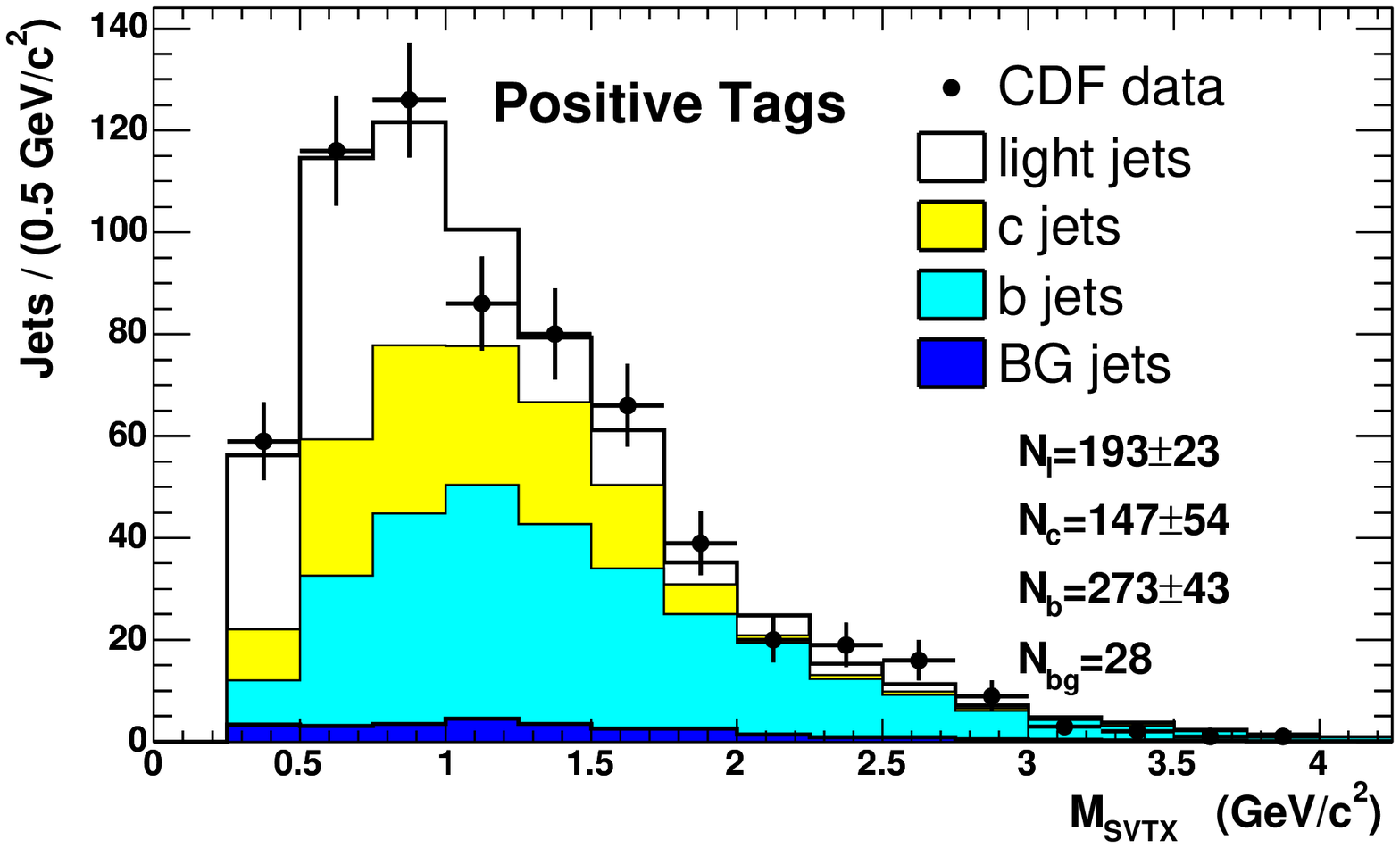} \hfill
\includegraphics[width=0.32\textwidth,height=6cm]{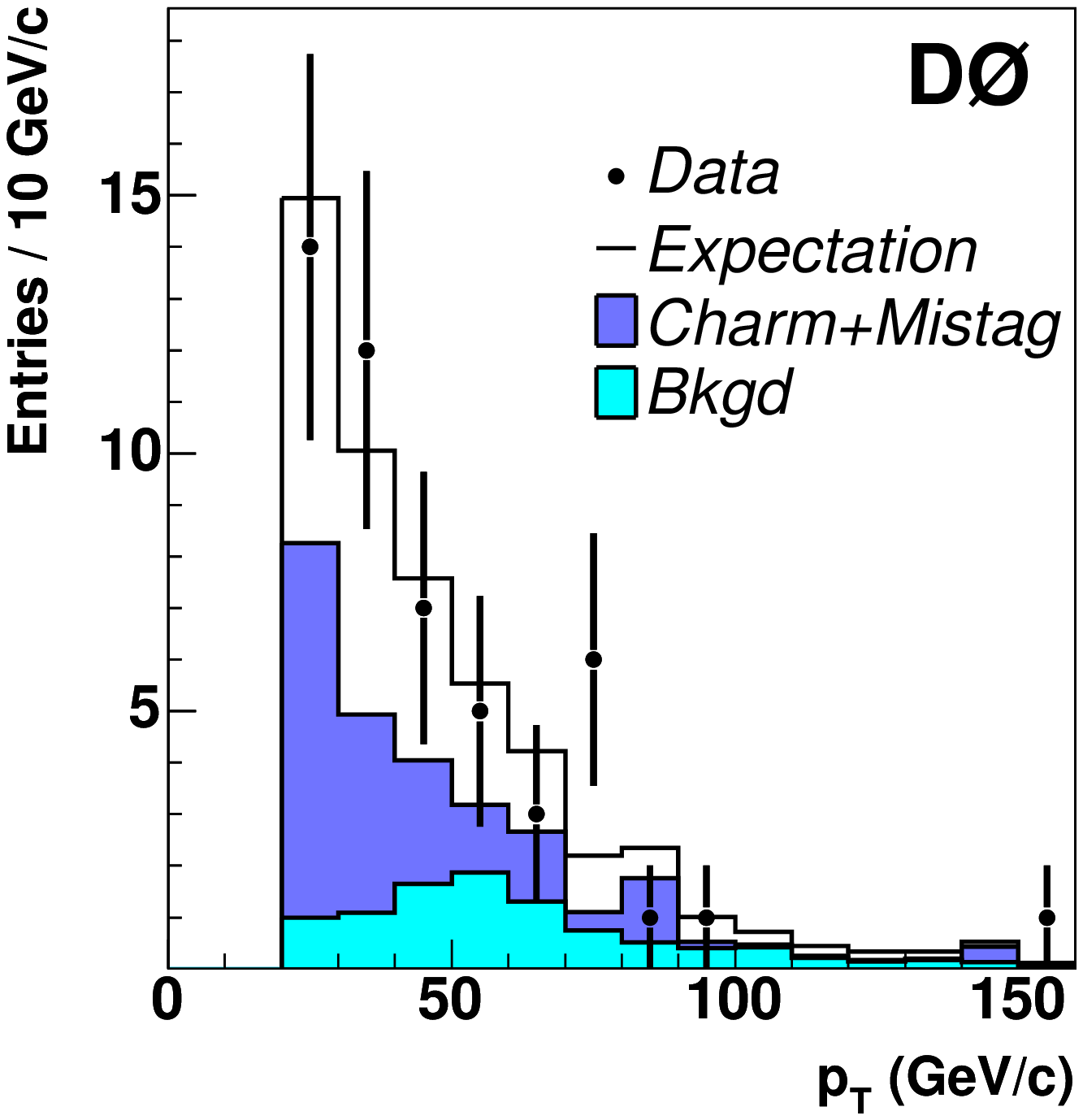}
\vspace*{-0.4cm}
\caption{
Left: CDF $\rm Z\to \bb$ signal extracted in double-b-tagged data,
relevant for $\rm H\to \bb$ searches.
Center: CDF invariant mass of tracks at the secondary vertex for positively tagged jets.
Right: D\O\ $P_t$ distribution for b-tagged jets of Z+jets events.
}
\label{fig:zbb}
 \vspace*{-0.2cm}
\end{figure}

Efficient B hadron tagging has already been demonstrated in data with $\rm Z \to\bb$
events. These measurements contribute to the energy resolution 
and energy scale determinations. 
Figure~\ref{fig:zbb} (left plot from~\cite{cdf-gghbb},
center plot from~\cite{cdf-gghbb_tag} and right plot from~\cite{d0-gghbb}) 
shows the reconstruction of the $\rm Z\to \bb$ mass and 
the good agreement between data and simulation for b-tagged events.

\section{Gluon Fusion $\rm gg\rightarrow H \to WW$}

\begin{figure}[hcbp]
\vspace*{-0.4cm}
\begin{minipage}{0.39\textwidth}
\includegraphics[width=\textwidth]{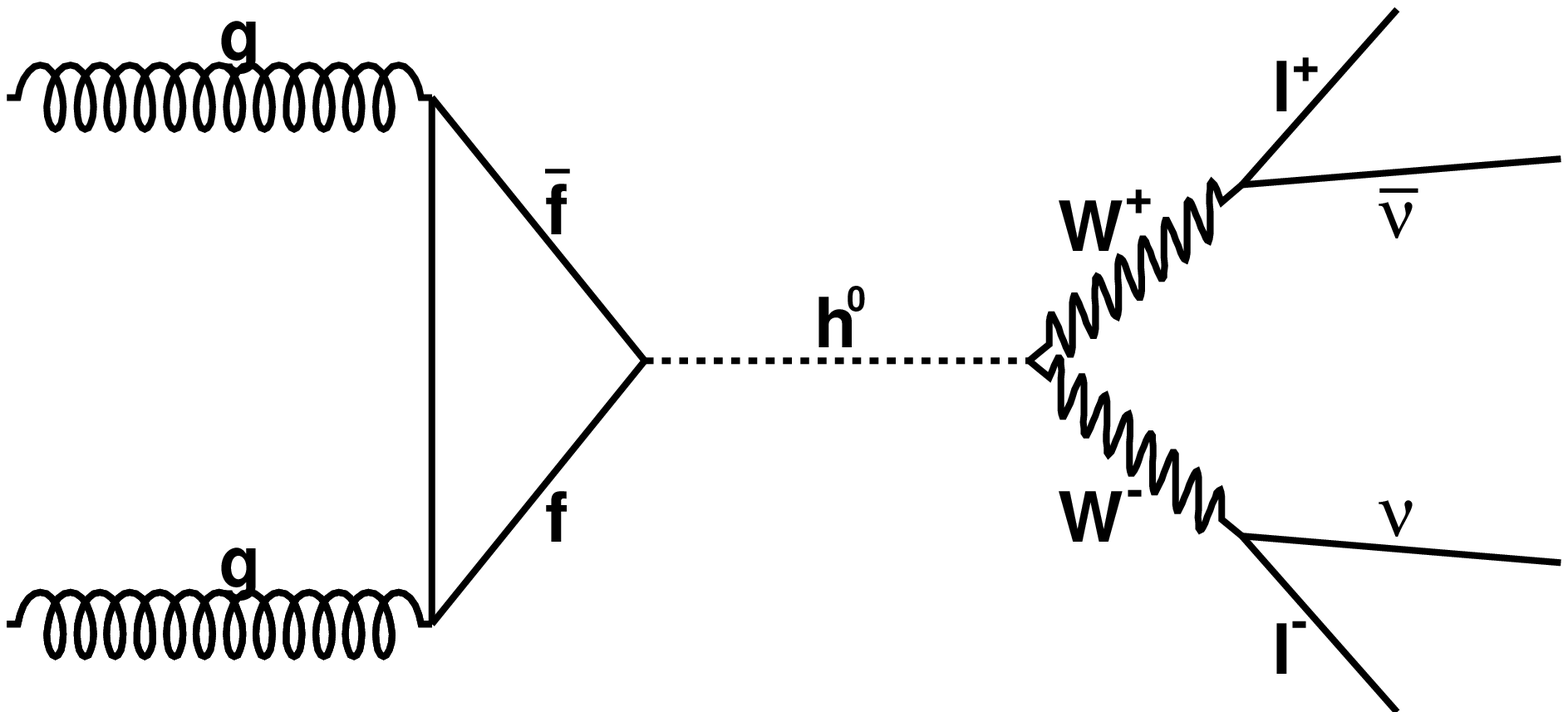} \hfill
\includegraphics[width=\textwidth]{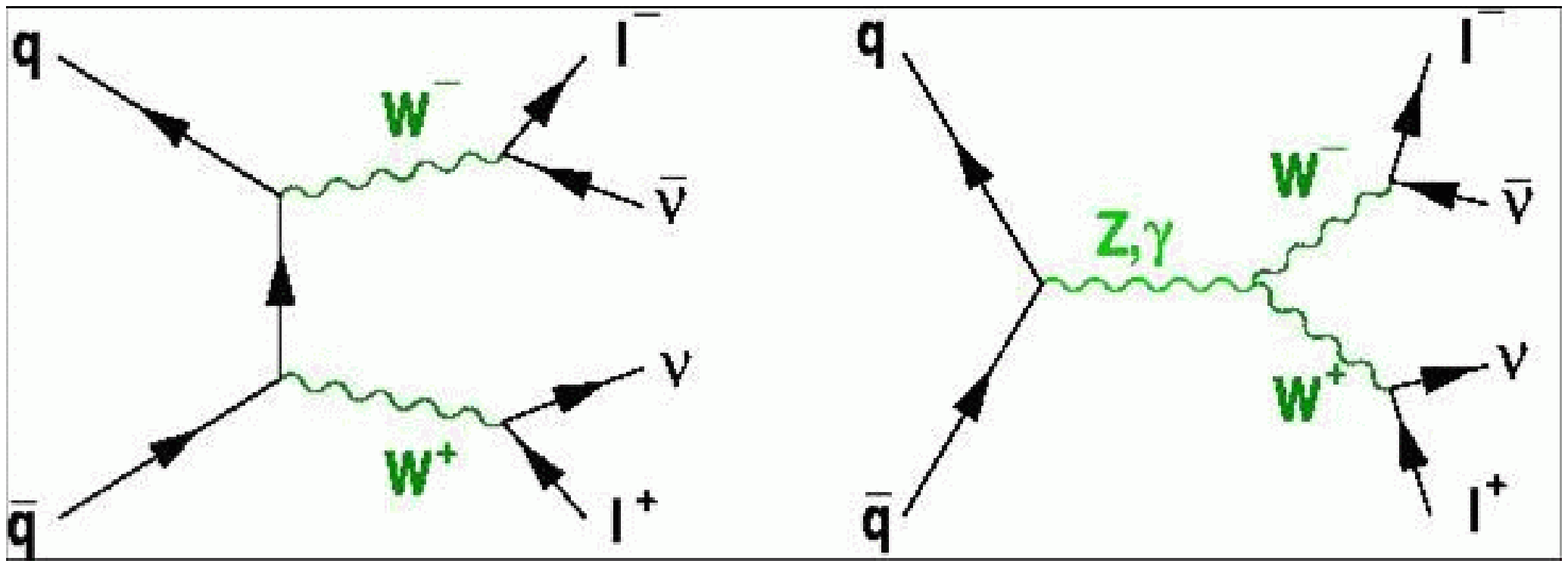}
\end{minipage}\hfill
\begin{minipage}{0.2\textwidth}
\includegraphics[width=\textwidth]{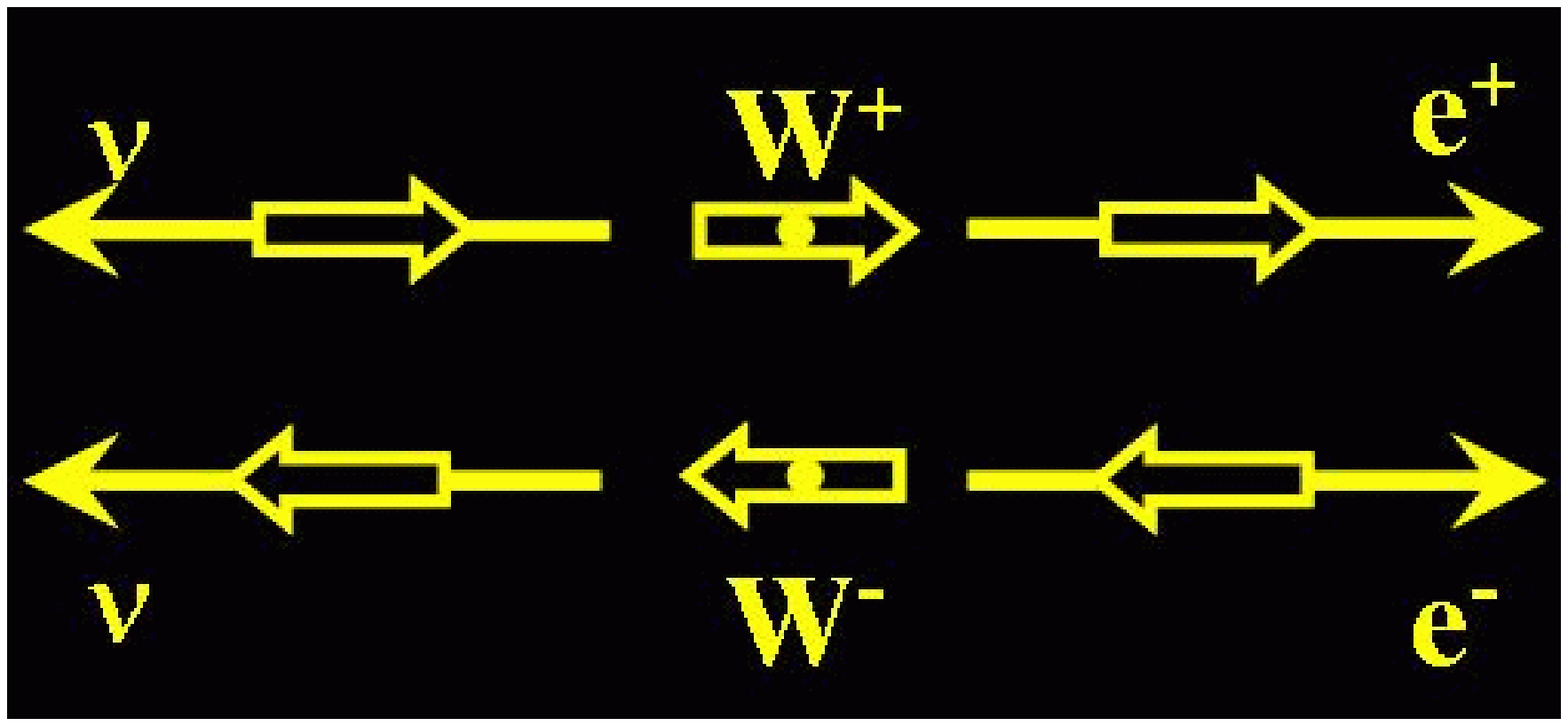}
\end{minipage}
\begin{minipage}{0.39\textwidth}
\includegraphics[width=\textwidth,height=5.8cm]{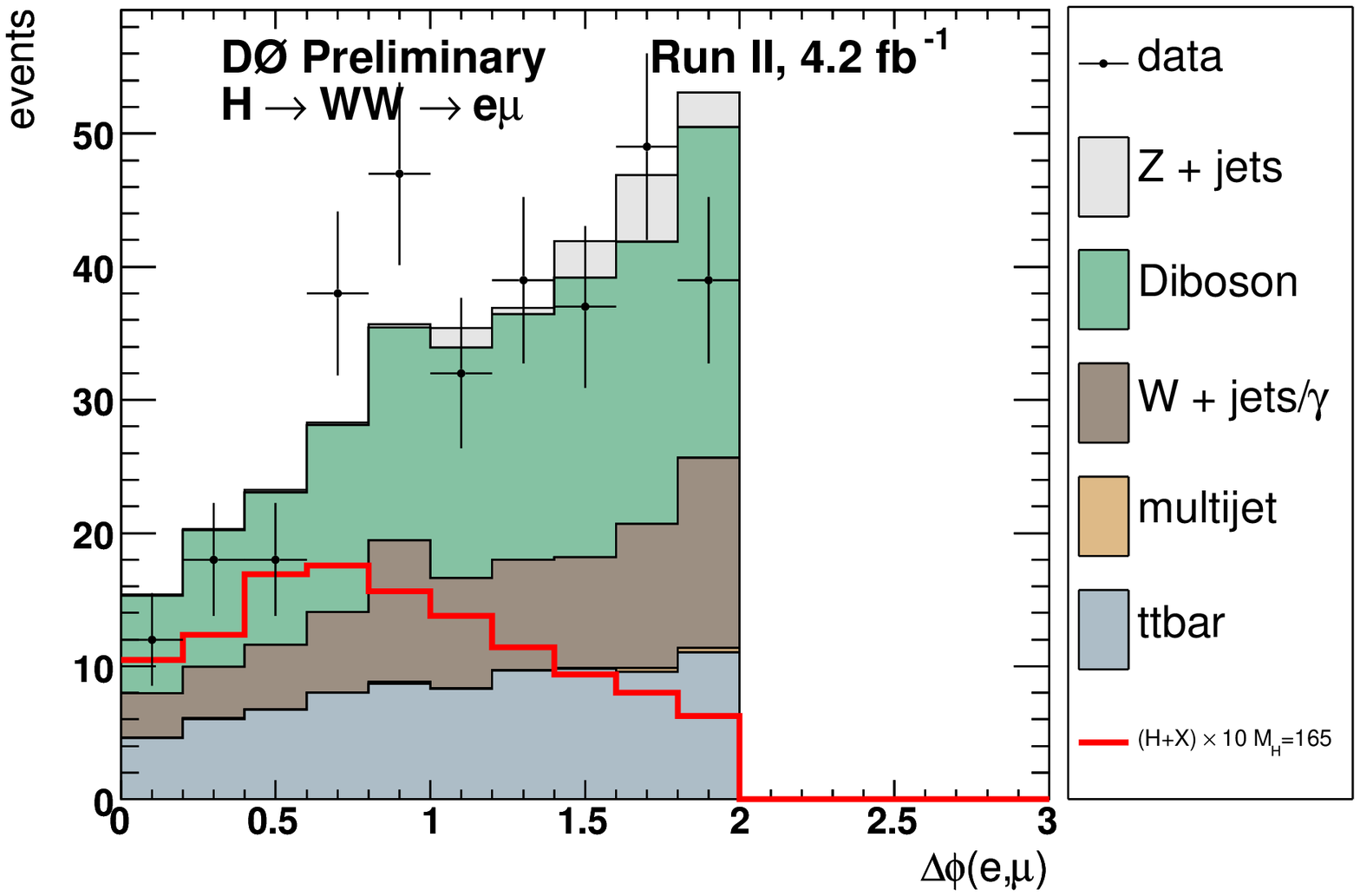}
\end{minipage}
\vspace*{-0.15cm}
\caption{
Left: $\rm gg\rightarrow H$($\rm H\to WW$) signal and background processes.
Center: indication of spin correlations between final state leptons and W pairs, 
        which lead to different dilepton azimuthal angular ($\Delta \Phi_{\rm ll}$)
        distributions for signal and background.
Right: D\O\ $\Delta \Phi_{\rm ll}$ distribution for data, and simulated signal and background. 
$\Delta \Phi_{\rm ll}$ is predicted to be smaller for the signal.
}
\label{fig:d0-ggHww1}
\vspace*{-0.5cm}
\end{figure}

\begin{figure}[hcp]
\includegraphics[width=0.48\textwidth,height=5cm]{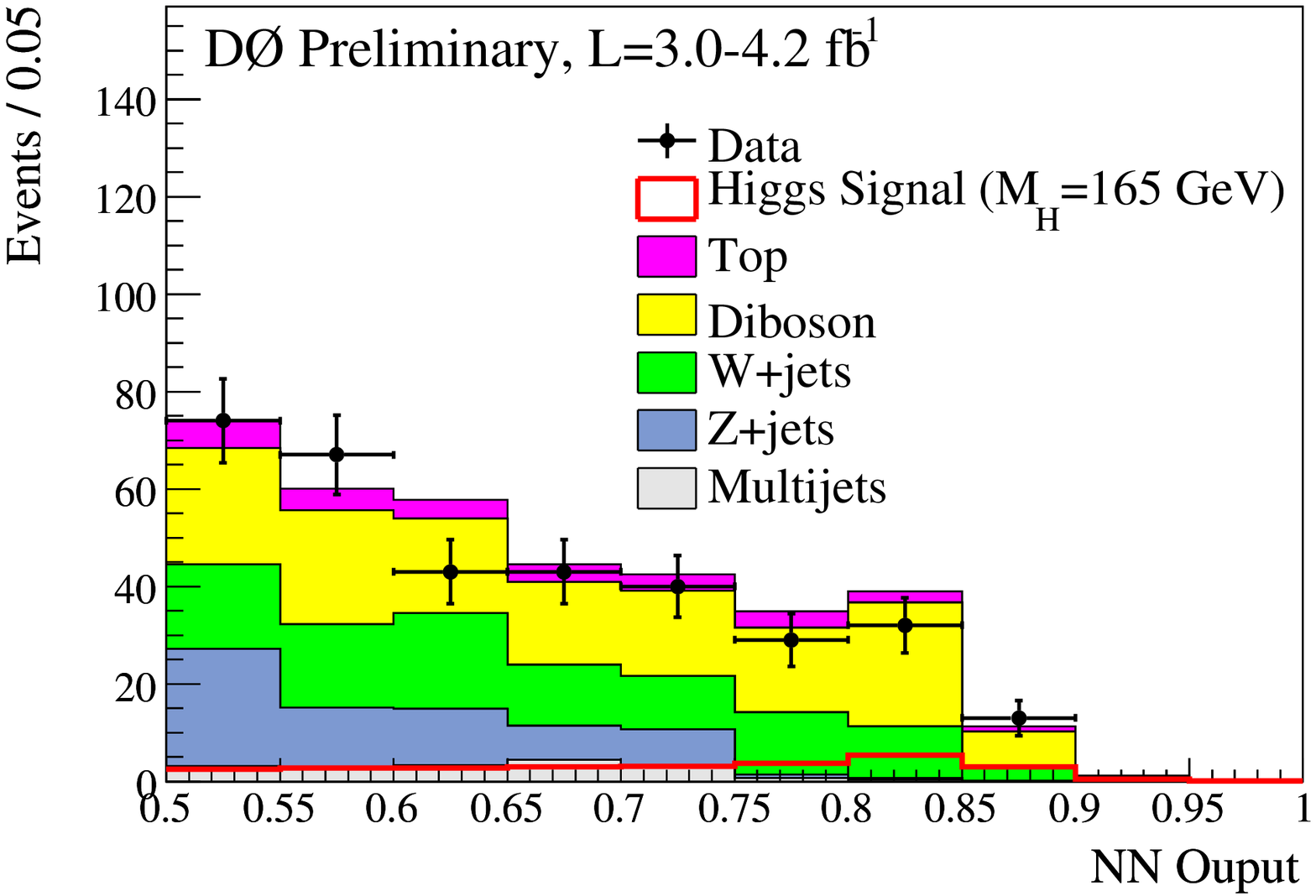}\hfill
\includegraphics[width=0.48\textwidth,height=5cm]{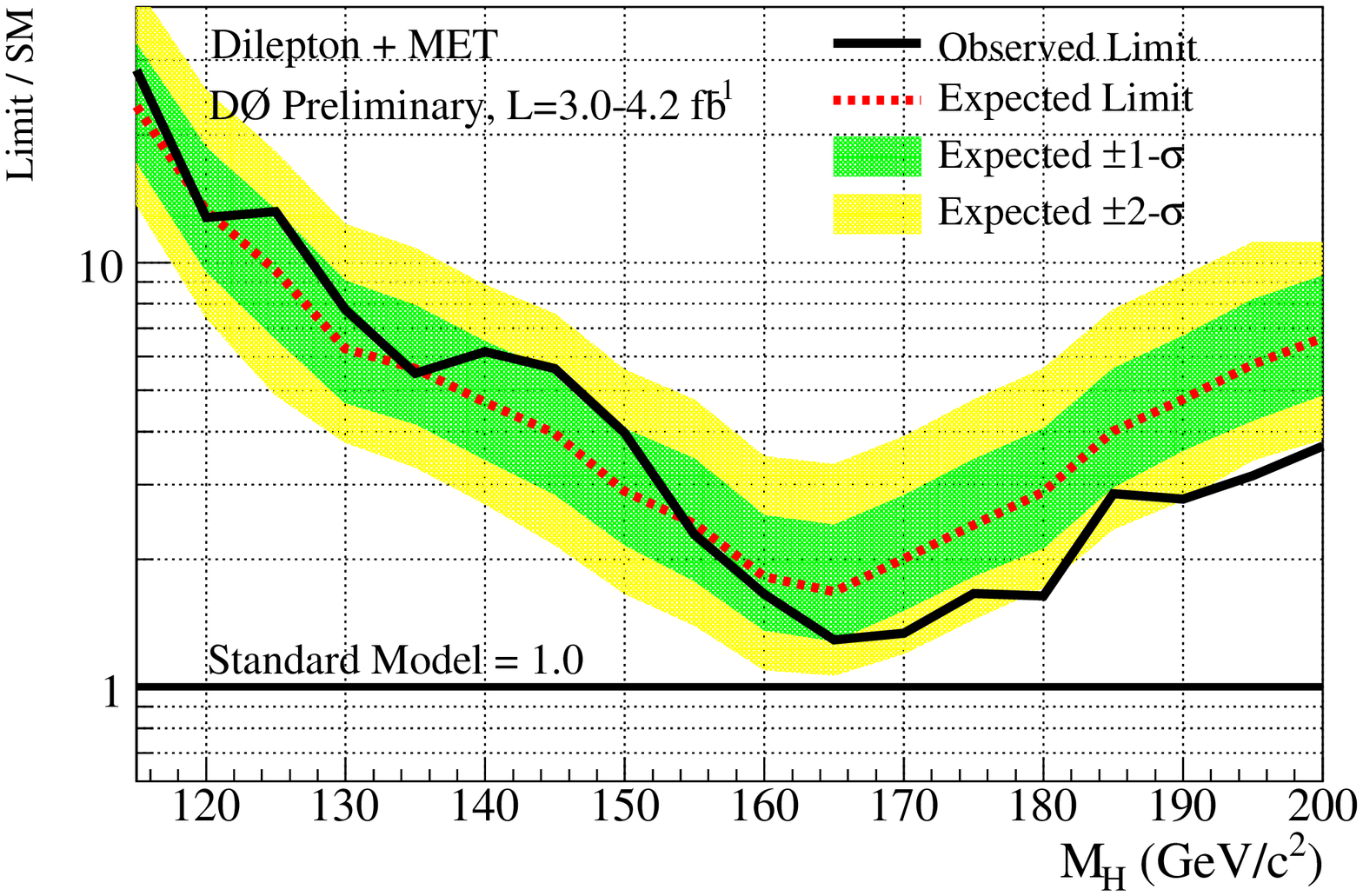}
\vspace*{-0.4cm}
\caption{
D\O\ $\rm gg\rightarrow H$($\rm H\to WW$).
Left: Neural network output.
Right: limit at 95\% CL.
}
\label{fig:d0-ggHww}
\end{figure}

\begin{figure}[hcp]
\includegraphics[width=0.48\textwidth,height=5cm]{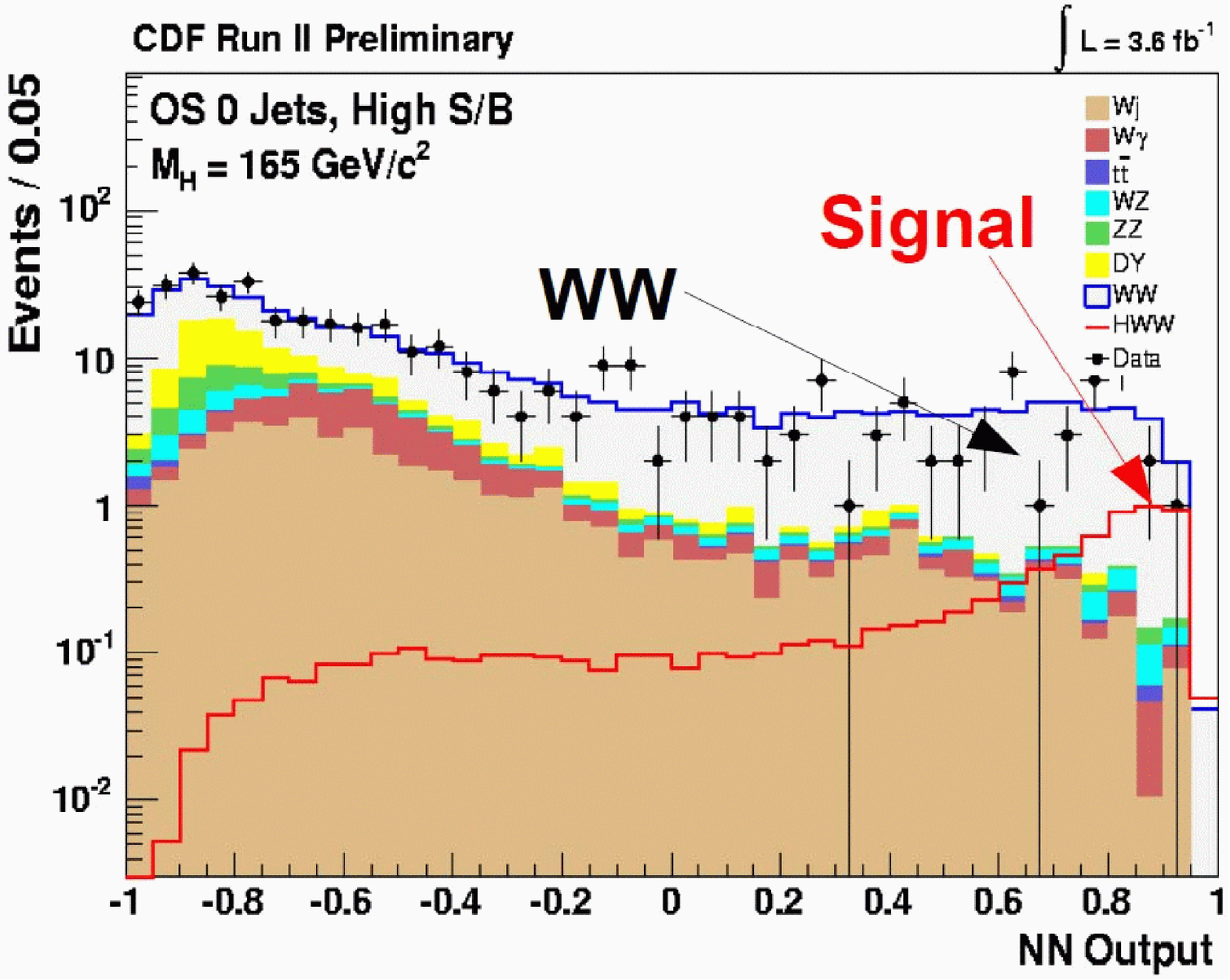}\hfill
\includegraphics[width=0.48\textwidth,height=5cm]{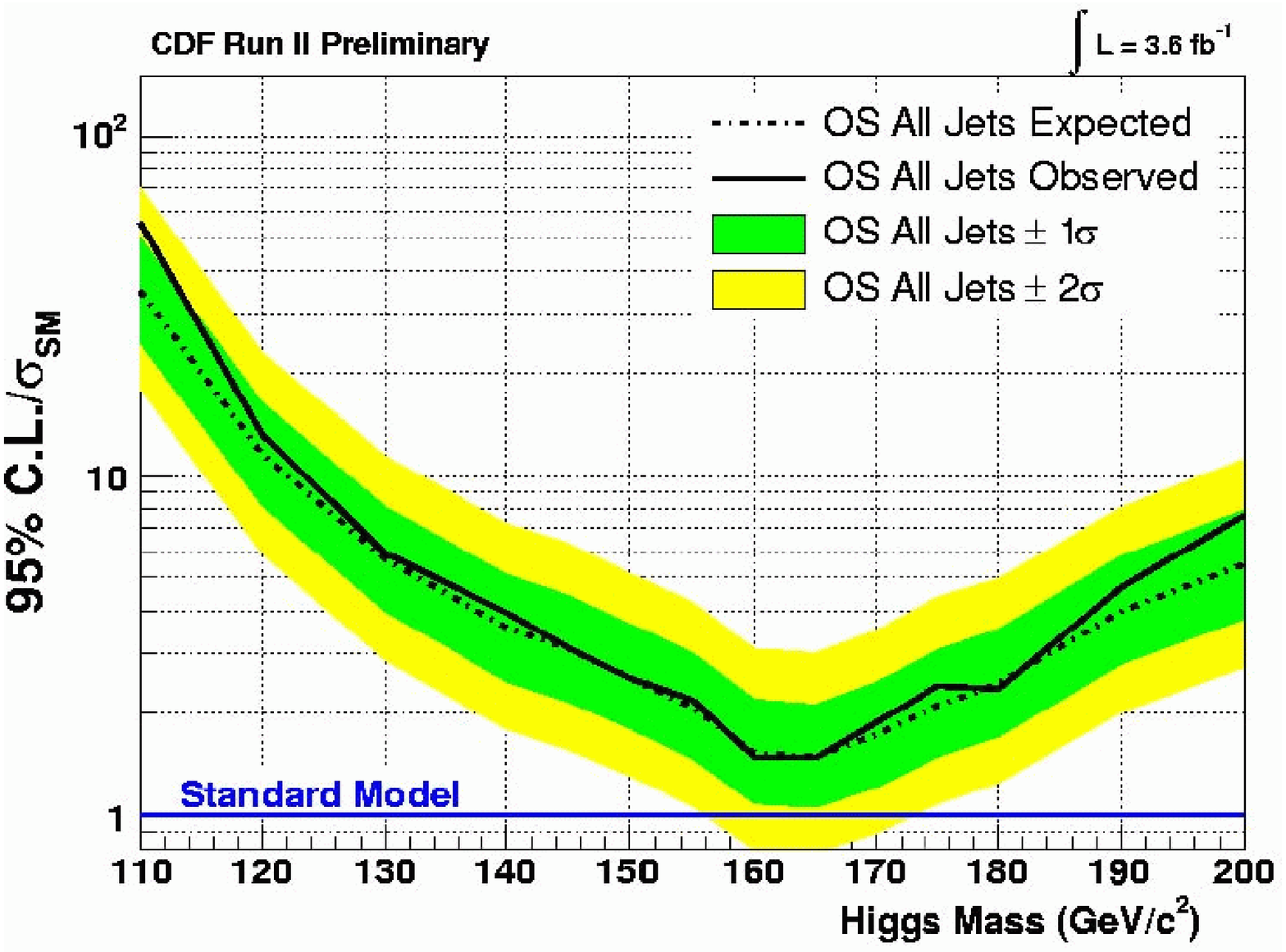}
\vspace*{-0.4cm}
\caption{CDF $\rm gg\rightarrow H$($\rm H\to WW$).
Left: Neural network output.
Right: limit at 95\% CL.
}
\label{fig:cdf-ggHww}
\end{figure}

For Higgs boson masses above about 135 GeV the process
$\rm gg\rightarrow H \to WW$ becomes important.
The production and decay process is illustrated in Fig.~\ref{fig:d0-ggHww1}, also shown
is a background process leading to the same final state particles. The spin information allows 
separation of signal and background. 
The angle between the opposite charged leptons $\Delta\Phi_{\rm ll}$
tends to be smaller for the signal than for the background as shown in 
Fig.~\ref{fig:d0-ggHww1} (from~\cite{d0-ggHww}). Based on 3.0-4.2~fb$^{-1}$
total luminosity the neural network output for $\rm gg\rightarrow H \to WW$
process and limits are shown in Fig.~\ref{fig:d0-ggHww} (from~\cite{d0-ggHww})
and based on 3.6~fb$^{-1}$ in
Fig.~\ref{fig:cdf-ggHww} (from~\cite{cdf-ggHww}).
Owing to the overwhelming $\rm b \bar b$ background, 
the $\rm gg\rightarrow H$($ \rm H\to \bb$) 
channel is not feasible at the Tevatron.

\section{Associated Production}

\subsection{$\rm WH( H\to \bb)$}

\begin{figure}[hbcp]
\vspace*{1cm}
\includegraphics[width=0.32\textwidth,height=5cm]{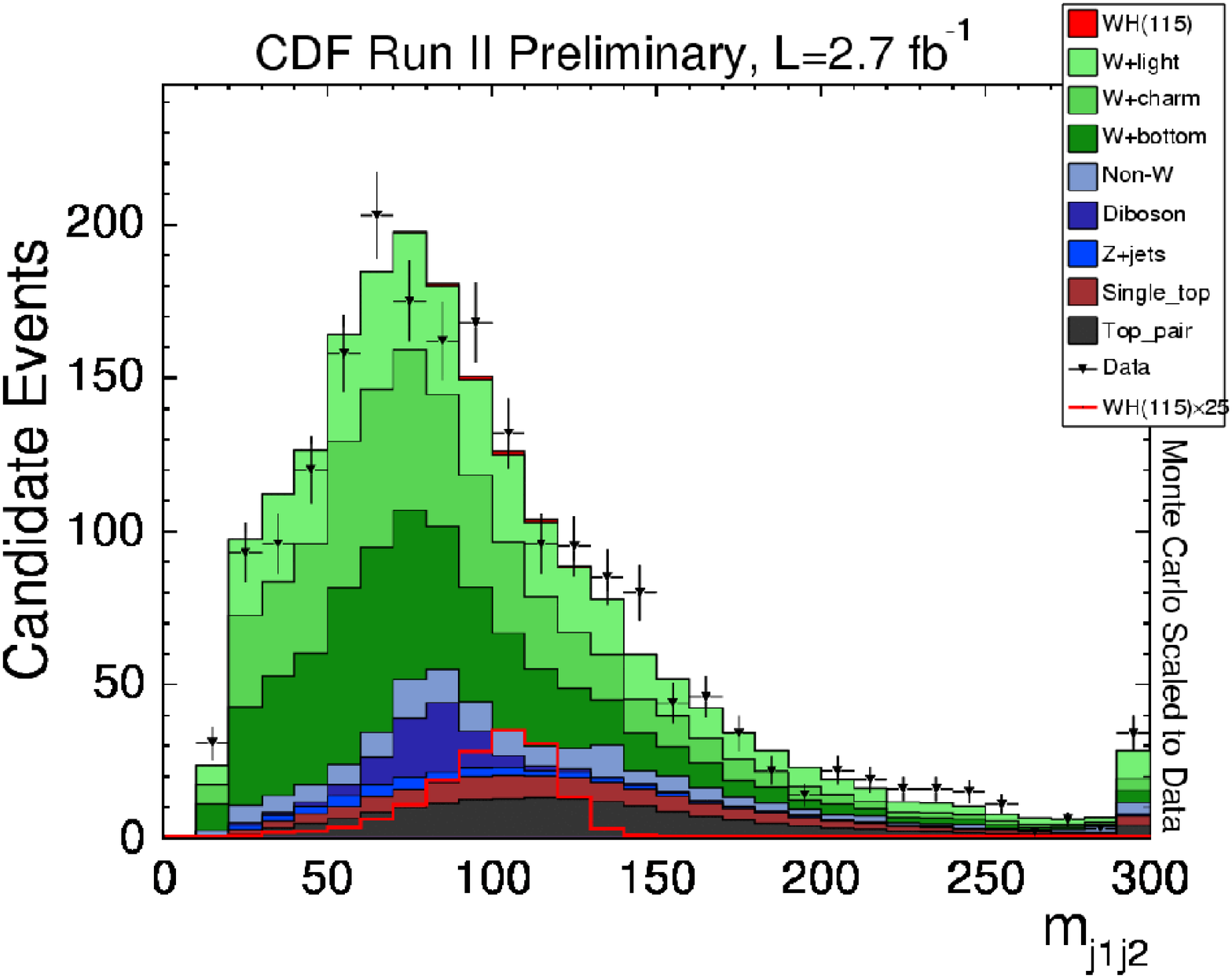} \hfill
\includegraphics[width=0.32\textwidth,height=5cm]{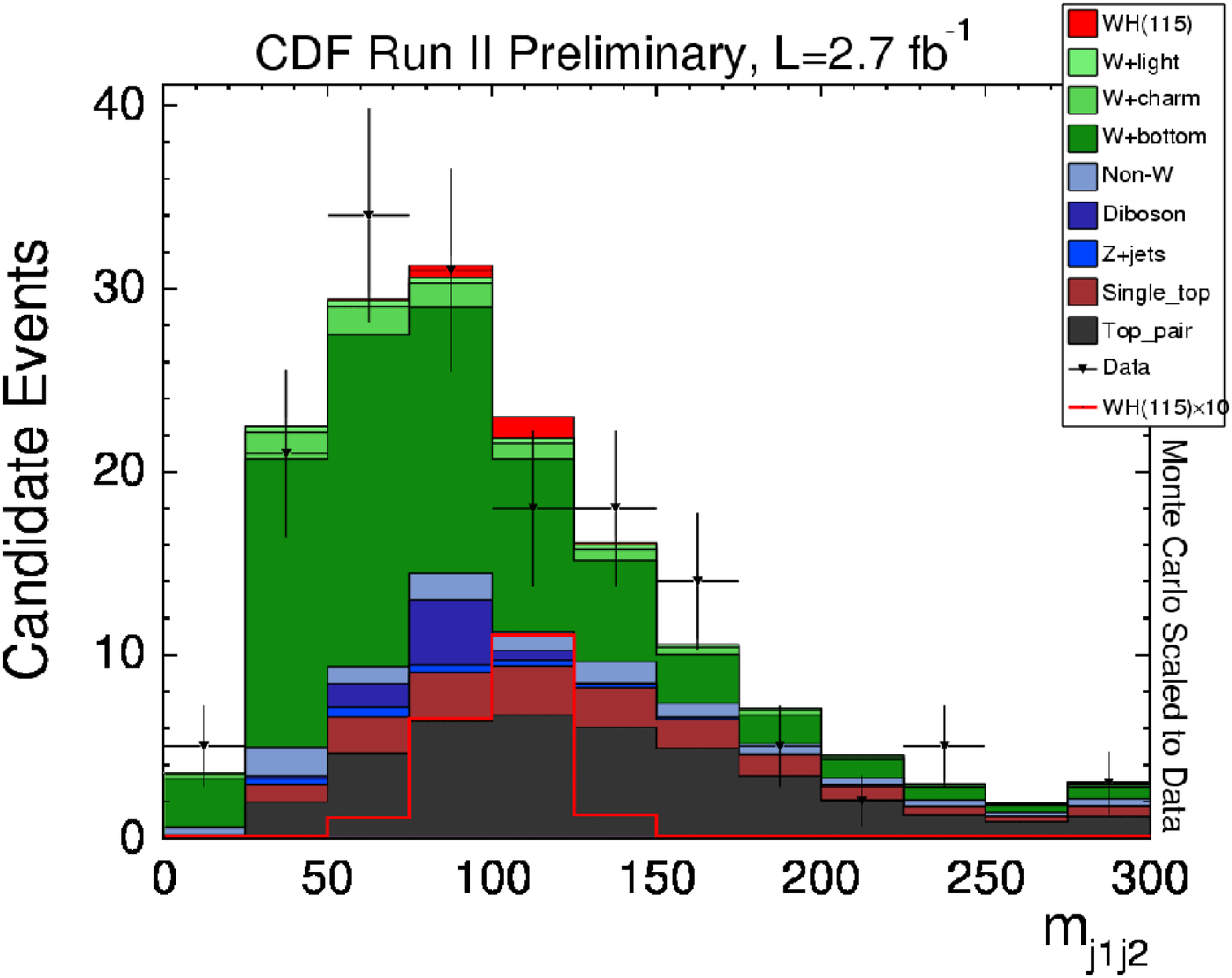} \hfill
\includegraphics[width=0.32\textwidth,height=5cm]{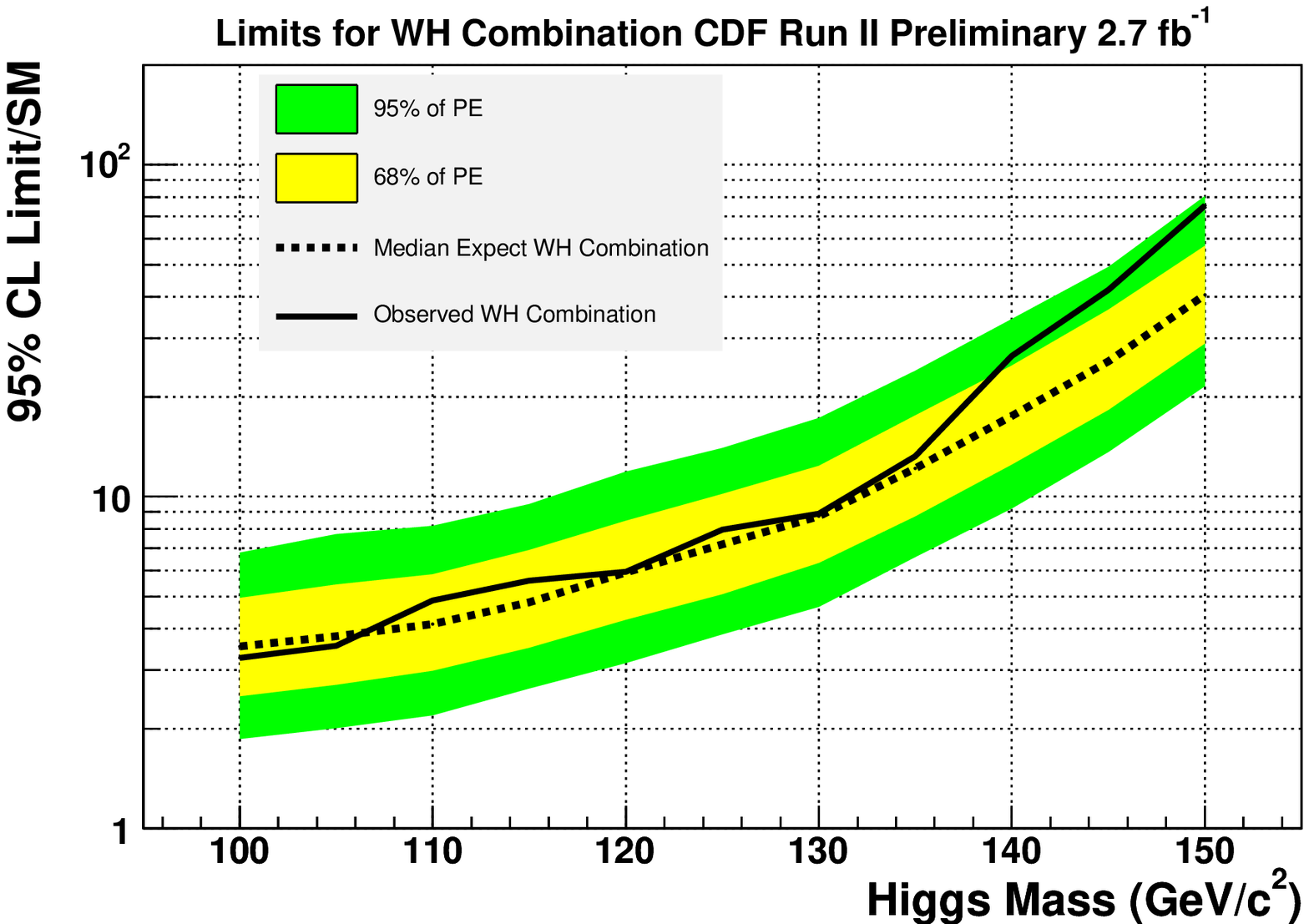}
\vspace*{-0.4cm}
\caption{
CDF WH($\rm H\to \bb$). Matrix Element and Boosted Decision Tree Techniques (ME+BDT).
Left: single b-tagging.
Center: double b-tagging.
Right: limit at 95\% CL from combined ME+BDT and neural network analysis results.
}
\label{fig:cdf-wHbb}
\end{figure}

\begin{figure}[hbcp]
\includegraphics[width=0.32\textwidth,height=5cm]{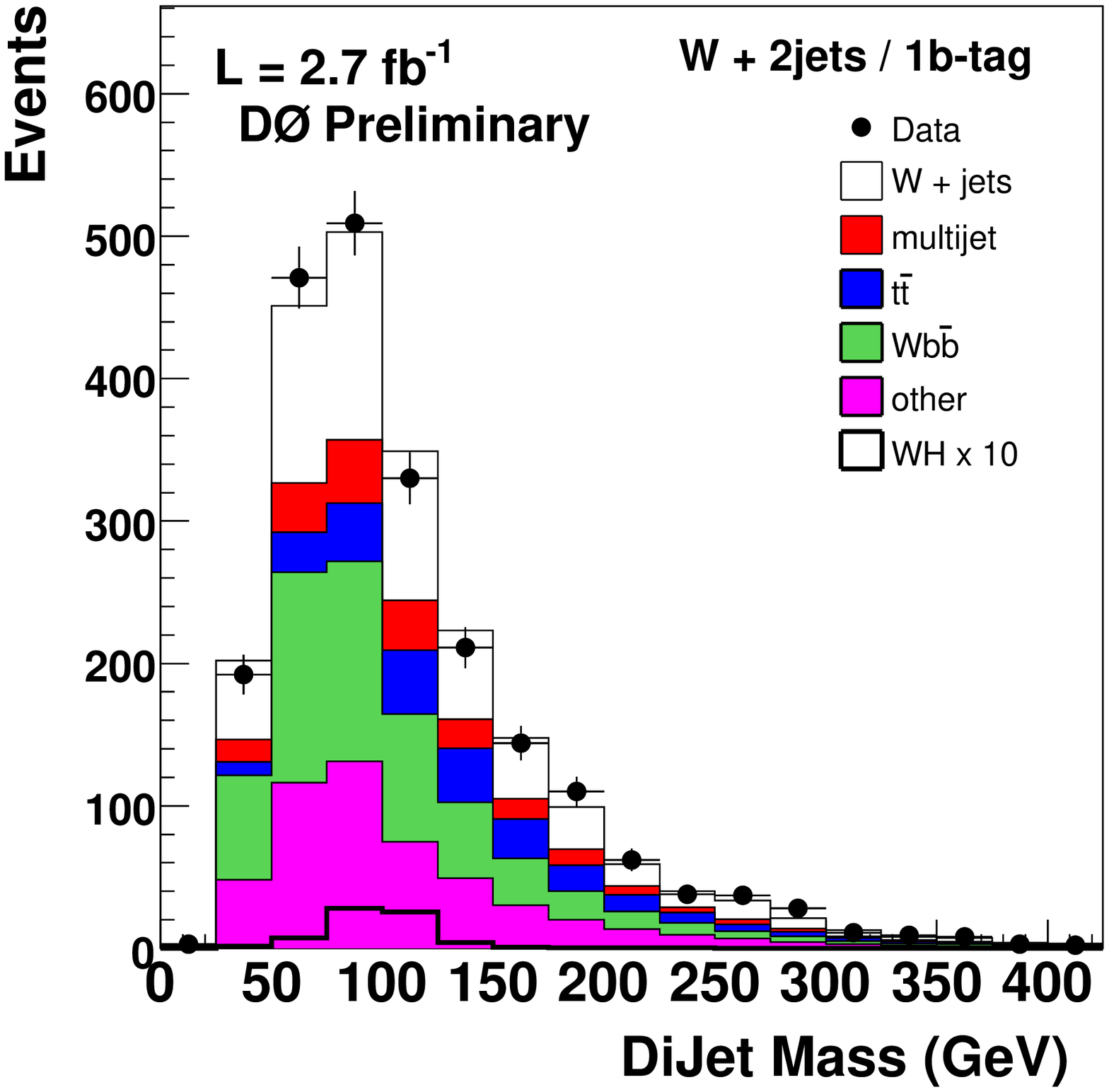} \hfill
\includegraphics[width=0.32\textwidth,height=5cm]{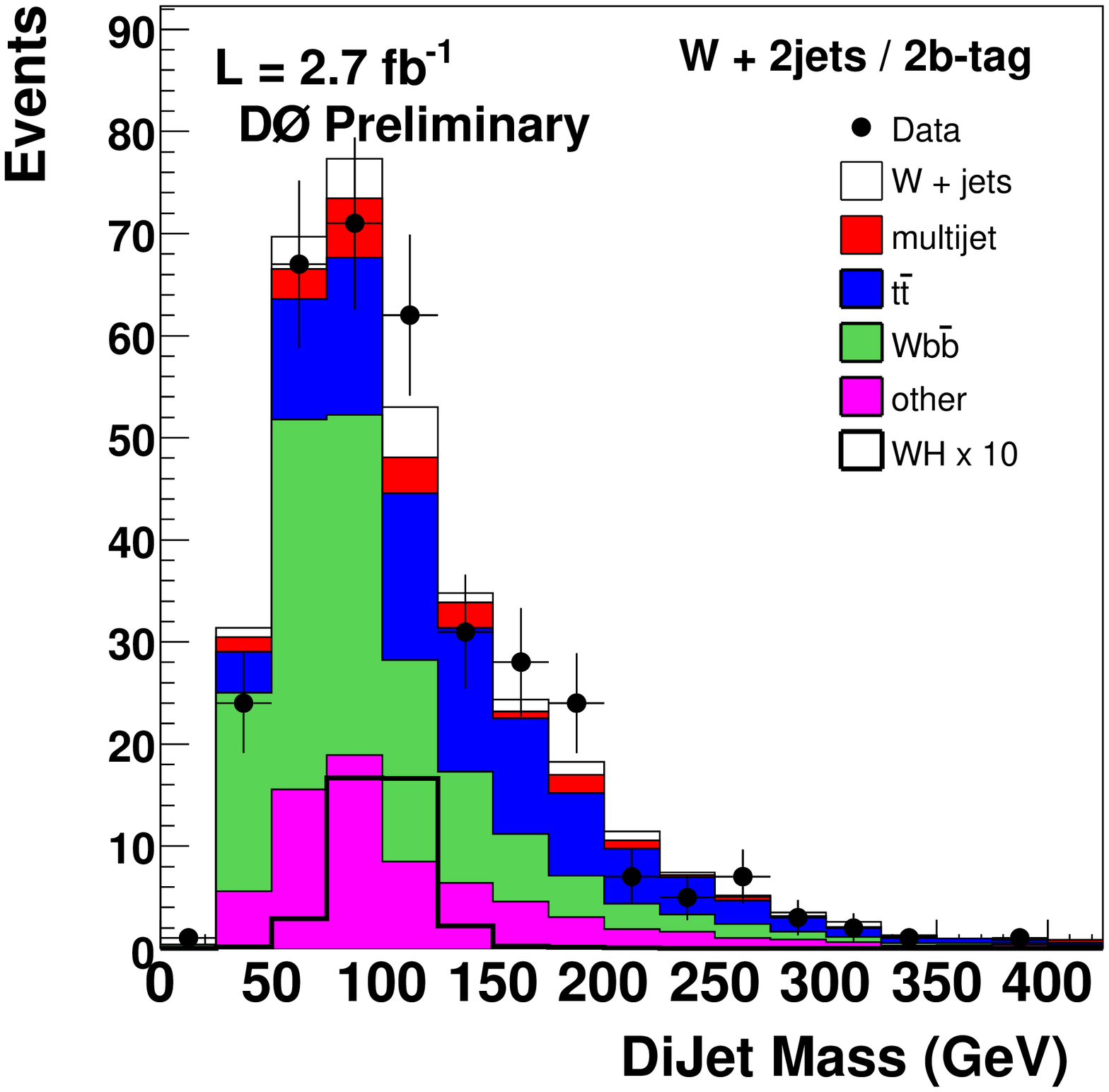} \hfill
\includegraphics[width=0.32\textwidth,height=5cm]{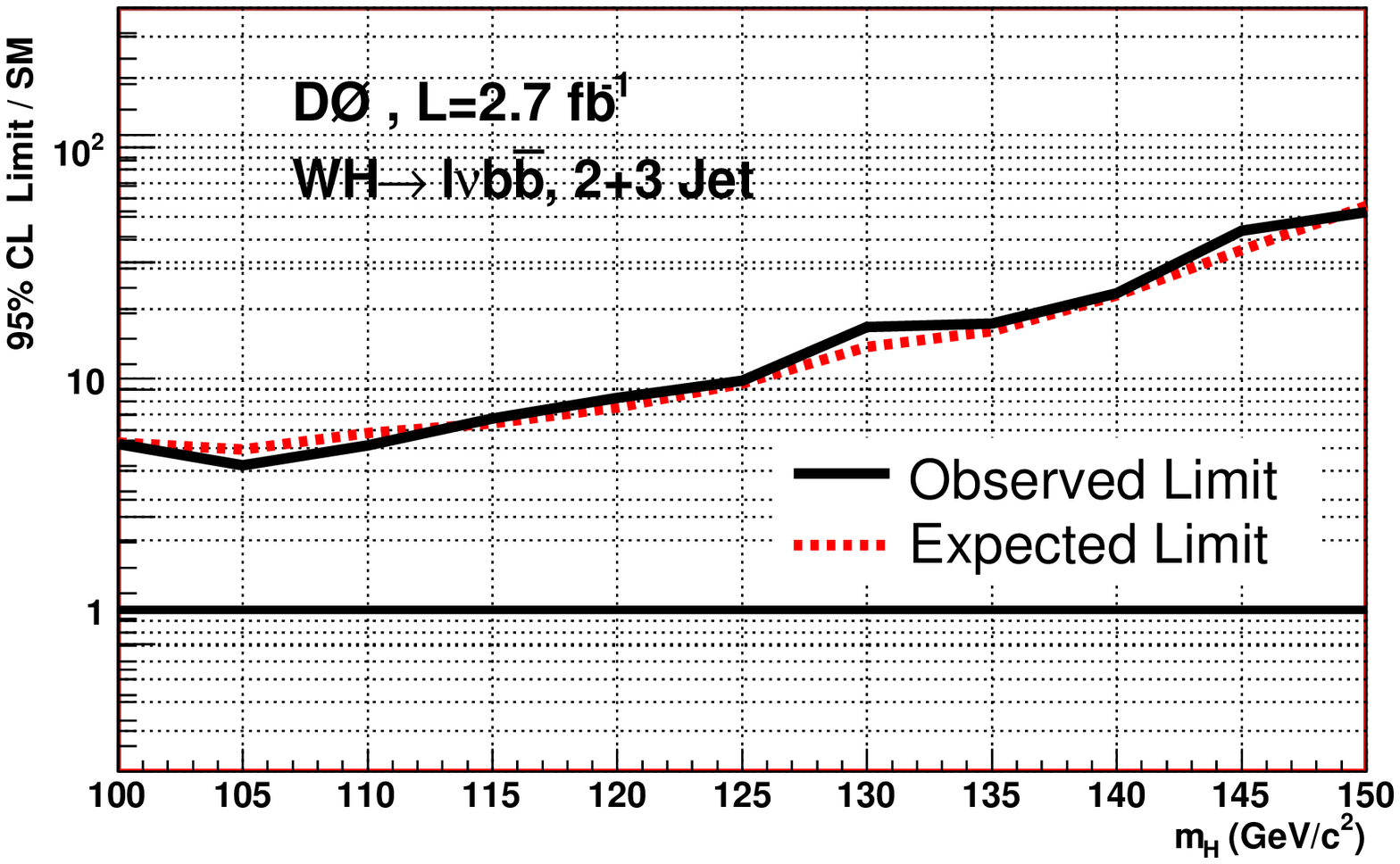}
\vspace*{-0.4cm}
\caption{D\O\ WH($\rm H\to \bb$).
Left: single b-tagging.
Center: double b-tagging.
Right: limit at 95\% CL.}
\label{fig:d0-wHbb}
\end{figure}

An important discovery channel is the reaction WH($\rm H\to \bb$), where the
W decays either to $\rm e\nu$ or $\rm \mu\nu$. The tagging of two b-quarks
improves the signal to background ratio as shown in 
Fig.~\ref{fig:cdf-wHbb} (from~\cite{cdf-WHbb})
and 
Fig.~\ref{fig:d0-wHbb} (from~\cite{d0-WHbb}).

\clearpage
\subsection{$\rm WH(H\to WW)$}

\begin{figure}[hbtp]
\vspace*{-0.2cm}
\begin{center}
\includegraphics[width=0.32\textwidth,height=5cm]{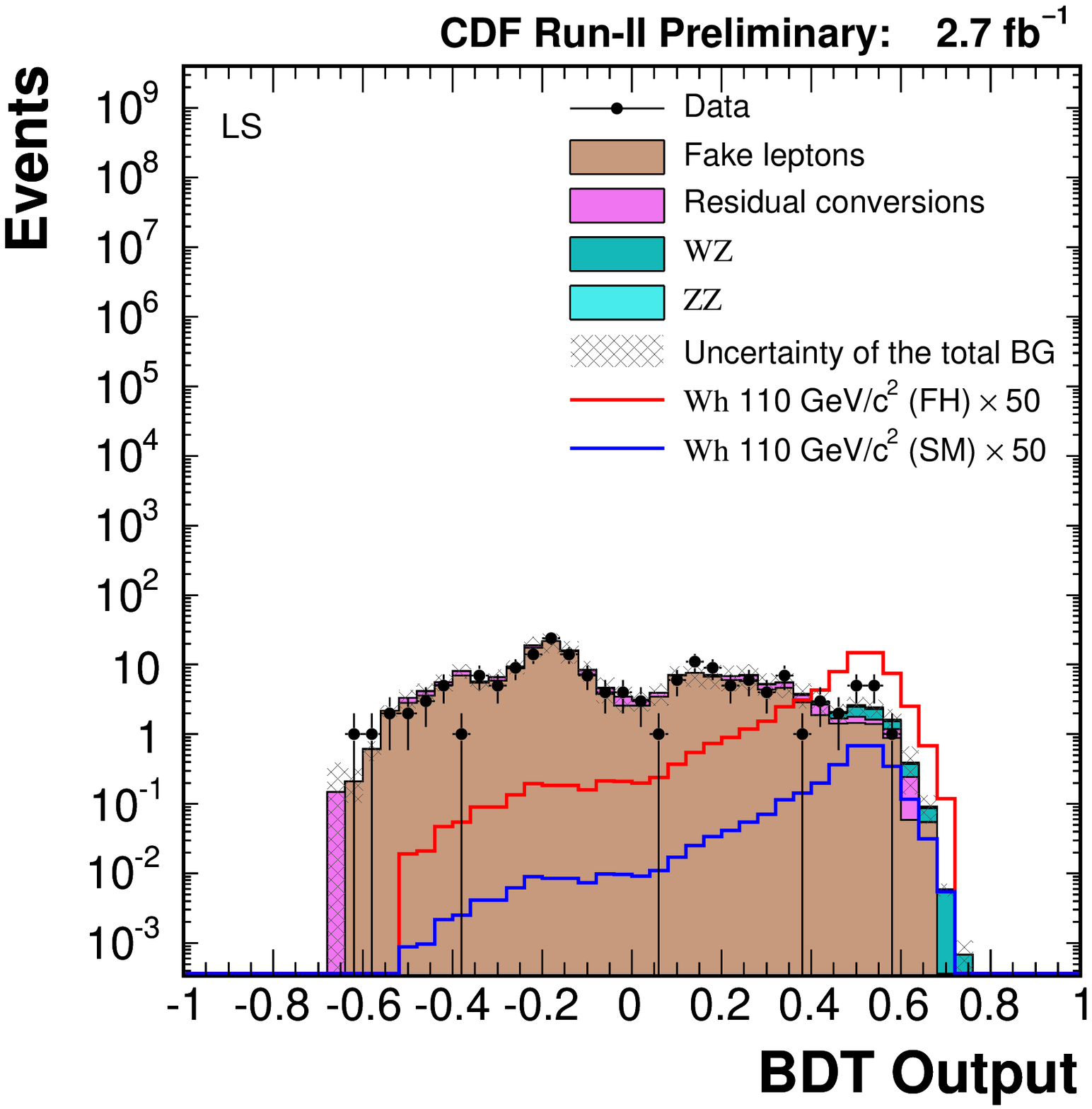} \hfill
\includegraphics[width=0.32\textwidth,height=5cm]{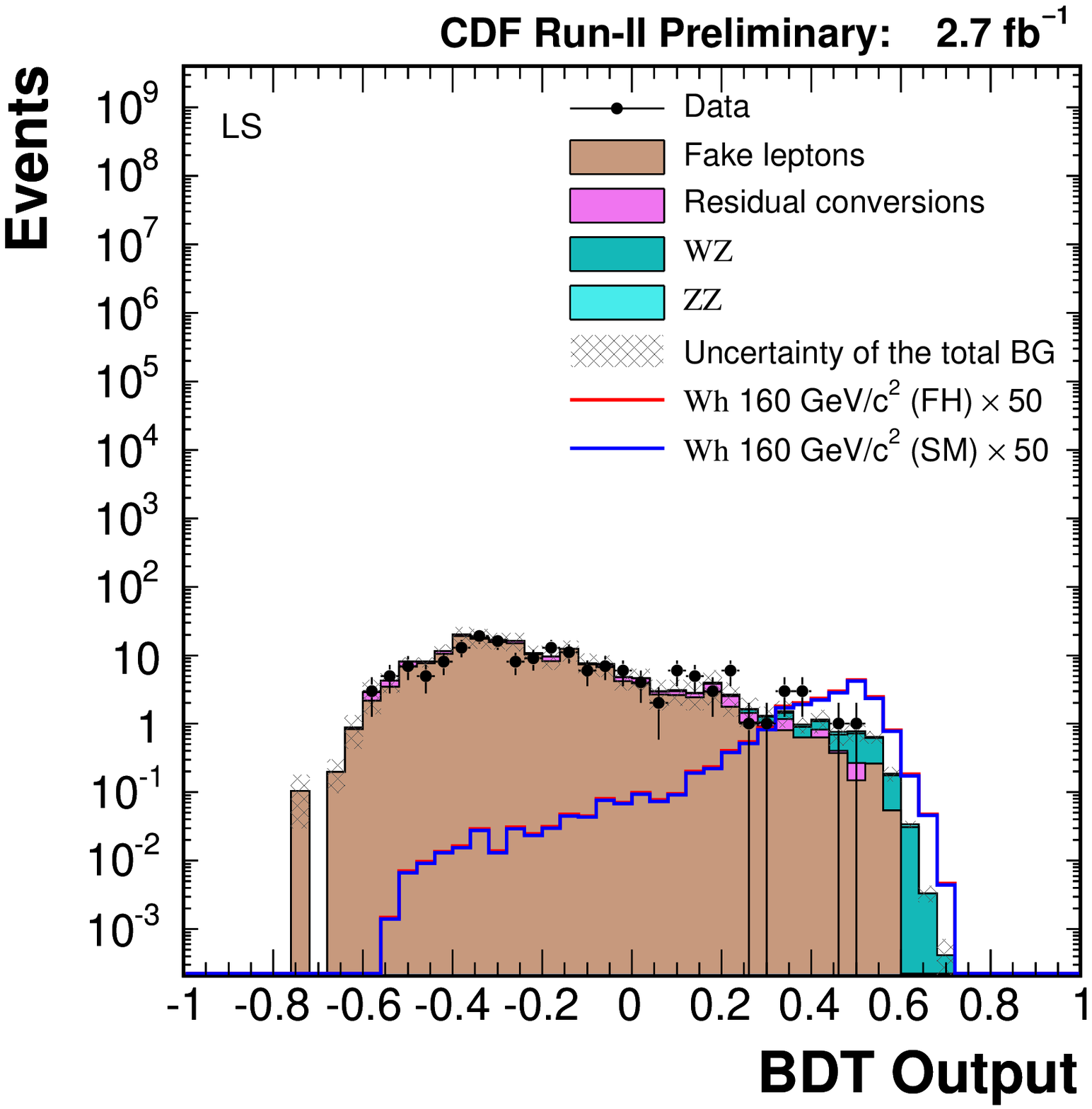} \hfill
\includegraphics[width=0.32\textwidth,height=5cm]{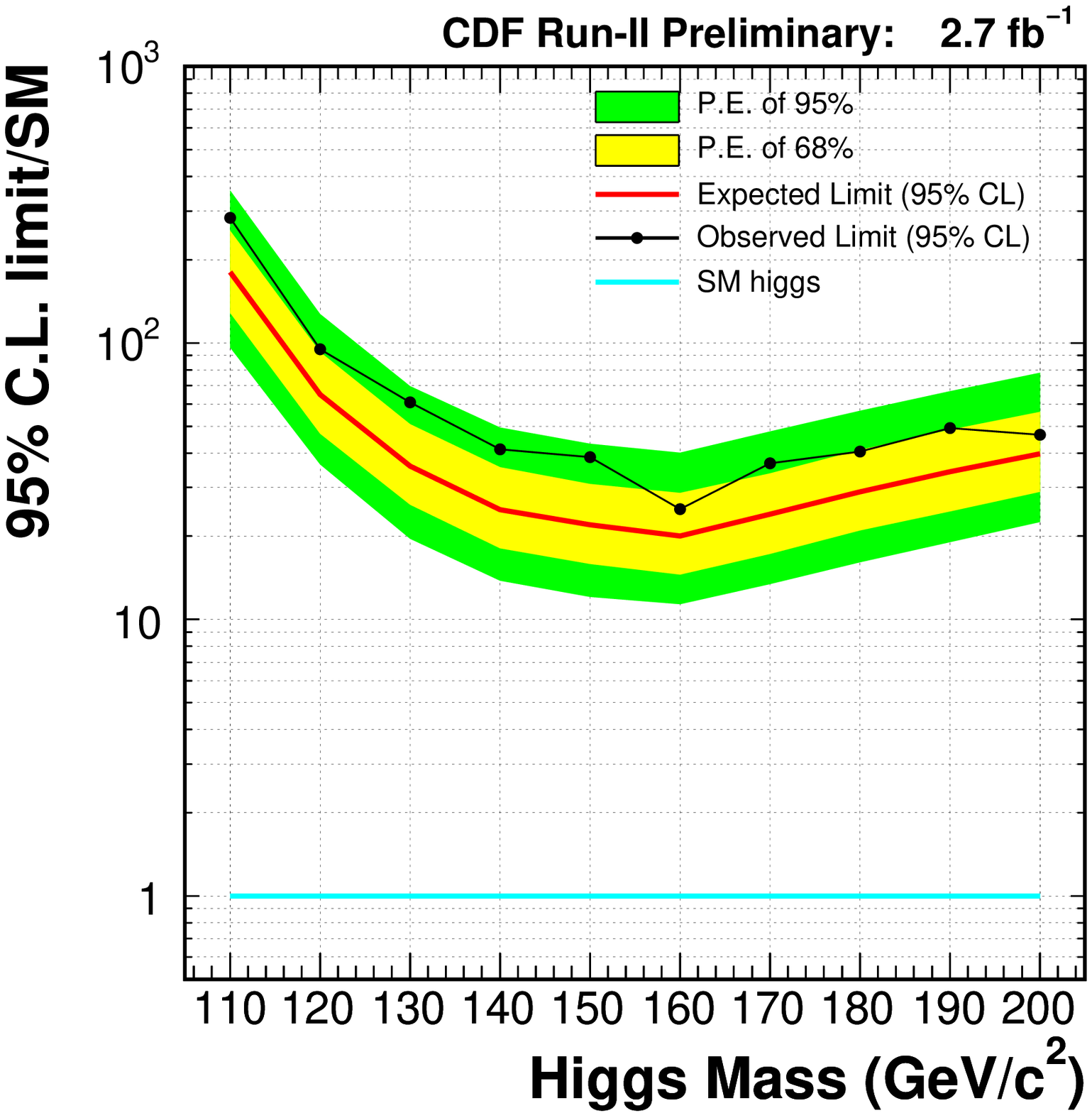}
\end{center}
\vspace*{-0.5cm}
\caption{CDF WH($\rm H\to WW$).
Left: comparison of simulated background and observed number of events 
(110 GeV Higgs boson selection) for the Boosted Decision Tree (BDT) output.
Center: comparison of simulated background and observed number of events 
(160 GeV Higgs boson selection).
Right: limit at 95\% CL.
}
\label{fig:cdf-wHww}
\vspace*{-0.3cm}
\end{figure}

\begin{figure}[hp]
\vspace*{0.2cm}
\includegraphics[width=0.32\textwidth,height=5cm]{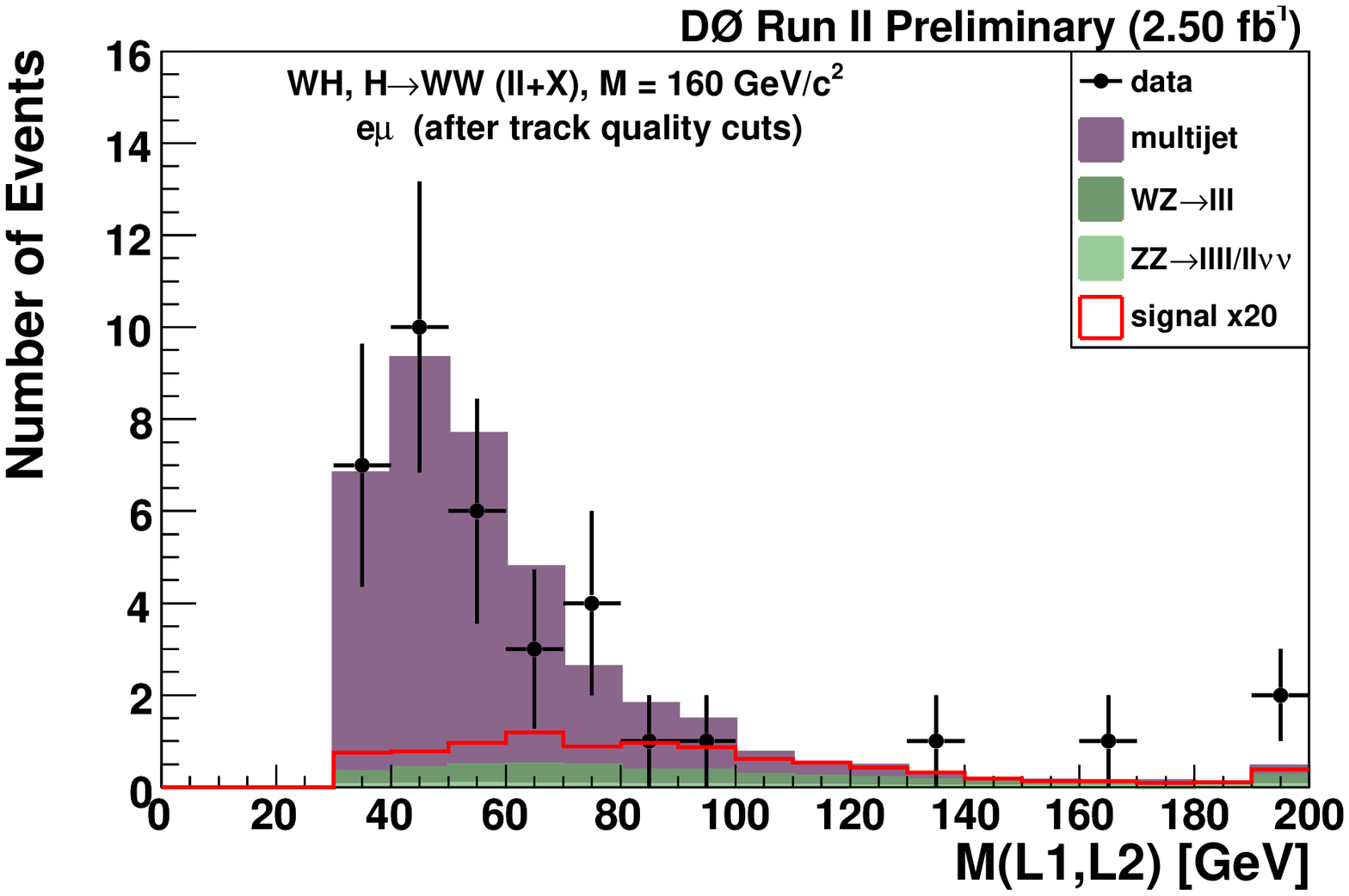} \hfill
\includegraphics[width=0.32\textwidth,height=5cm]{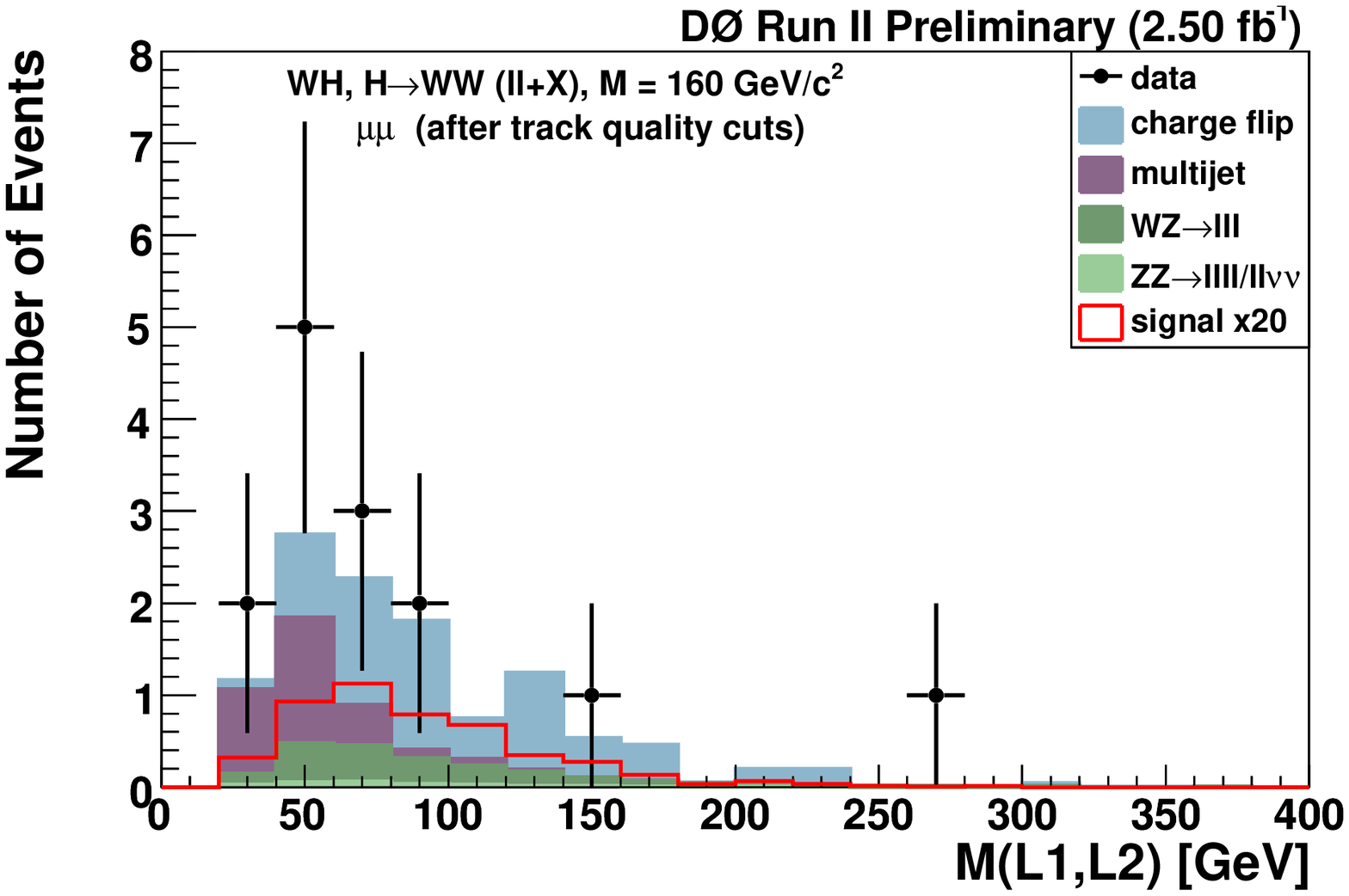} \hfill
\includegraphics[width=0.32\textwidth,height=5cm]{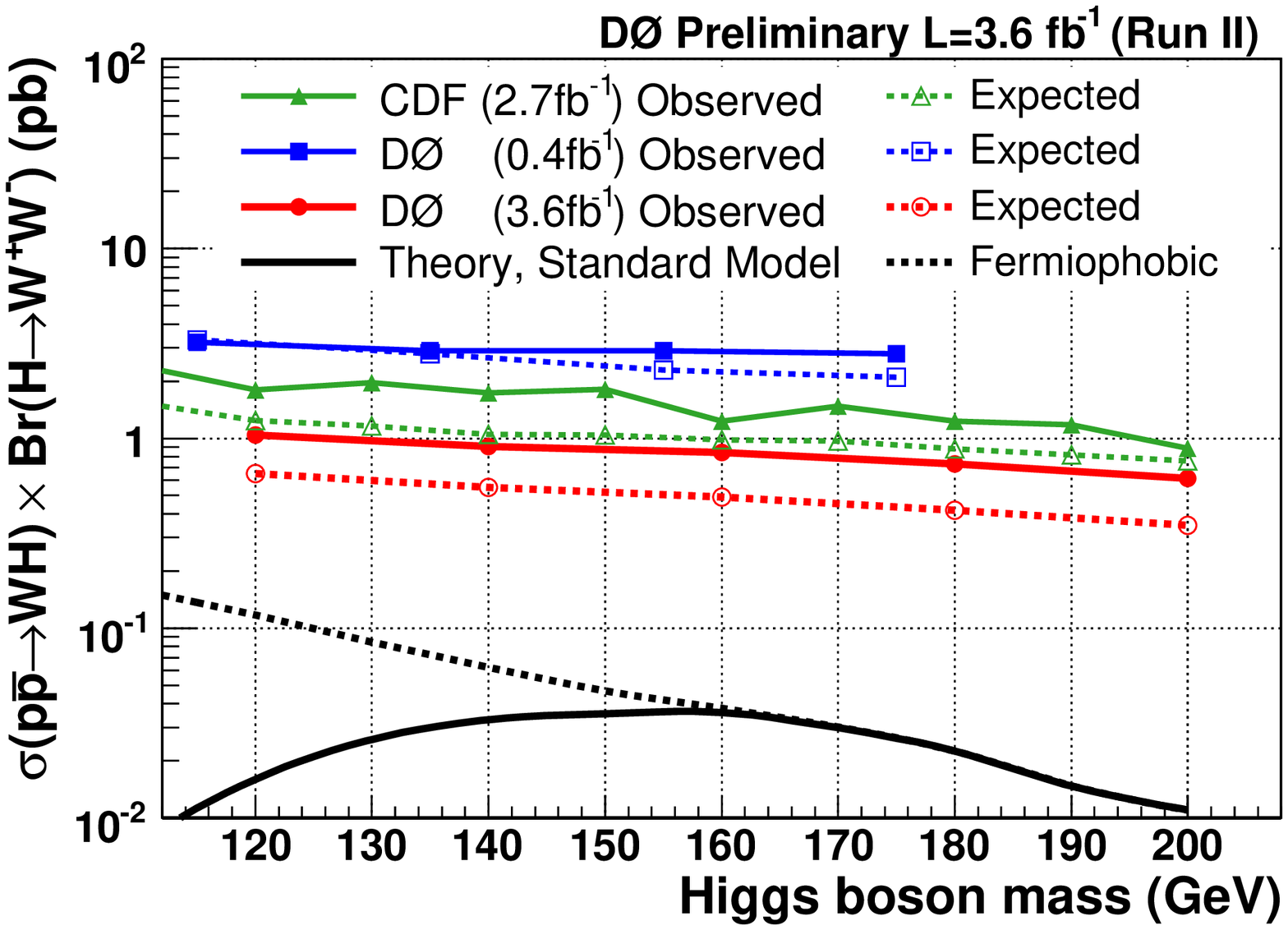} 
\vspace*{-0.4cm}
\caption{D\O\ WH($\rm H\to WW$).
Left: e$\mu$ invariant mass.
Center: $\mu\mu$ invariant mass.
Right: limit at 95\% CL.
}
\label{fig:d0-wHww}
\vspace*{-0.2cm}
\end{figure}

Recent results for the search WH($\rm H\to WW$) in the like-signed charged lepton final state
are shown in 
Fig.~\ref{fig:cdf-wHww} (from~\cite{cdf-WHww}, 2.7~fb$^{-1}$ luminosity) and
Fig.~\ref{fig:d0-wHww} (from~\cite{d0-WHww}, 3.6~fb$^{-1}$ luminosity),
which lead to weaker sensitivity as in the $\rm H\to \bb$ decay mode.

\subsection{$\rm ZH\to \ell\ell\bb$}

\begin{figure}[hp]
\vspace*{-0.4cm}
\includegraphics[width=0.32\textwidth,height=5cm]{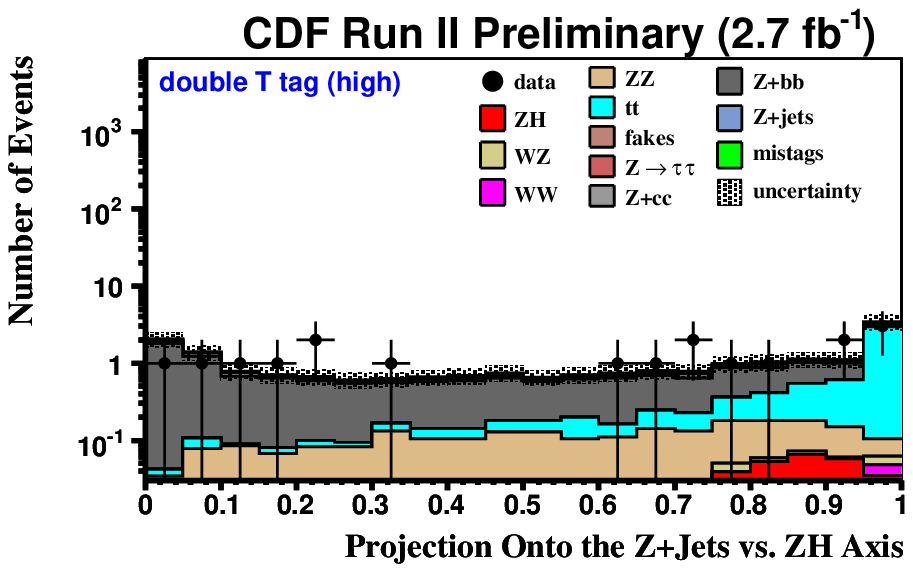} \hfill
\includegraphics[width=0.32\textwidth,height=5cm]{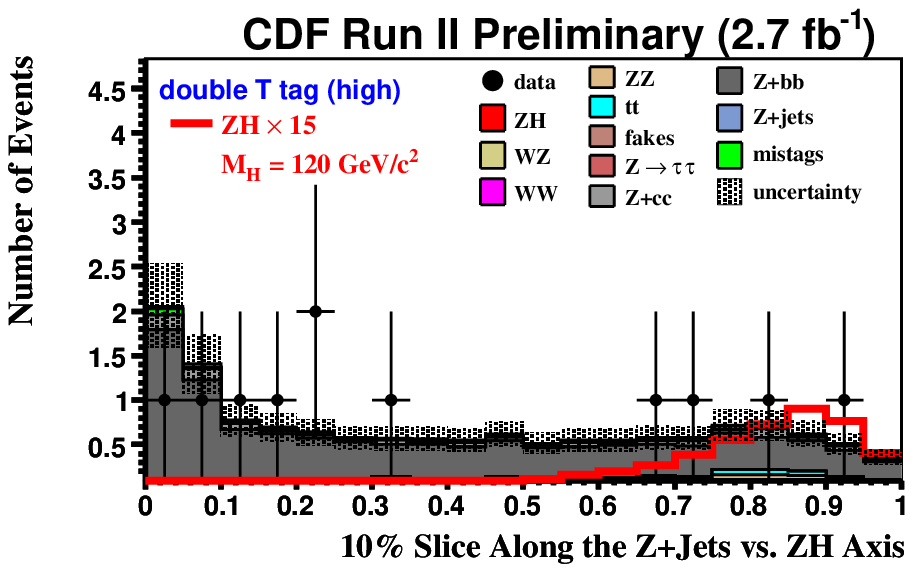} \hfill
\includegraphics[width=0.32\textwidth,height=5cm]{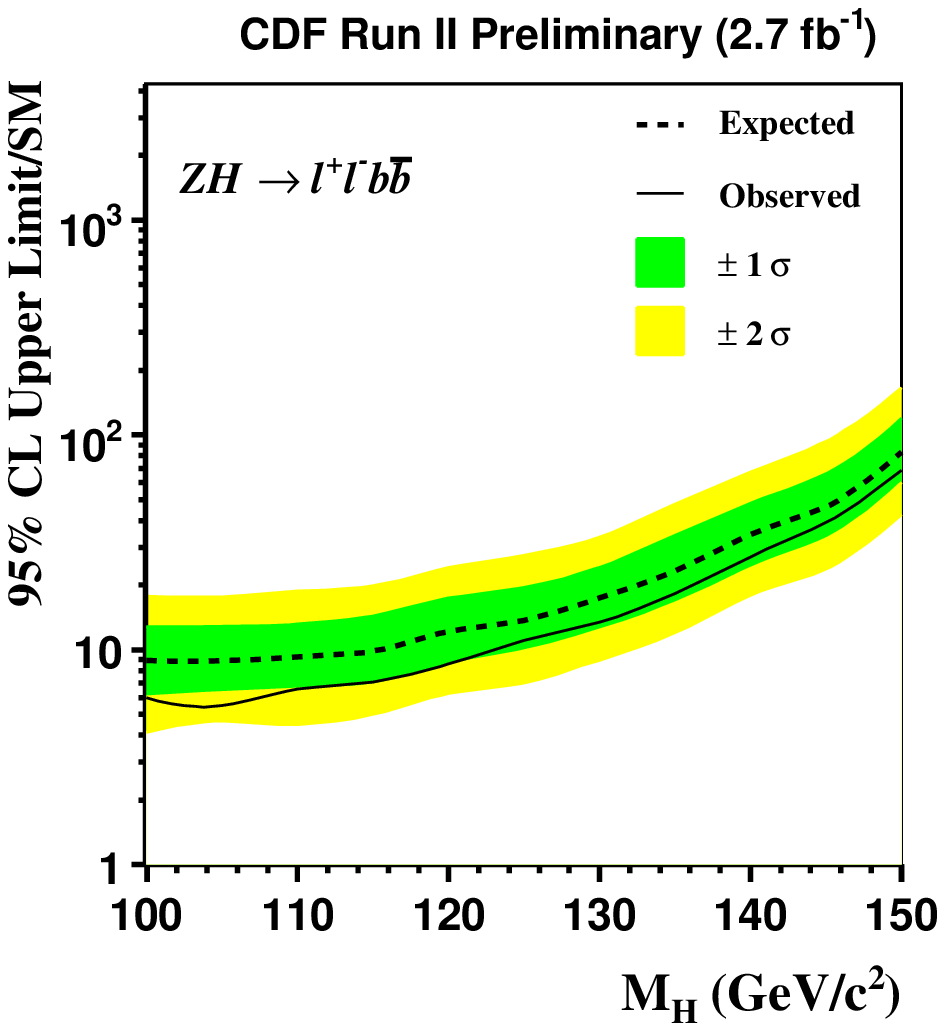}
\vspace*{-0.4cm}
\caption{CDF ZH($\rm Z\to\ell\ell$)($\rm H\to\bb$).
Left: $\rm Z\to \ell^+\ell^-$ neural network output (2-dimensional) projection.
Mostly Z+jets and $\rm t\bar t$ background.
Center: $\rm Z\to \ell^+\ell^-$ neural network output (2-dimensional) projection.
Right: limit at 95\% CL.
}
\label{fig:cdf-llbb}
\end{figure}

\begin{figure}[hp]
\vspace*{-0.2cm}
\includegraphics[width=0.32\textwidth,height=5cm]{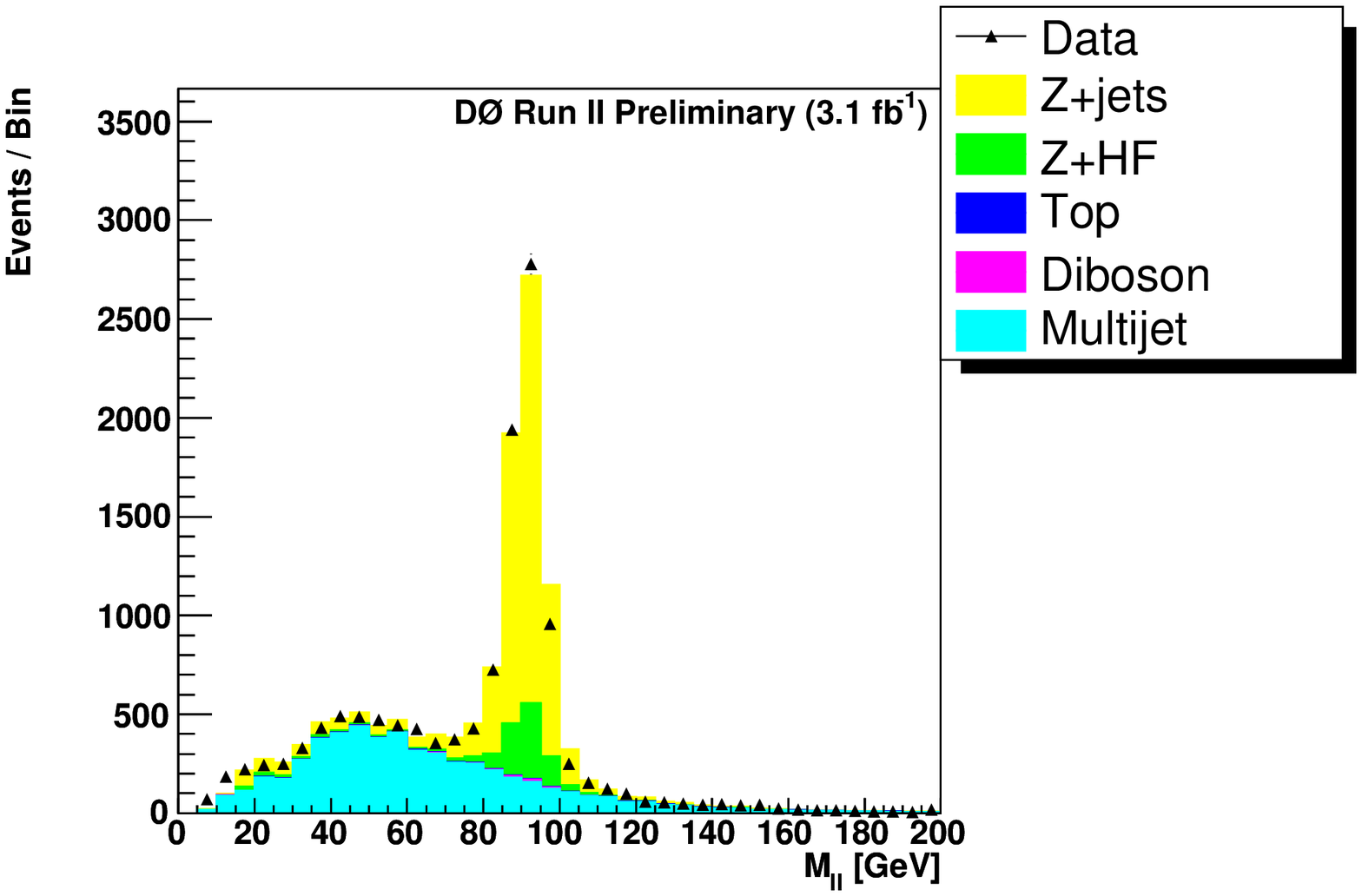} \hfill
\includegraphics[width=0.32\textwidth,height=5cm]{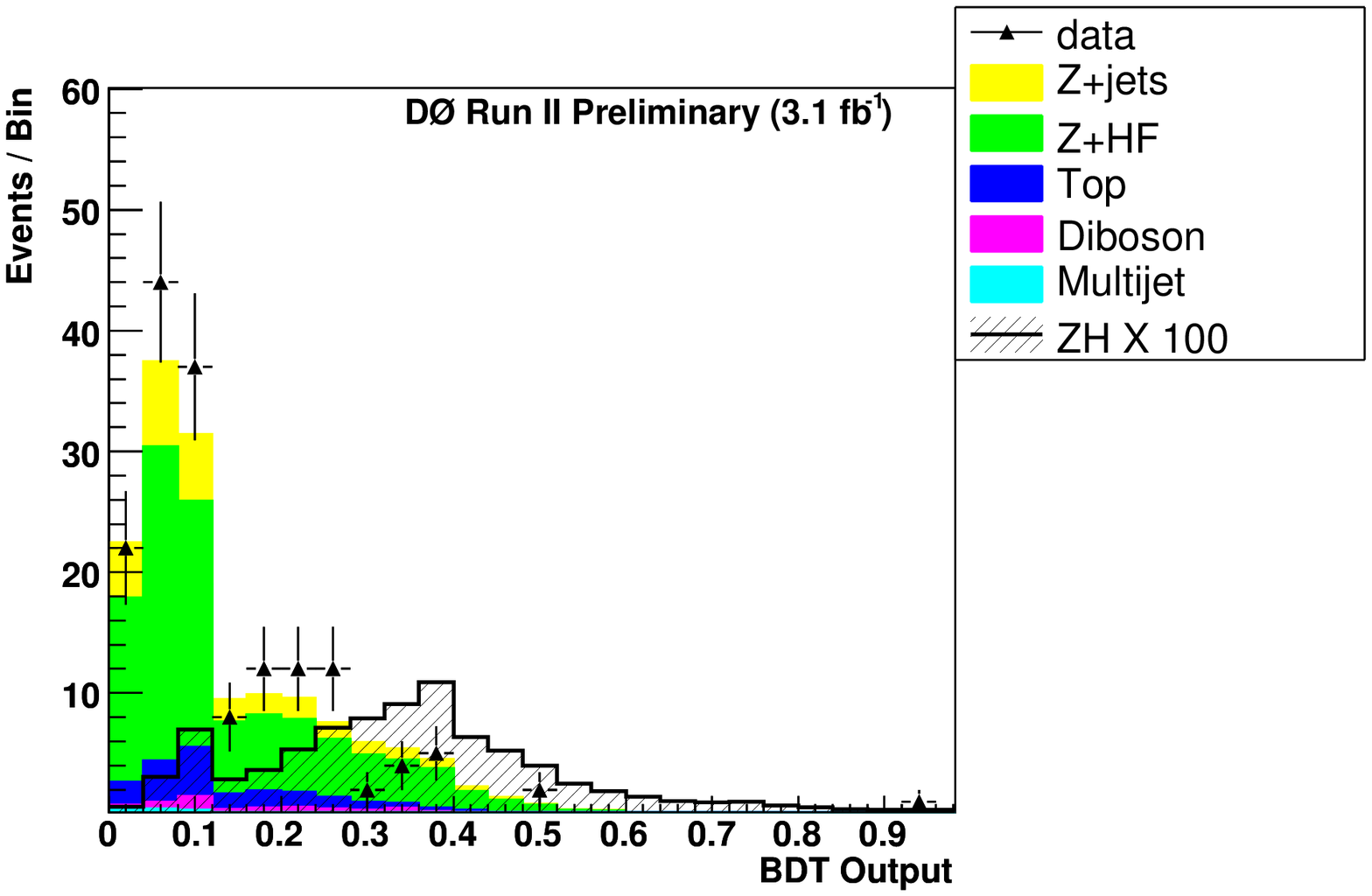} \hfill
\includegraphics[width=0.32\textwidth,height=5cm]{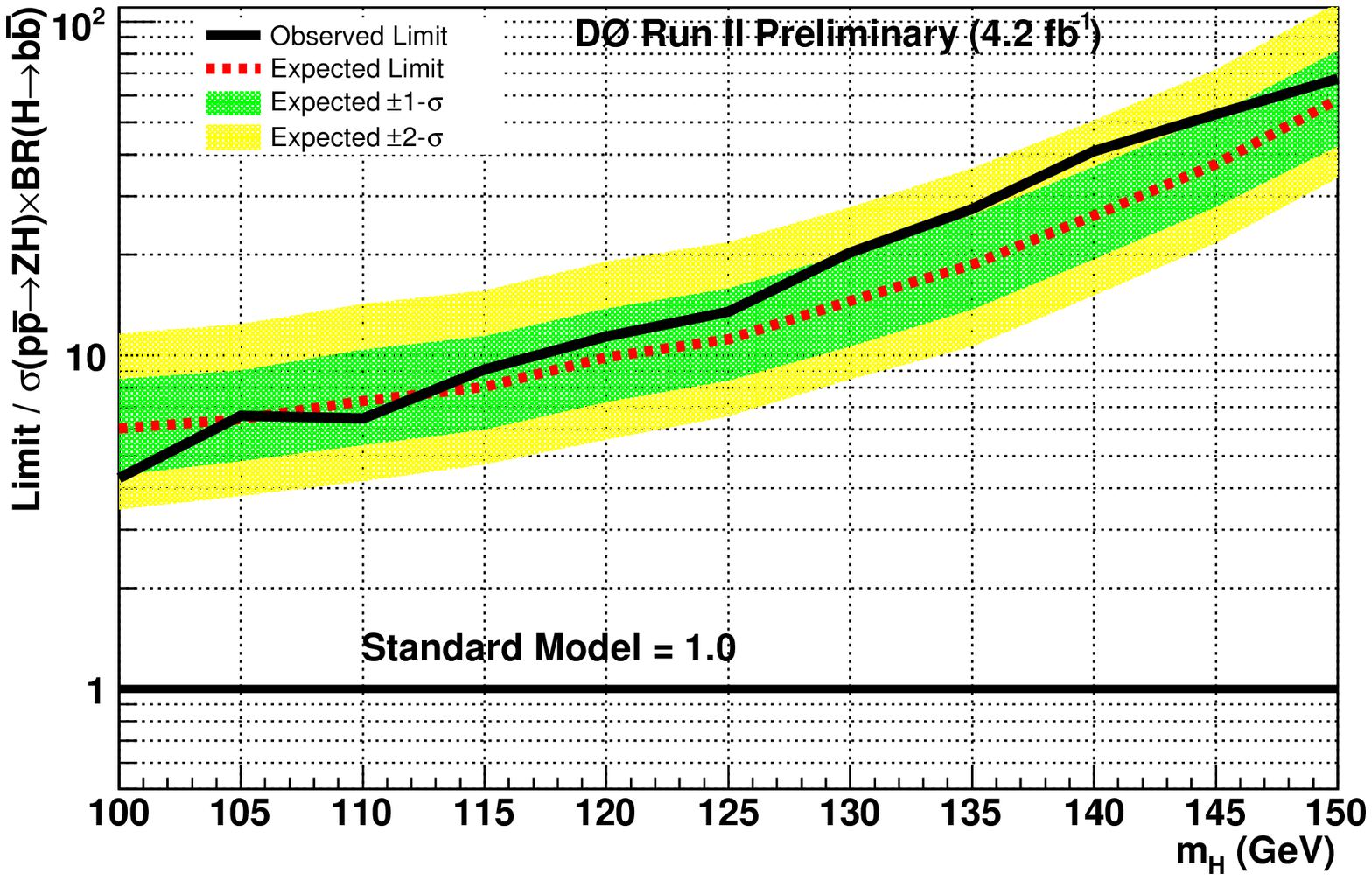}
\vspace*{-0.4cm}
\caption{D\O\ ZH ($\rm Z\to\ell\ell$)($\rm H\to\bb$).
%Left: $\rm Z\to e^+e^-$ neural network output. 
Left: $\rm Z\to e^+e^-$ invariant mass. 
Center: $\rm Z\to\mu^+\mu^-$ boosted decision tree (BDT) output.
Right: cross-section limit combined with previous $\rm Z\to\ell\ell$ results.
}
\label{fig:d0-llbb}
%\vspace*{-0.2cm}
\end{figure}

The CDF and D\O\ collaborations have searched for $\rm ZH\to e^+e^-\bb~~and~~ \mu^+\mu^-\bb$ signals.
These signals are very clean, however, they have a small production cross-section.
The results are shown in 
Fig.~\ref{fig:cdf-llbb} (from~\cite{cdf-llbb}) and
Fig.~\ref{fig:d0-llbb} (from~\cite{d0-llbb}).

\subsection{$\rm ZH\to \nn\bb$}

\begin{figure}[hp]
\vspace*{-0.2cm}
\includegraphics[width=0.32\textwidth,height=5cm]{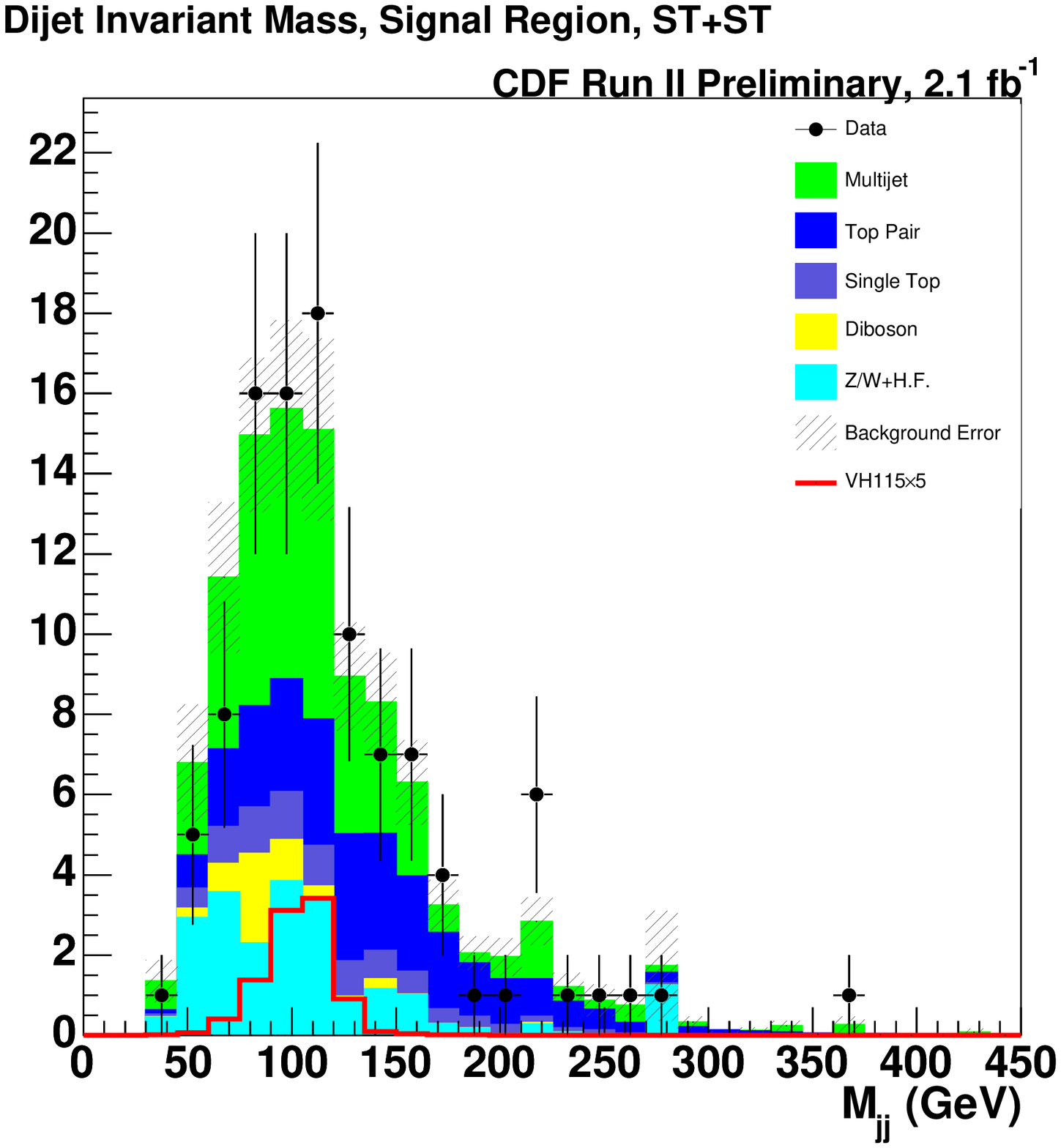} \hfill
\includegraphics[width=0.32\textwidth,height=5cm]{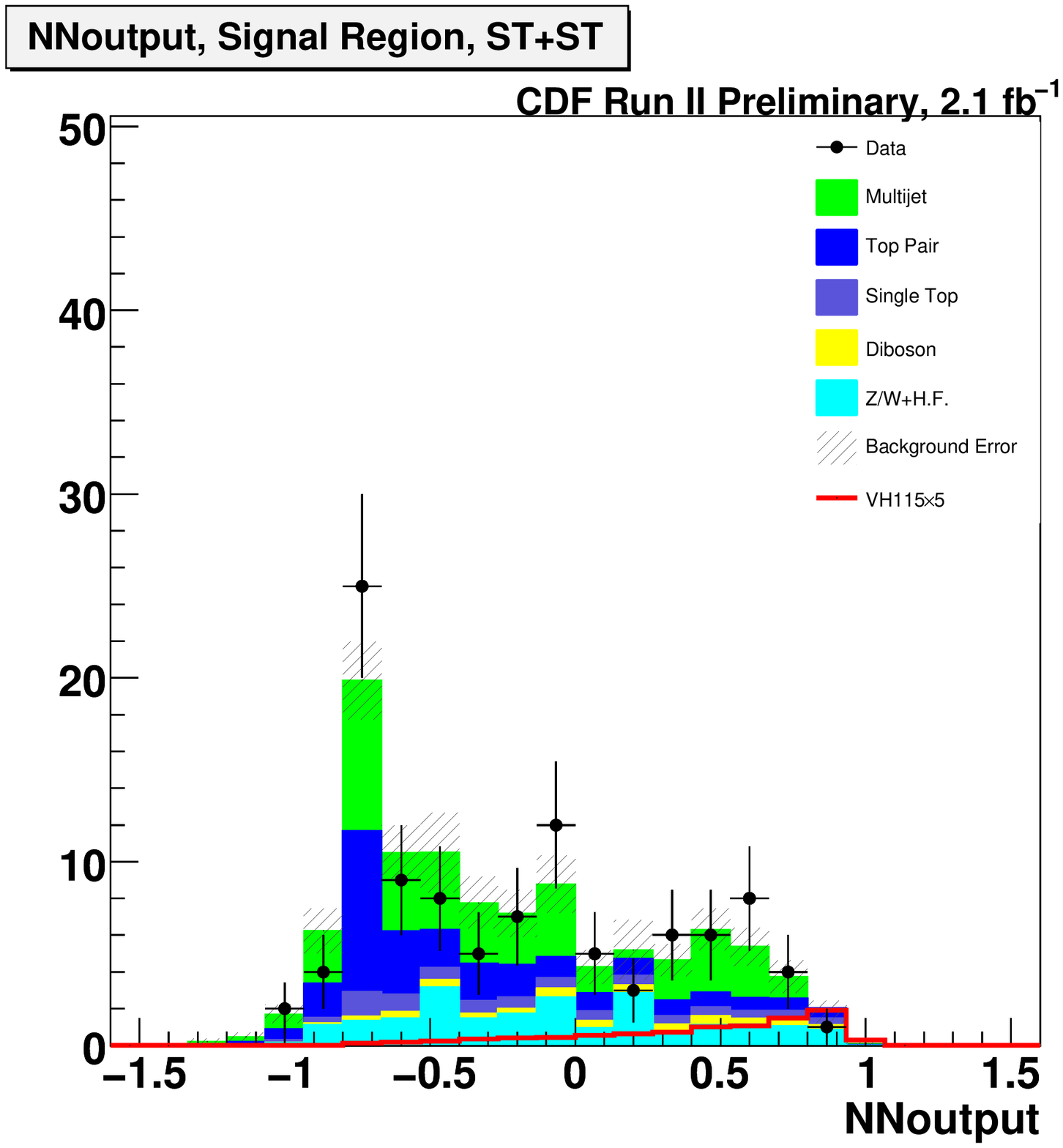} \hfill
\includegraphics[width=0.32\textwidth,height=5cm]{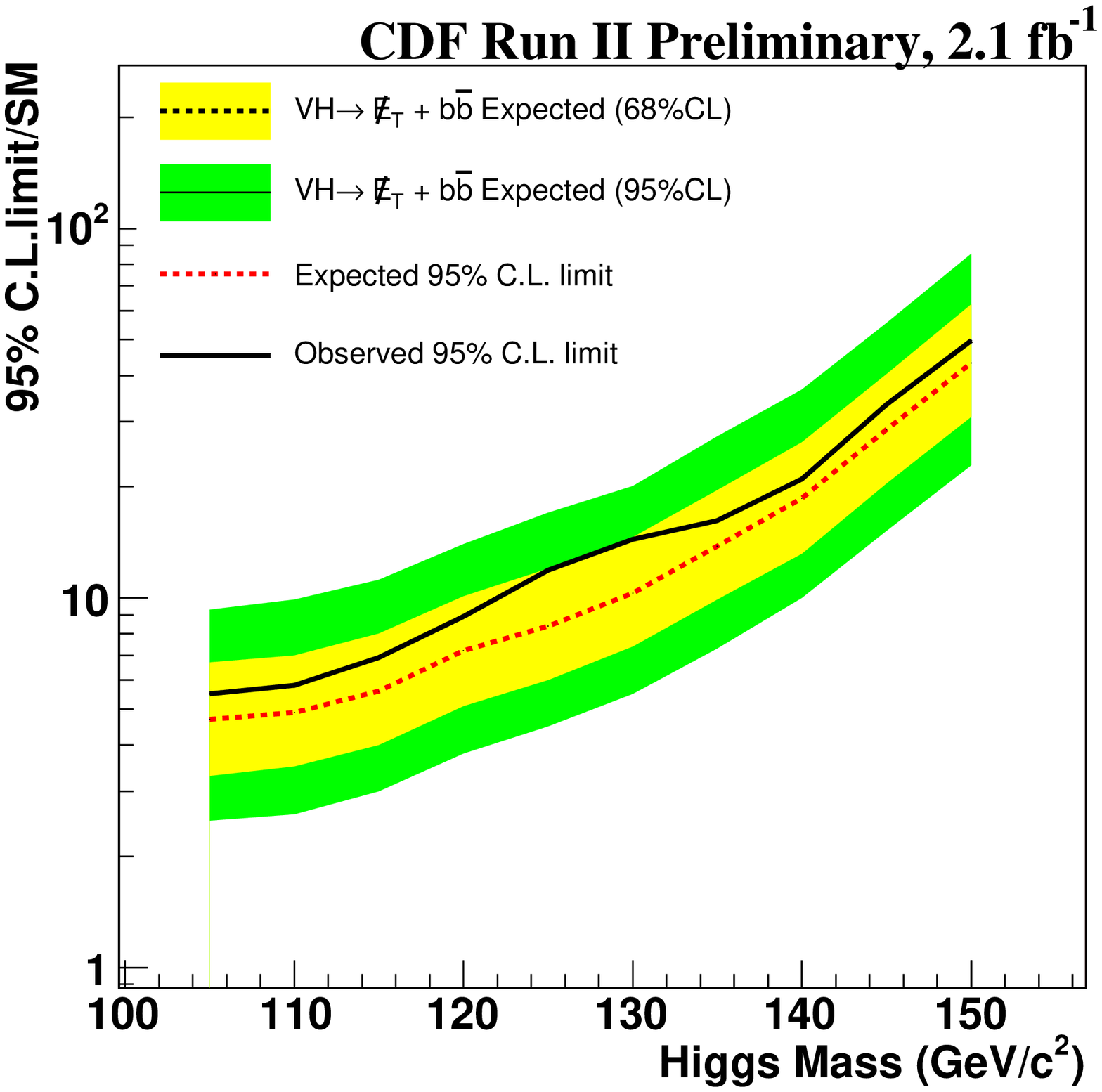}
\vspace*{-0.4cm}
\caption{CDF ZH($\rm Z\to\nu\nu$)($\rm H\to\bb$).
Left: invariant di-jet mass. 
Center: neural network.
Right: limit at 95\% CL.
}
\label{fig:cdf-zHbb}
\end{figure}

\begin{figure}[hp]
\vspace*{-0.2cm}
\includegraphics[width=0.32\textwidth,height=5cm]{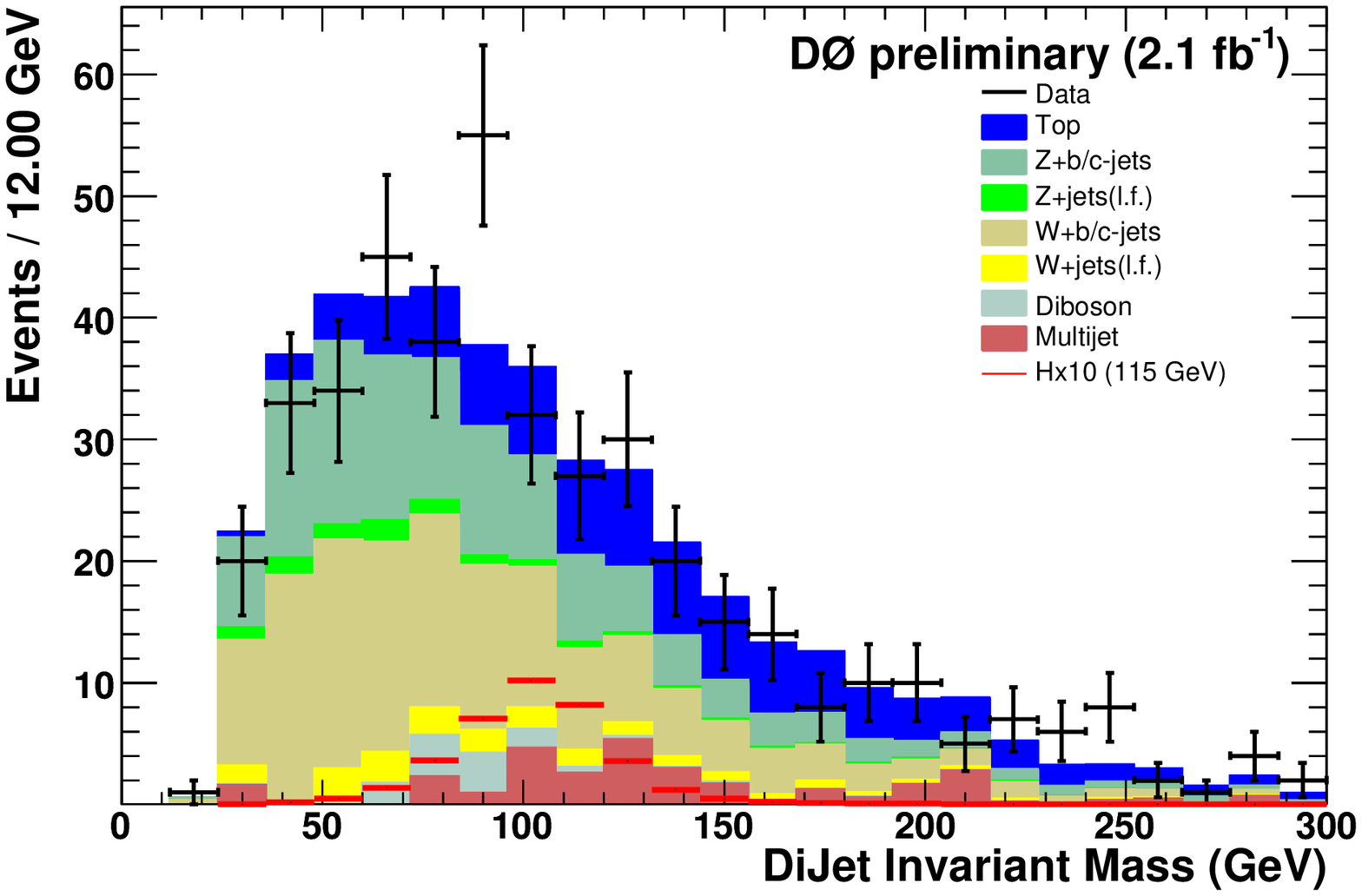} \hfill
\includegraphics[width=0.32\textwidth,height=5cm]{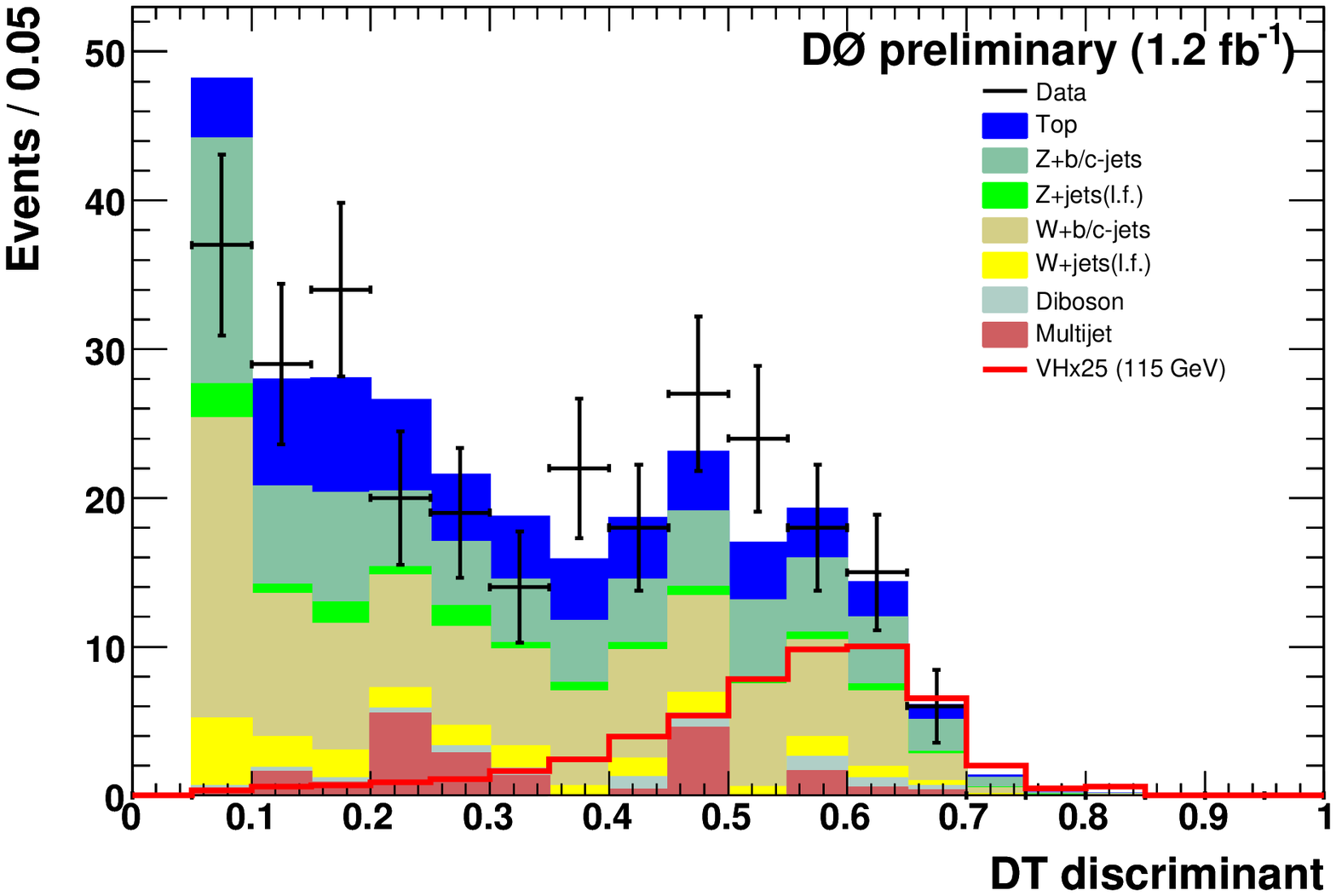} \hfill
\includegraphics[width=0.32\textwidth,height=5cm]{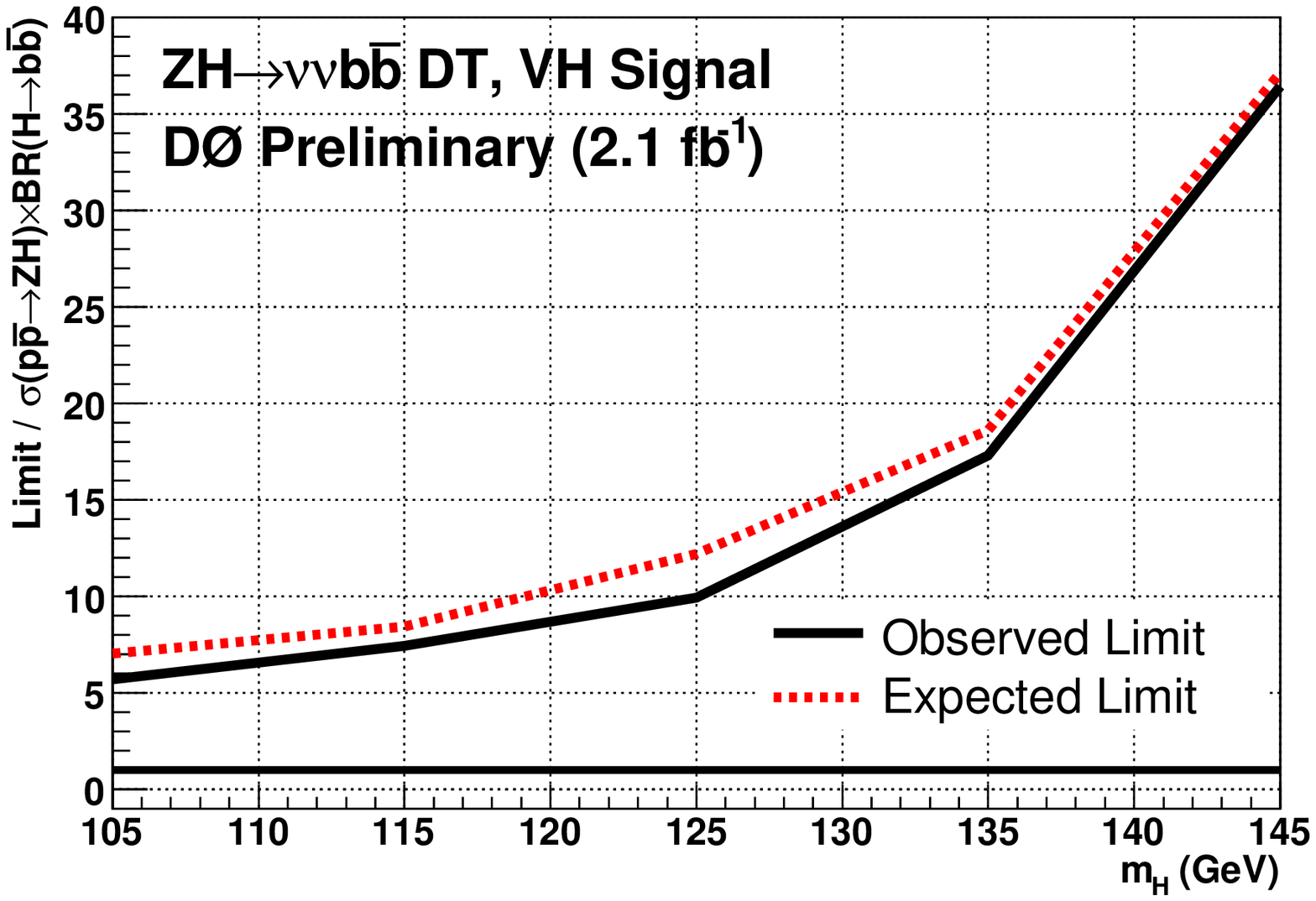}
\vspace*{-0.4cm}
\caption{D\O\ ZH($\rm Z\to\nu\nu$)($\rm H\to\bb$).
Left: invariant di-jet mass. 
Center: discriminant variable output.
Right: limit at 95\% CL.
}
\label{fig:d0-zHbb}
\end{figure}

Both Tevatron collaborations have searched for a $\rm ZH\to \nn\bb$ signal.
The results from the expected missing energy and b-jet signal 
are shown in 
Fig.~\ref{fig:cdf-zHbb} (from~\cite{cdf-zHbb}) and
Fig.~\ref{fig:d0-zHbb} (from~\cite{d0-zHbb}).

\clearpage
\section{$\rm H\to\tau^+\tau^-$}

\begin{figure}[hp]
\vspace*{-0.2cm}
\includegraphics[width=0.32\textwidth,height=5cm]{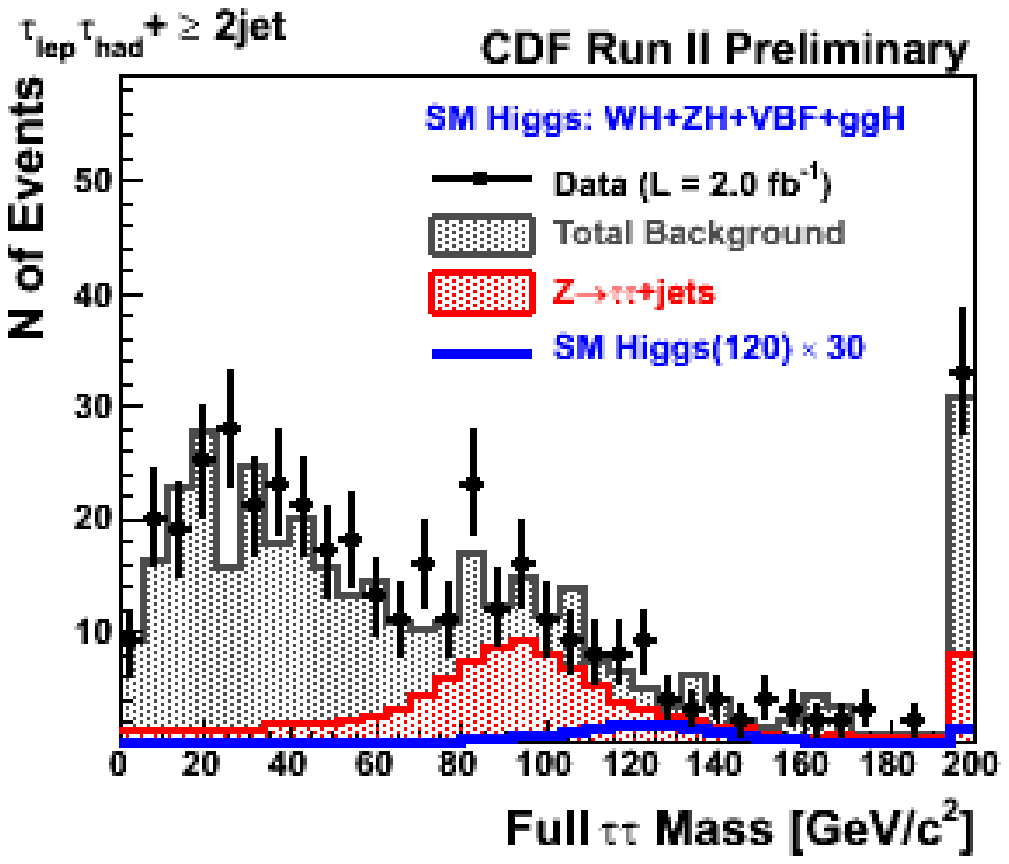} \hfill
\includegraphics[width=0.32\textwidth,height=5cm]{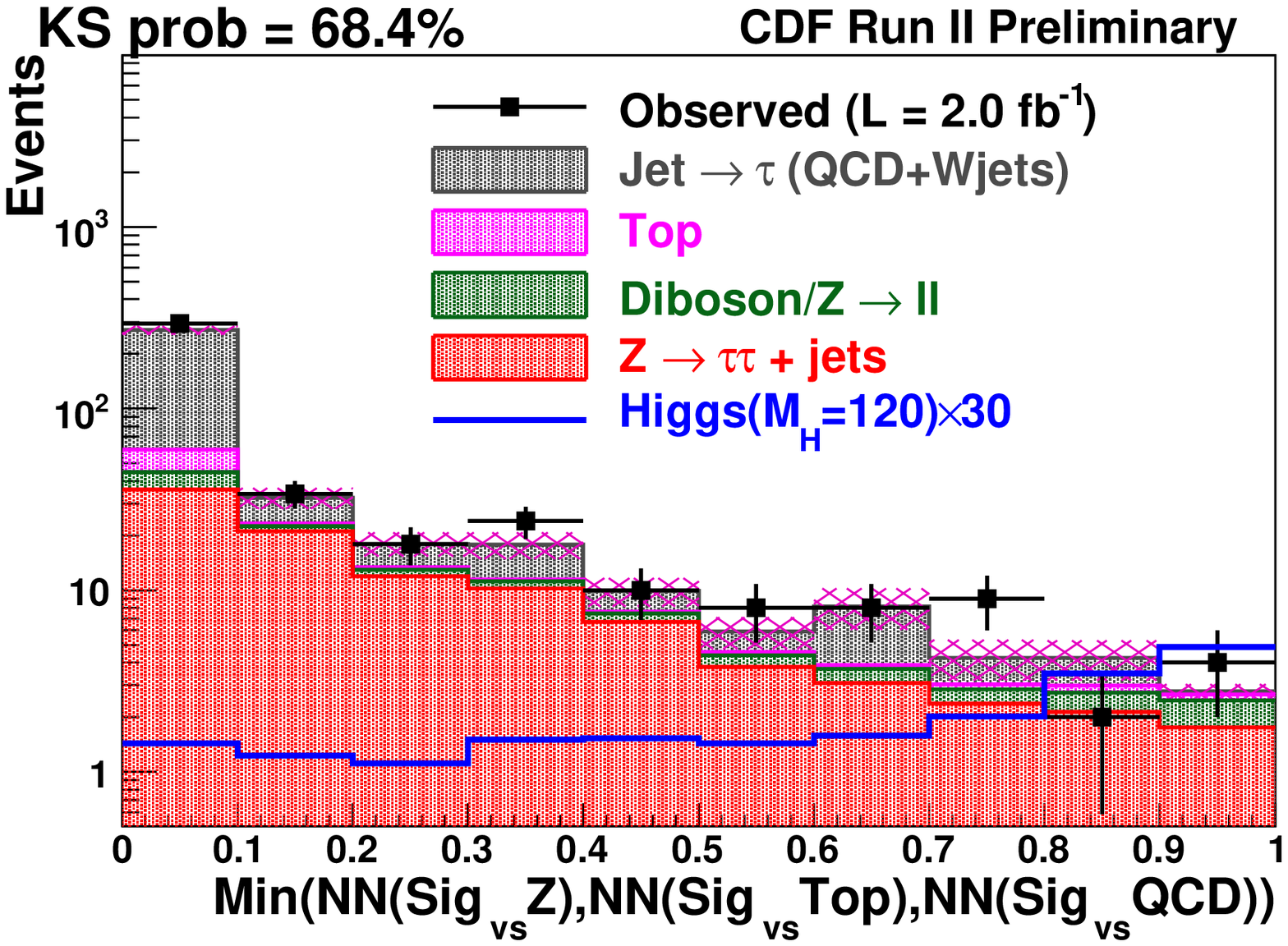} \hfill
\includegraphics[width=0.32\textwidth,height=5cm]{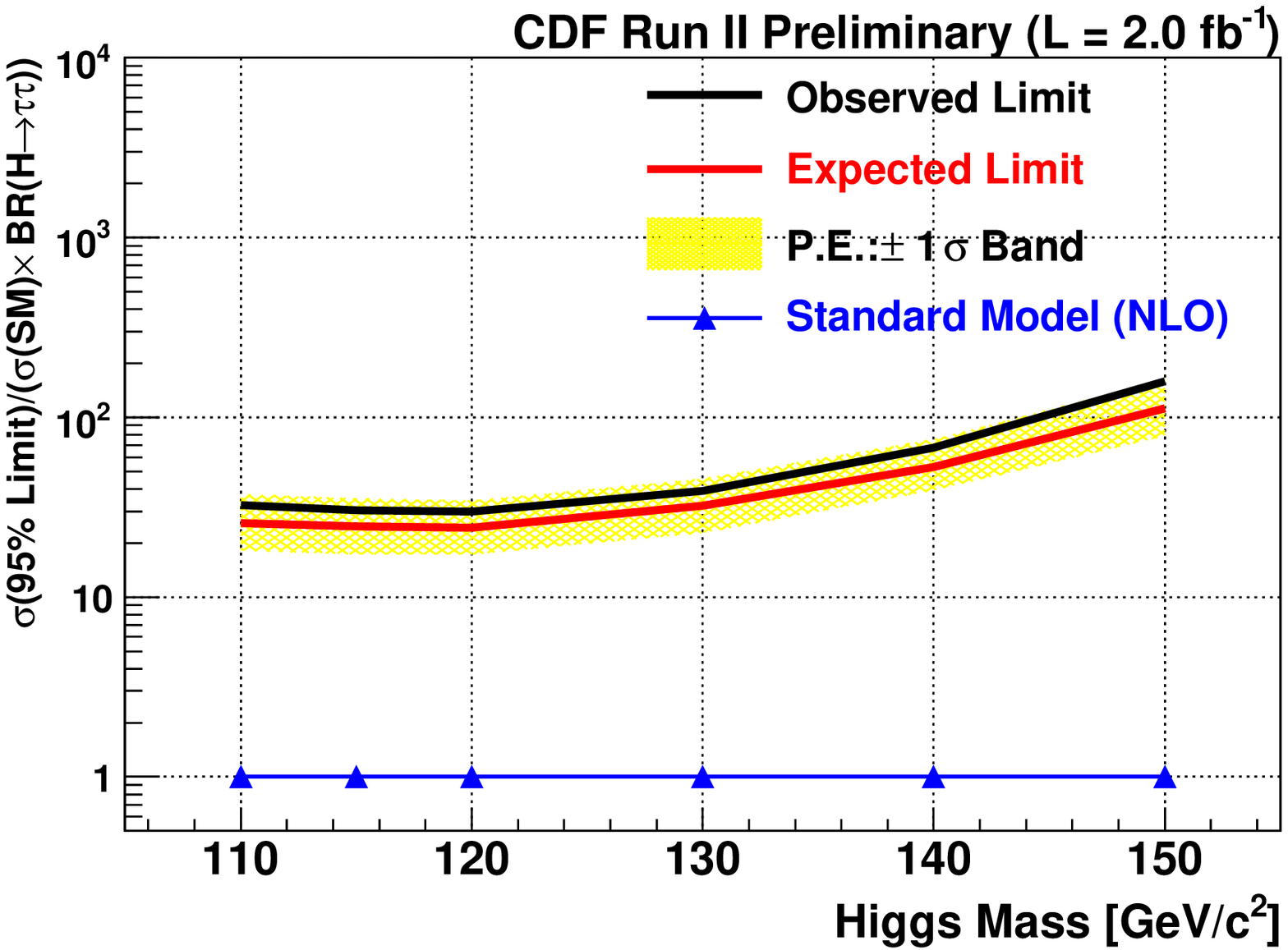}
\vspace*{-0.3cm}
\caption{CDF ($\rm H\to\tau^+\tau^-$).
Left: invariant mass. 
Center: neural network output.
Right: limit\,at\,95\%\,CL.
}
\label{fig:cdf_sm_tautau}
\end{figure}

\begin{figure}[hp]
\vspace*{-0.4cm}
\includegraphics[width=0.64\textwidth,height=5cm]{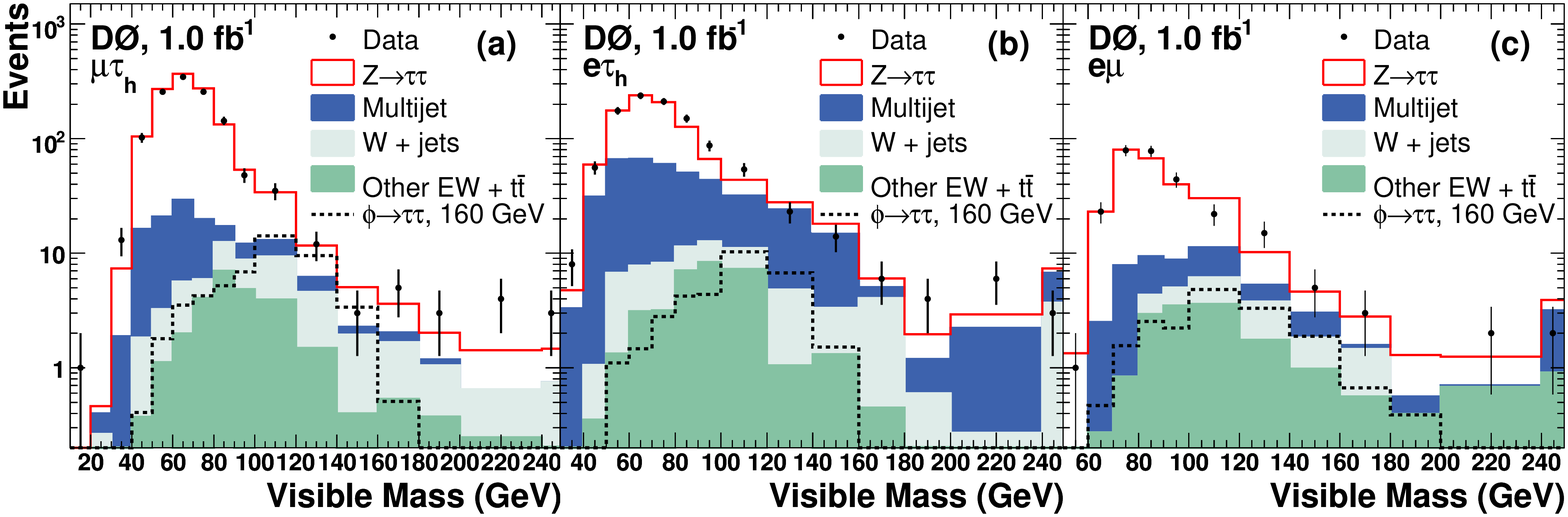} \hfill
\includegraphics[width=0.32\textwidth,height=5cm]{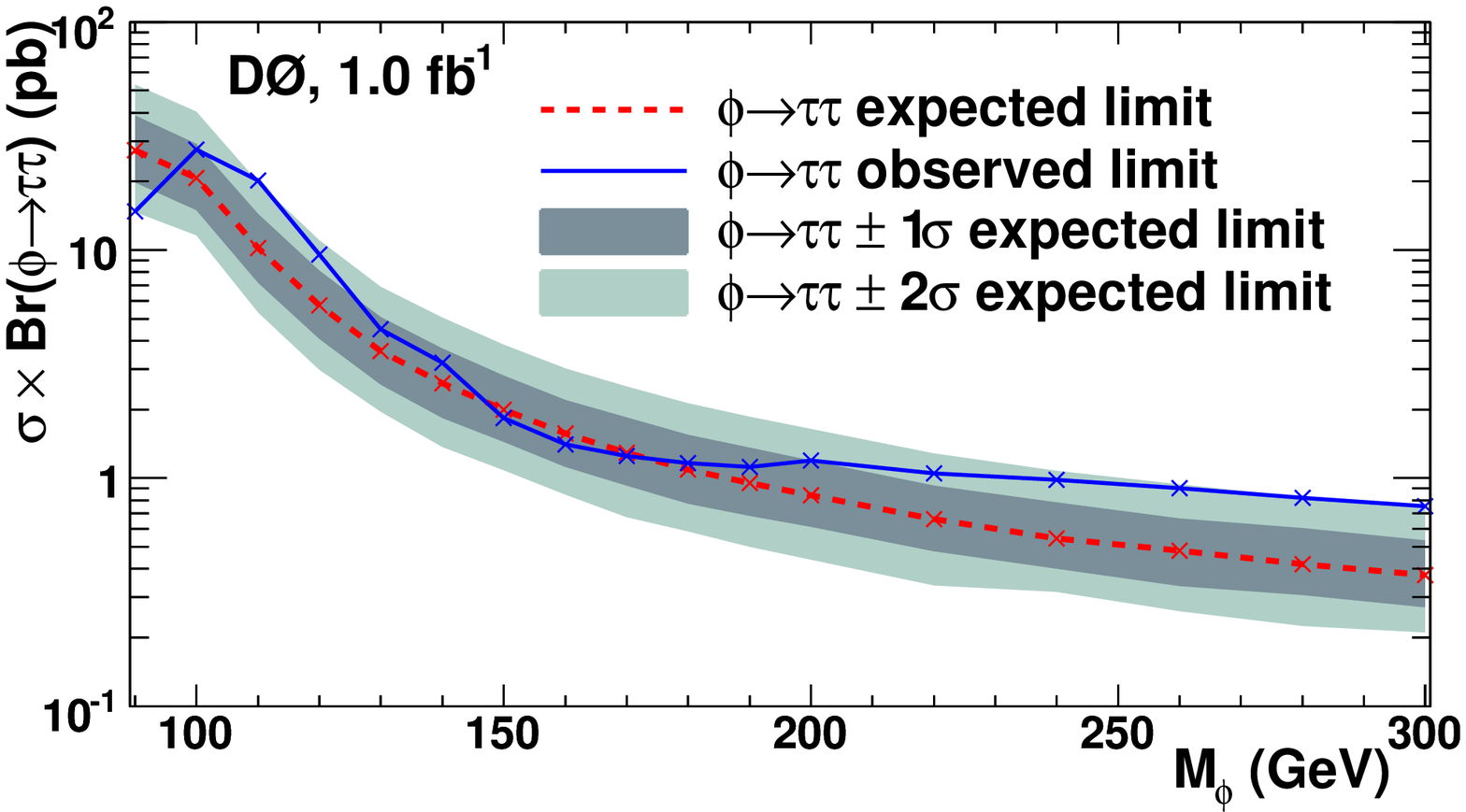}
\vspace*{-0.3cm}
\caption{D\O\ ($\rm H\to\tau^+\tau^-$).
Left: invariant masses for $\rm \mu\tau_h$, $\rm e\tau_h$ and $\rm e\mu$.
Right: limit at 95\% CL.
}
\label{fig:d0-sm_tautau}
\end{figure}

Both Tevatron collaborations have searched for a $\rm H\to \tau^+\tau^-$ signal.
Results are shown in Figs.~\ref{fig:cdf_sm_tautau} (from~\cite{cdf-sm_tautau})
and~\ref{fig:d0-sm_tautau} (from~\cite{d0-sm_tautau}).

\vspace*{-0.3cm}
\section{$\rm H\to\gamma\gamma$}
\vspace*{-0.1cm}

\begin{figure}[hp]
\vspace*{-0.4cm}
\includegraphics[width=0.32\textwidth,height=5cm]{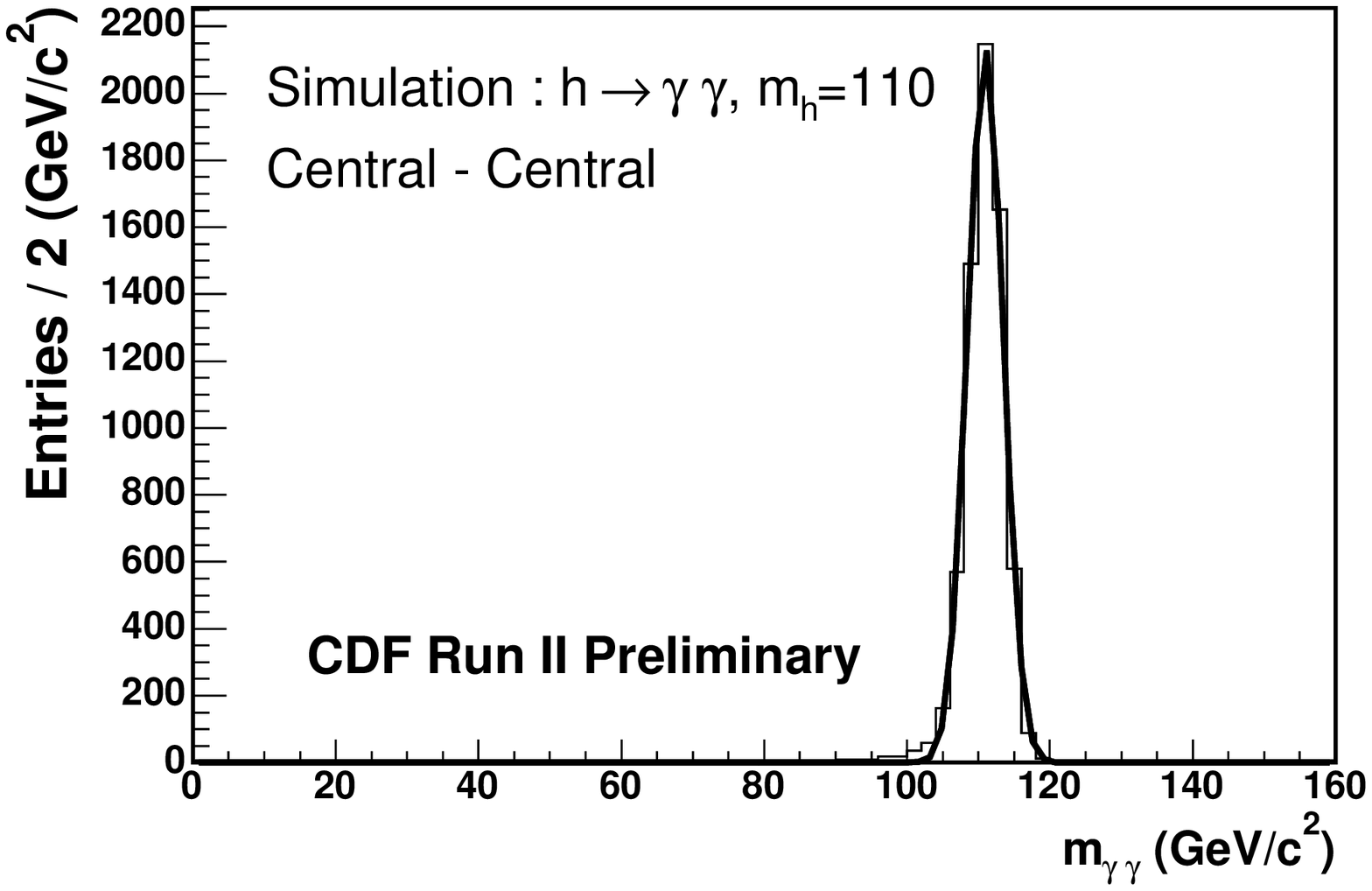} \hfill
\includegraphics[width=0.32\textwidth,height=5cm,bbllx=24pt,bblly=287pt,bburx=518pt,bbury=541pt,clip=]{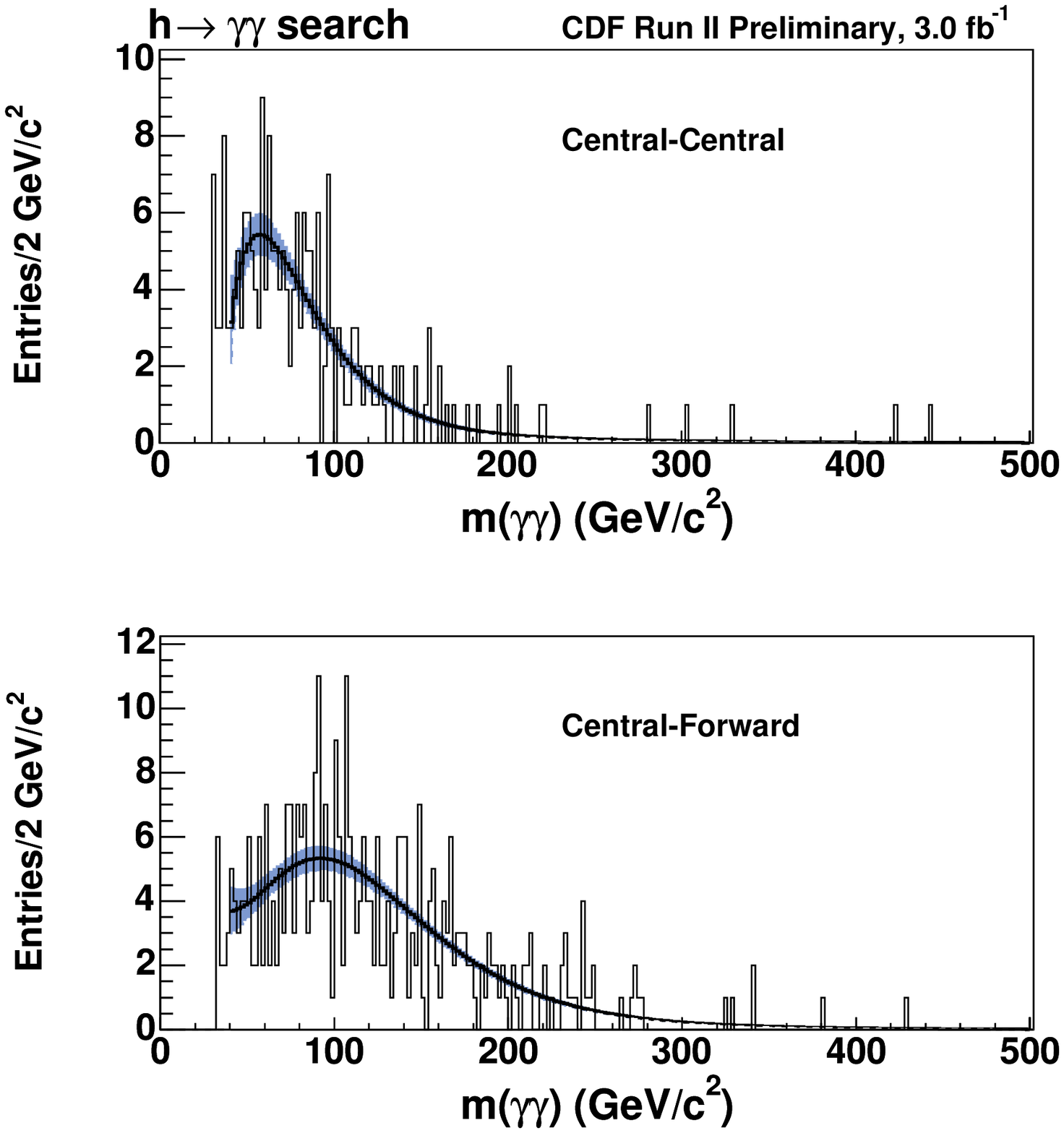}\hfill
\includegraphics[width=0.32\textwidth,height=5cm,bbllx=24pt,bblly=11pt,bburx=518pt,bbury=260pt,clip=]{plots/cdf_gamma_gamma_DataFitsErrorBands.ps}
\vspace*{-0.5cm}
\caption{CDF ($\rm H\to\gamma\gamma$).
Left: simulated $\rm H\to\gamma\gamma$ invariant mass. 
Center: invariant mass spectrum with both photons in the central region.
Right: invariant mass spectrum with one photon in the central region and one in the forward region.
}
\label{fig:cdf_sm_gammagamma}
\vspace*{-0.5cm}
\end{figure}

\begin{figure}[tp]
\vspace*{-0.4cm}
\includegraphics[width=0.32\textwidth,height=5cm]{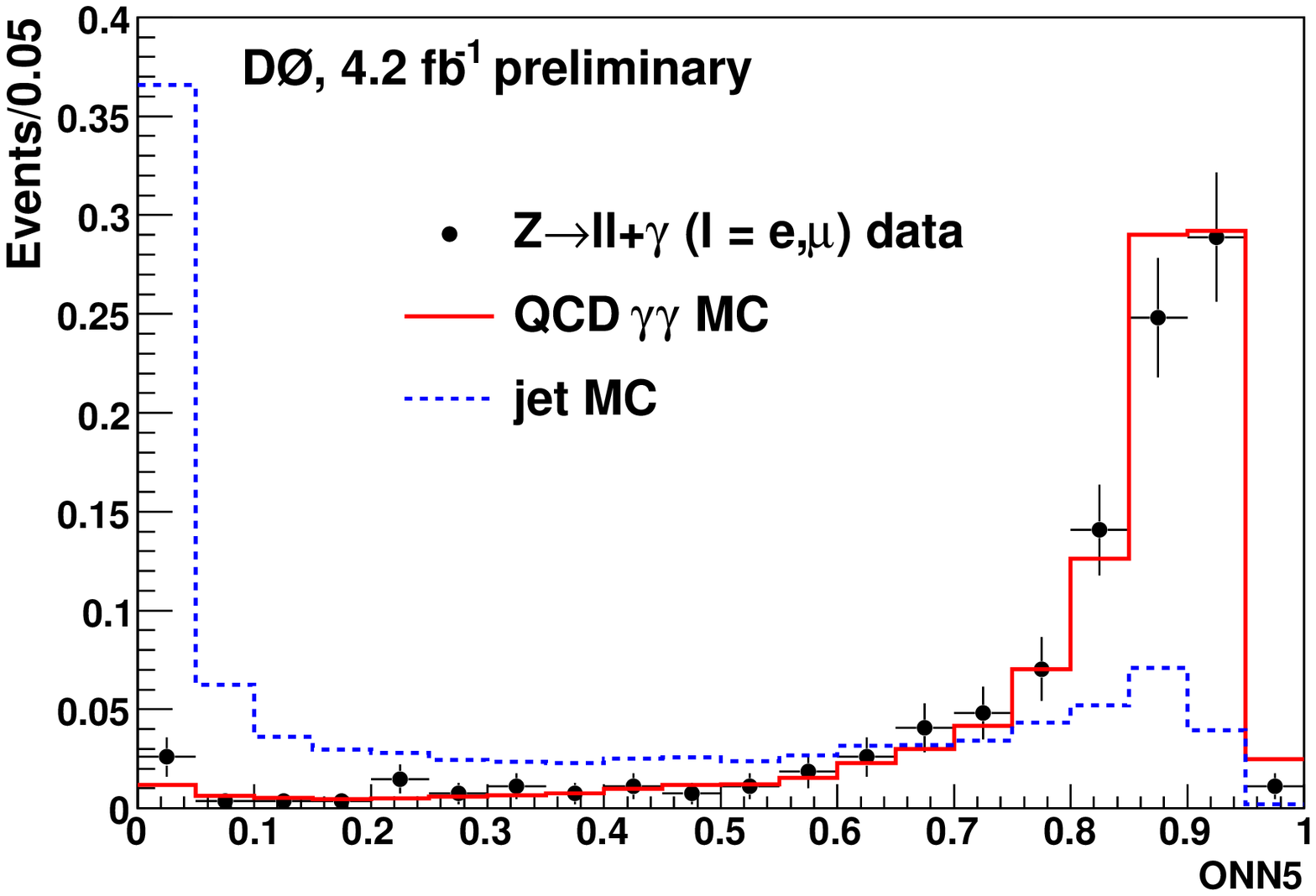} \hfill
\includegraphics[width=0.32\textwidth,height=5cm]{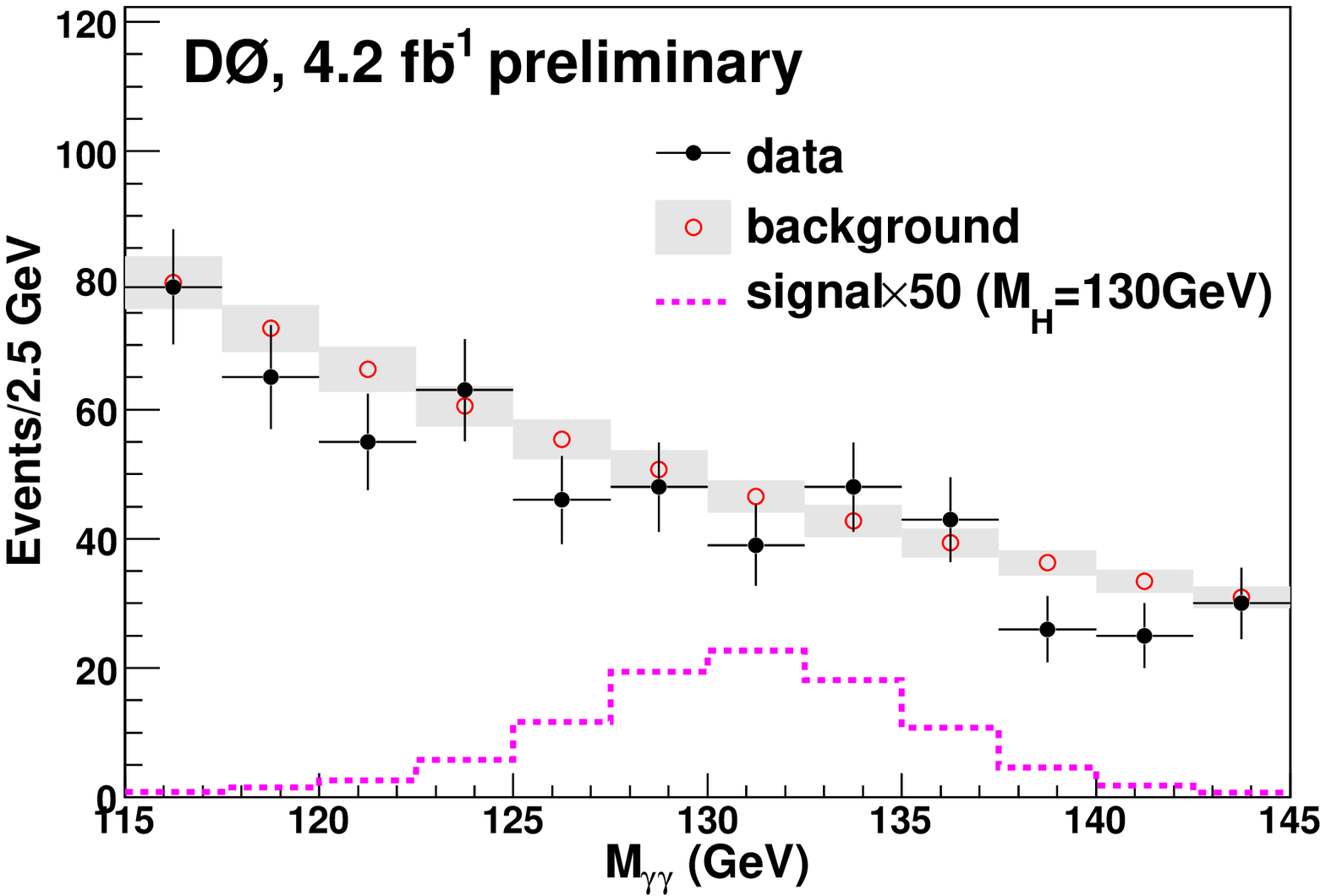} \hfill
\includegraphics[width=0.32\textwidth,height=5cm]{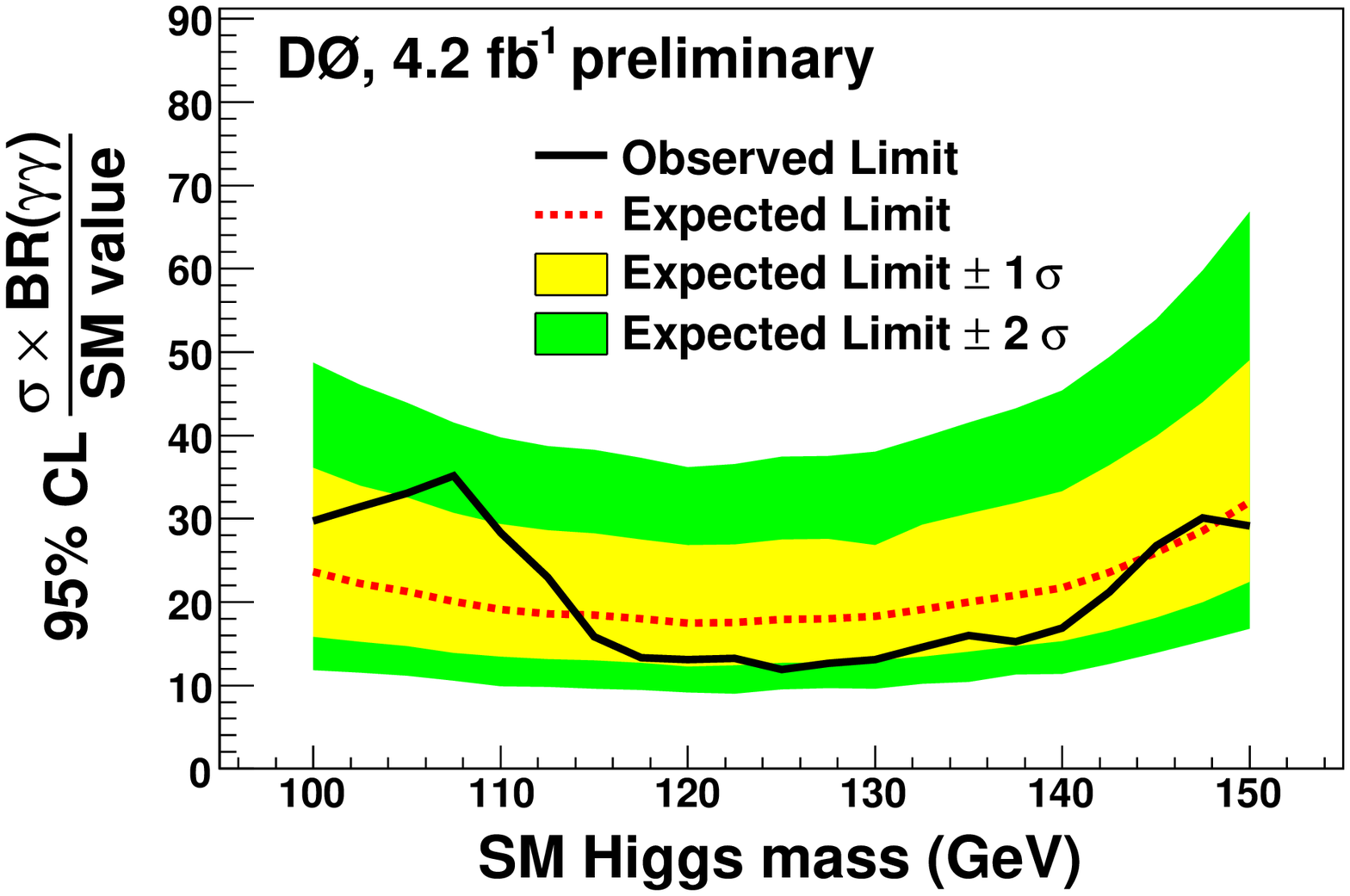}
\vspace*{-0.5cm}
\caption{D\O\ ($\rm H\to\gamma\gamma$).
Left: neural network output.
Center: invariant mass. 
Right: limit at 95\% CL.
}
\label{fig:d0-sm_gammagamma}
\vspace*{-0.4cm}
\end{figure}

Both Tevatron collaborations have searched for a $\rm H\to \gamma\gamma$ signal.
Results are shown in Figs.~\ref{fig:cdf_sm_gammagamma} (from~\cite{cdf-sm_gammagamma})
and~\ref{fig:d0-sm_gammagamma} (from~\cite{d0-sm_gammagamma}).

\section{$\rm t\bar tH$}
\vspace*{-0.1cm}

\begin{figure}[hp]
\vspace*{-0.4cm}
\includegraphics[width=0.32\textwidth,height=5cm]{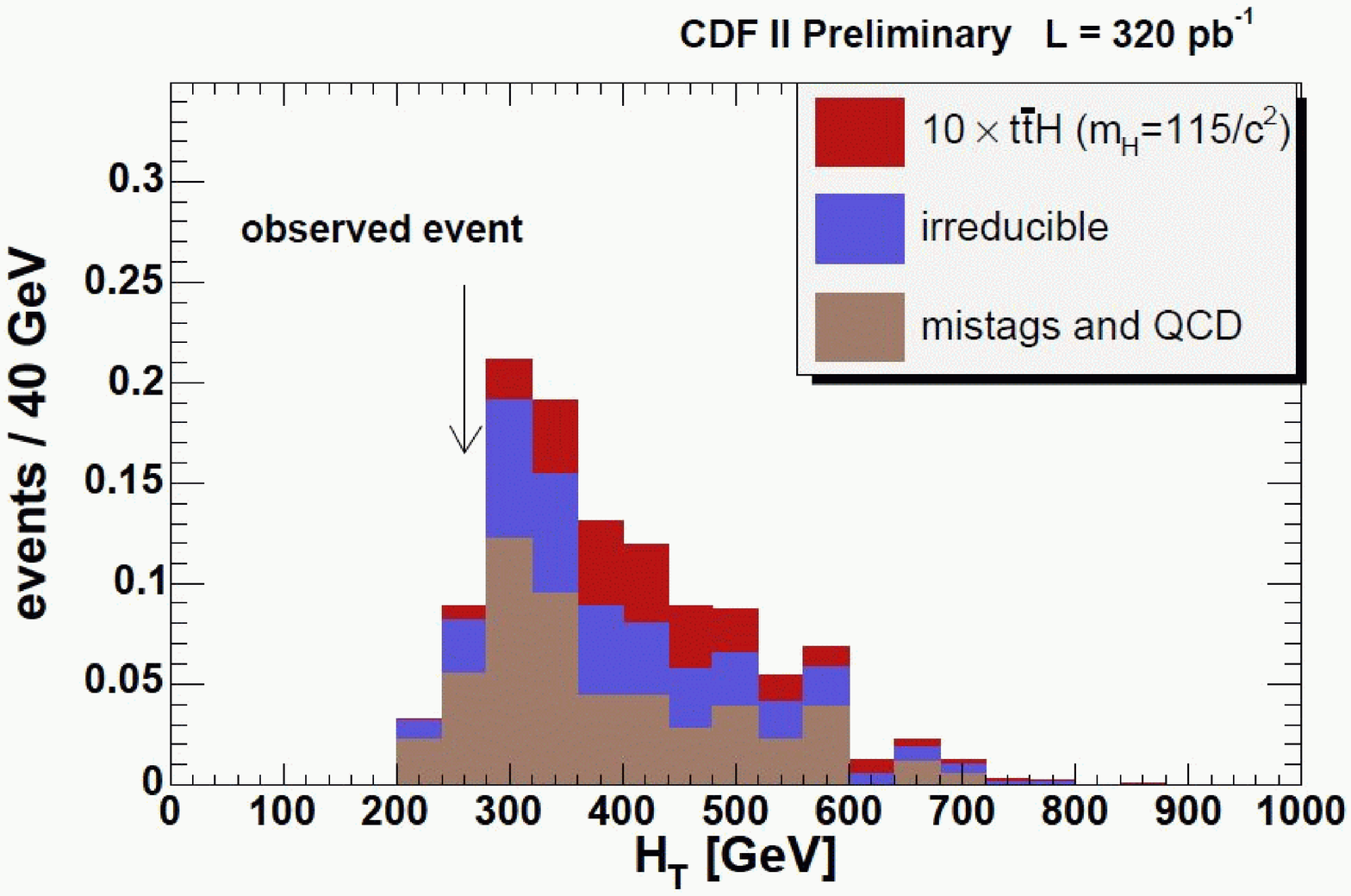} \hfill
\includegraphics[width=0.32\textwidth,height=5cm]{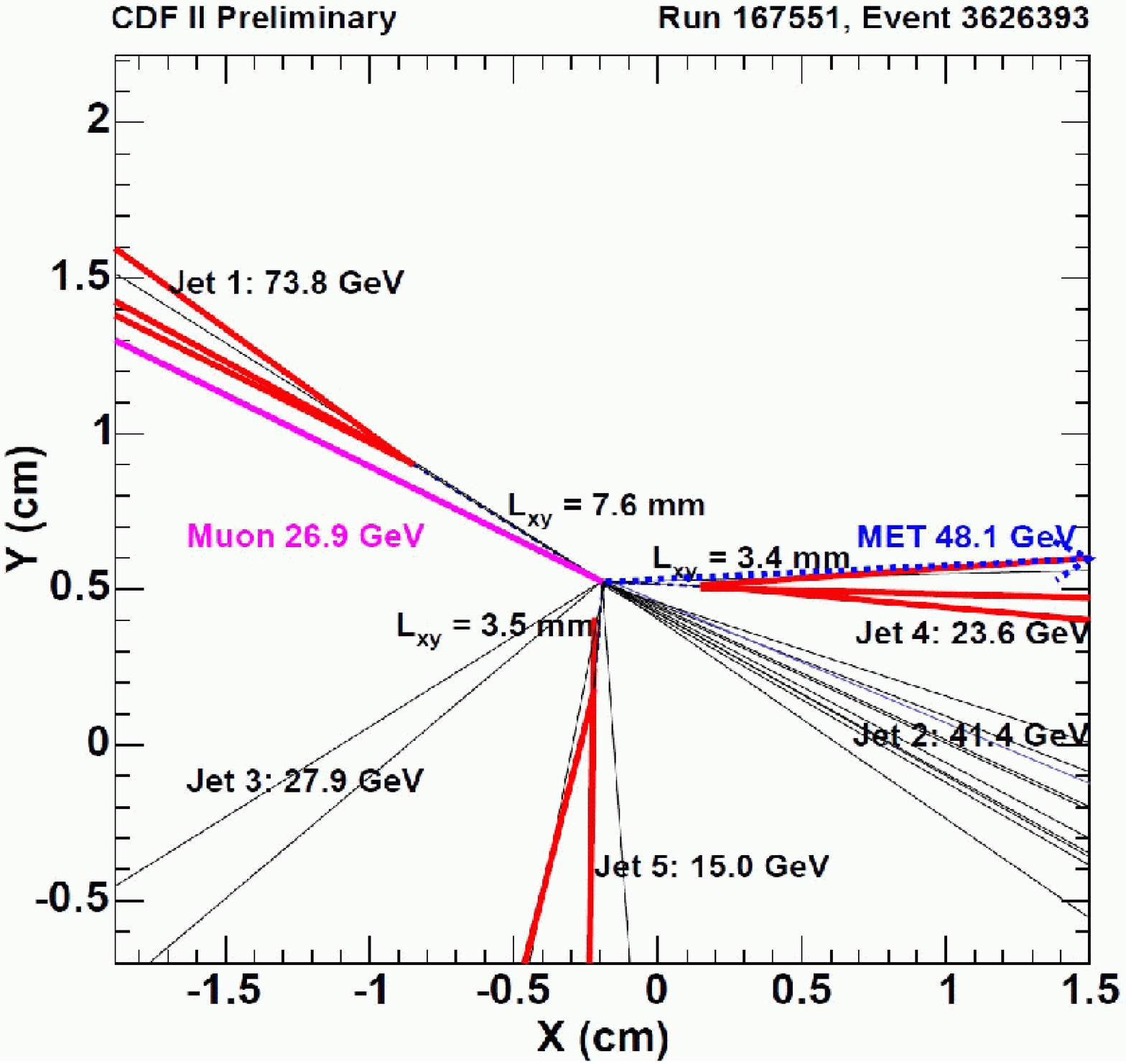} \hfill
\includegraphics[width=0.32\textwidth,height=5cm]{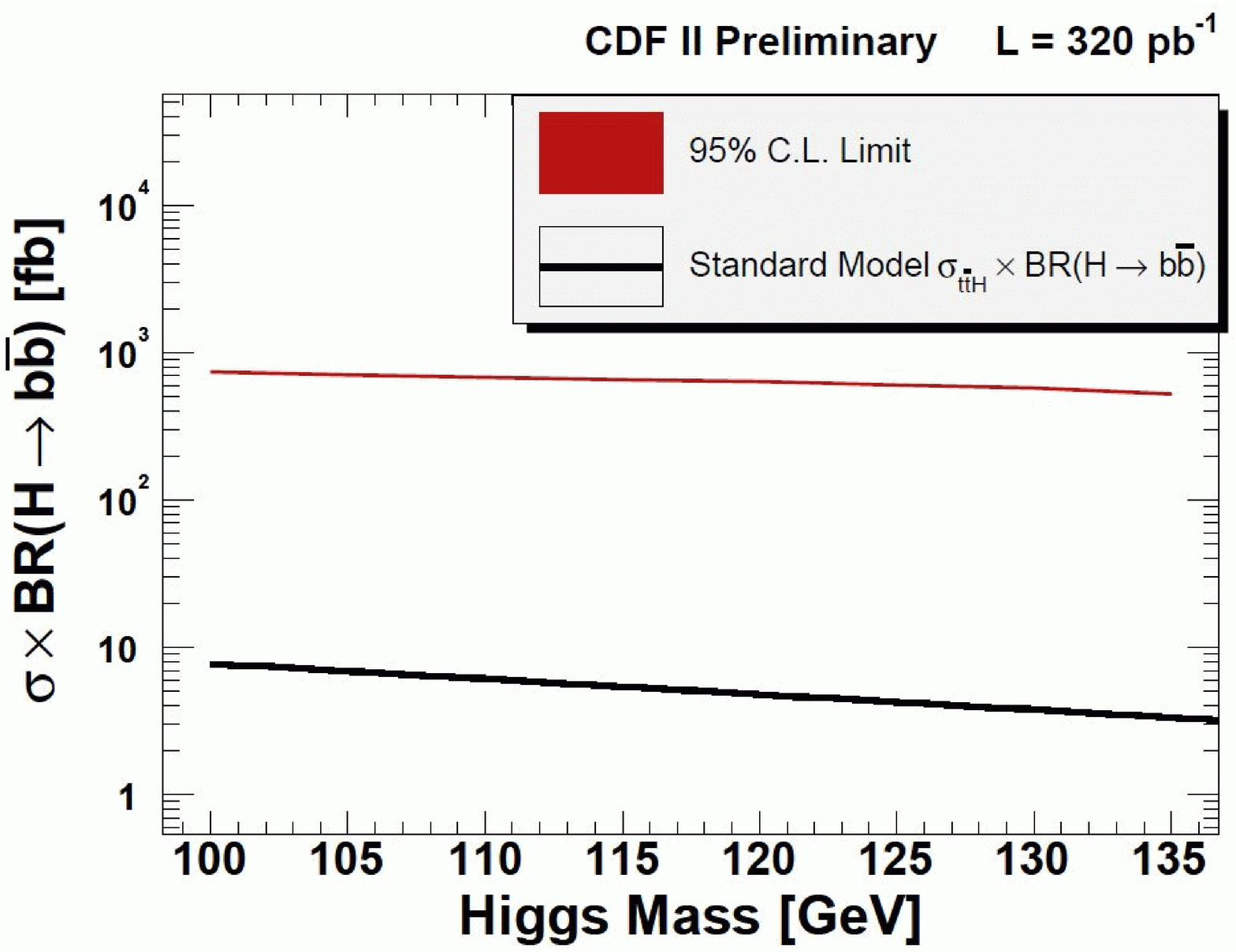}
\vspace*{-0.5cm}
\caption{CDF ($\rm t\bar t\to t\bar tH$).
Left: $H_{\rm T}$ distribution.
Center: candidate event.
Right: limit at 95\% CL.
}
\label{fig:cdf_sm_ttH}
\vspace*{-0.2cm}
\end{figure}

\begin{figure}[hp]
\vspace*{-0.4cm}
\includegraphics[width=0.32\textwidth,height=5cm]{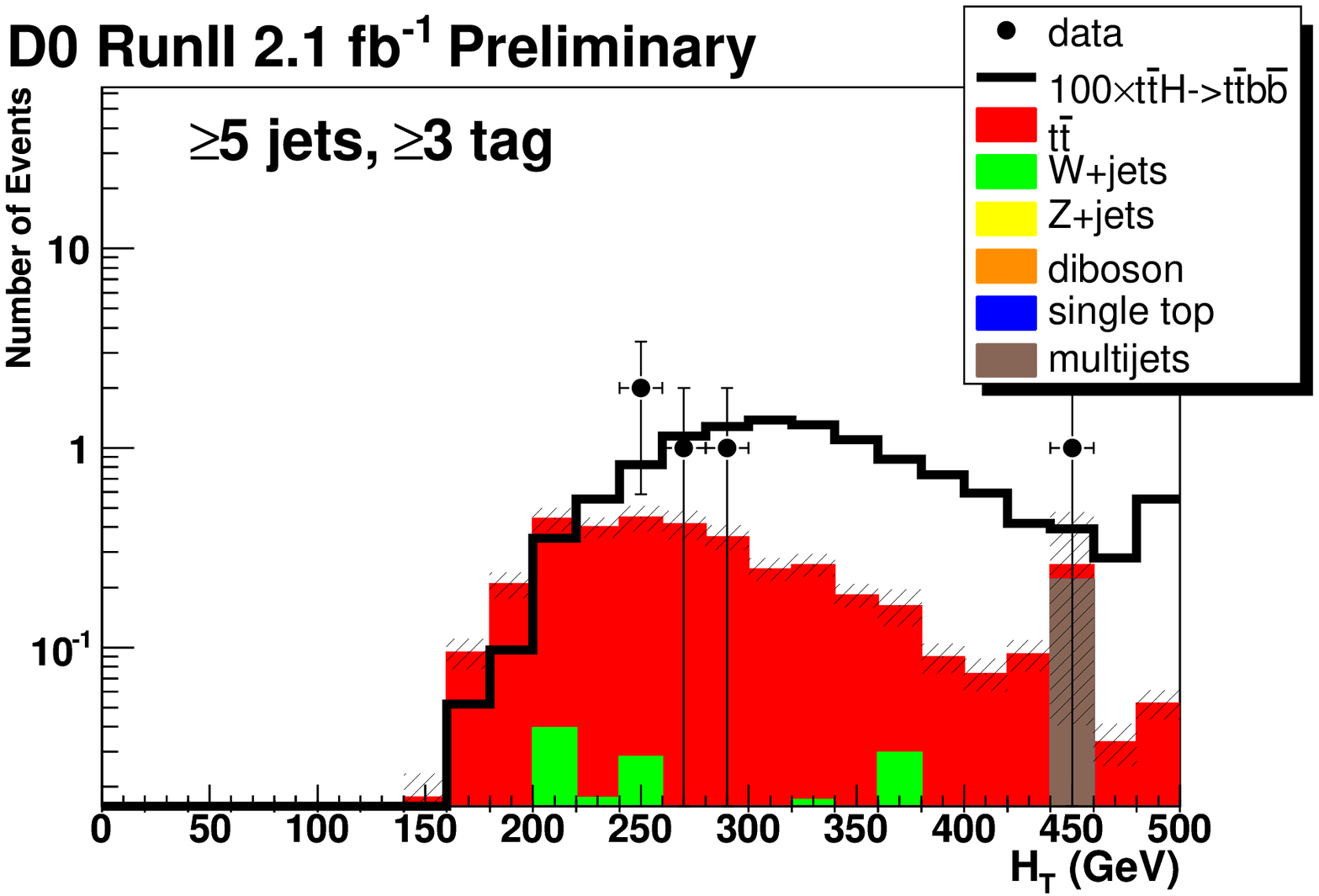} \hfill
\includegraphics[width=0.32\textwidth,height=5cm]{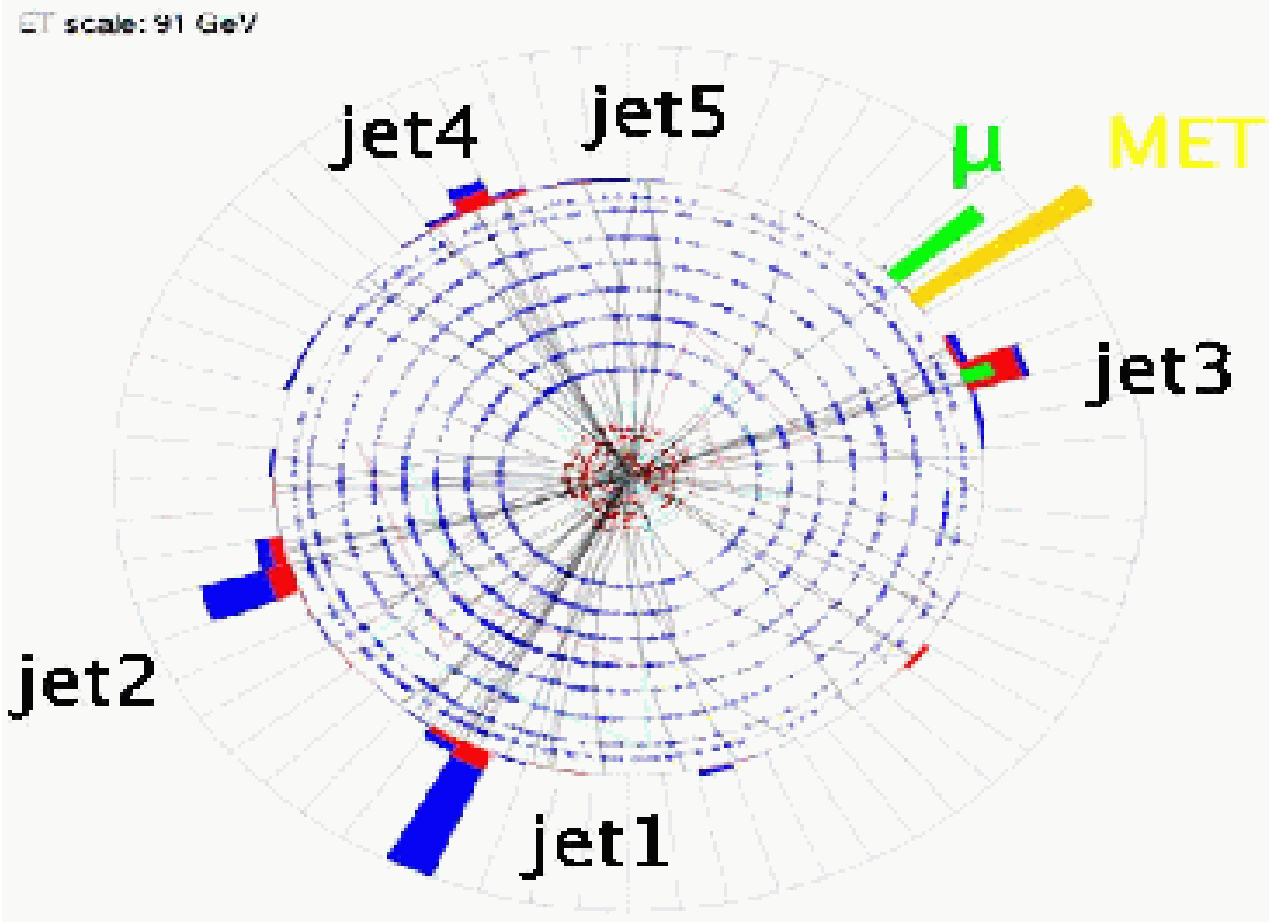} \hfill
\includegraphics[width=0.32\textwidth,height=5cm]{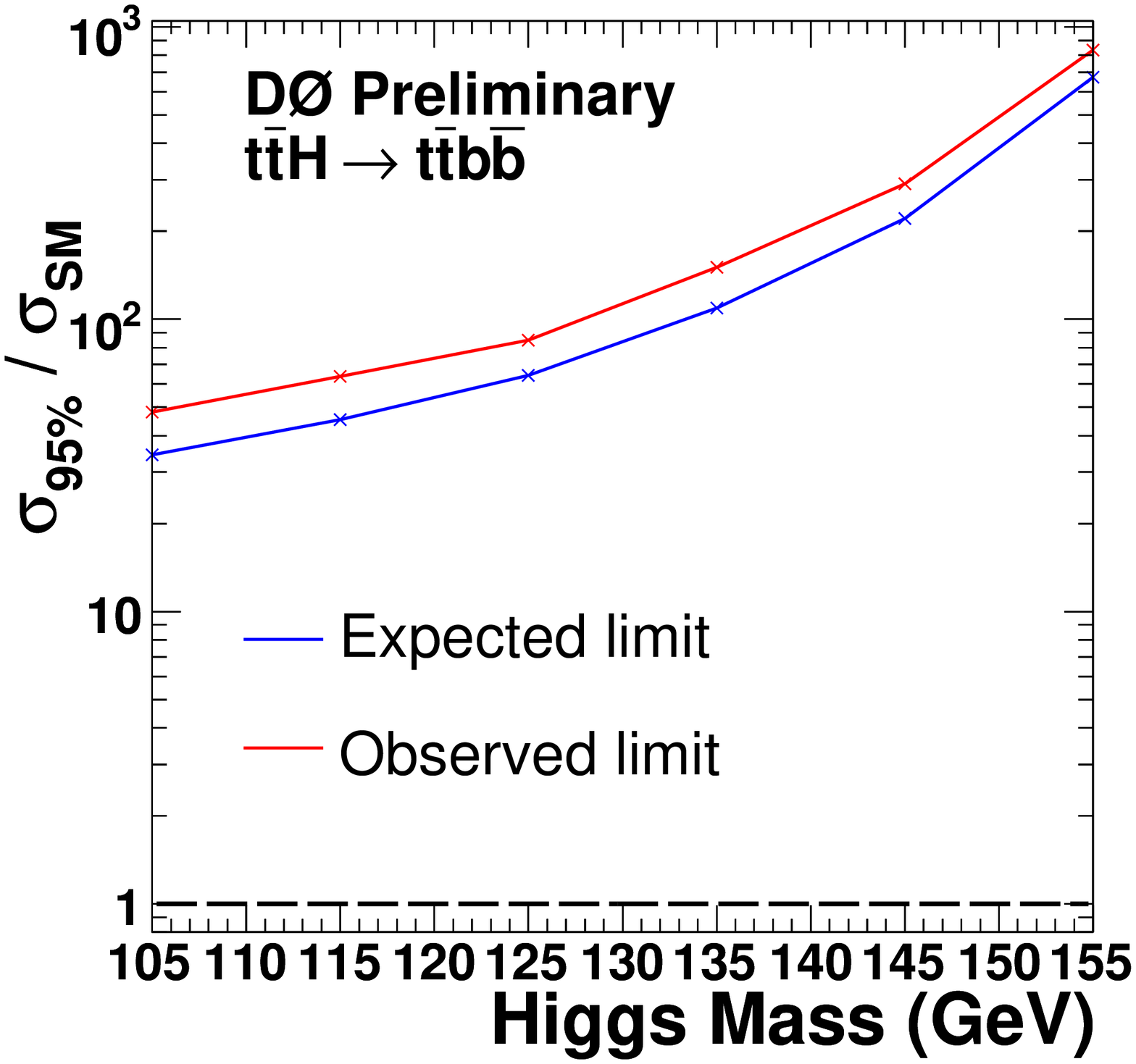}
\vspace*{-0.5cm}
\caption{D\O\ ($\rm t\bar t\to t\bar tH$).
Left: $H_{\rm T}$ distribution.
Center: candidate event.
Right: limit at 95\% CL.
}
\label{fig:d0-sm_ttH}
\vspace*{-0.5cm}
\end{figure}

Both Tevatron collaborations have searched for a $\rm t\bar t\to t\bar tH$ signal.
Results are shown in Figs.~\ref{fig:cdf_sm_ttH} (from~\cite{cdf-sm_ttH})
and~\ref{fig:d0-sm_ttH} (from~\cite{d0-sm_ttH}).

\section{Combined SM Higgs Boson Limits}
\vspace*{-0.1cm}

\begin{figure}[tp]
\begin{center}
\includegraphics[width=0.49\textwidth,height=6cm]{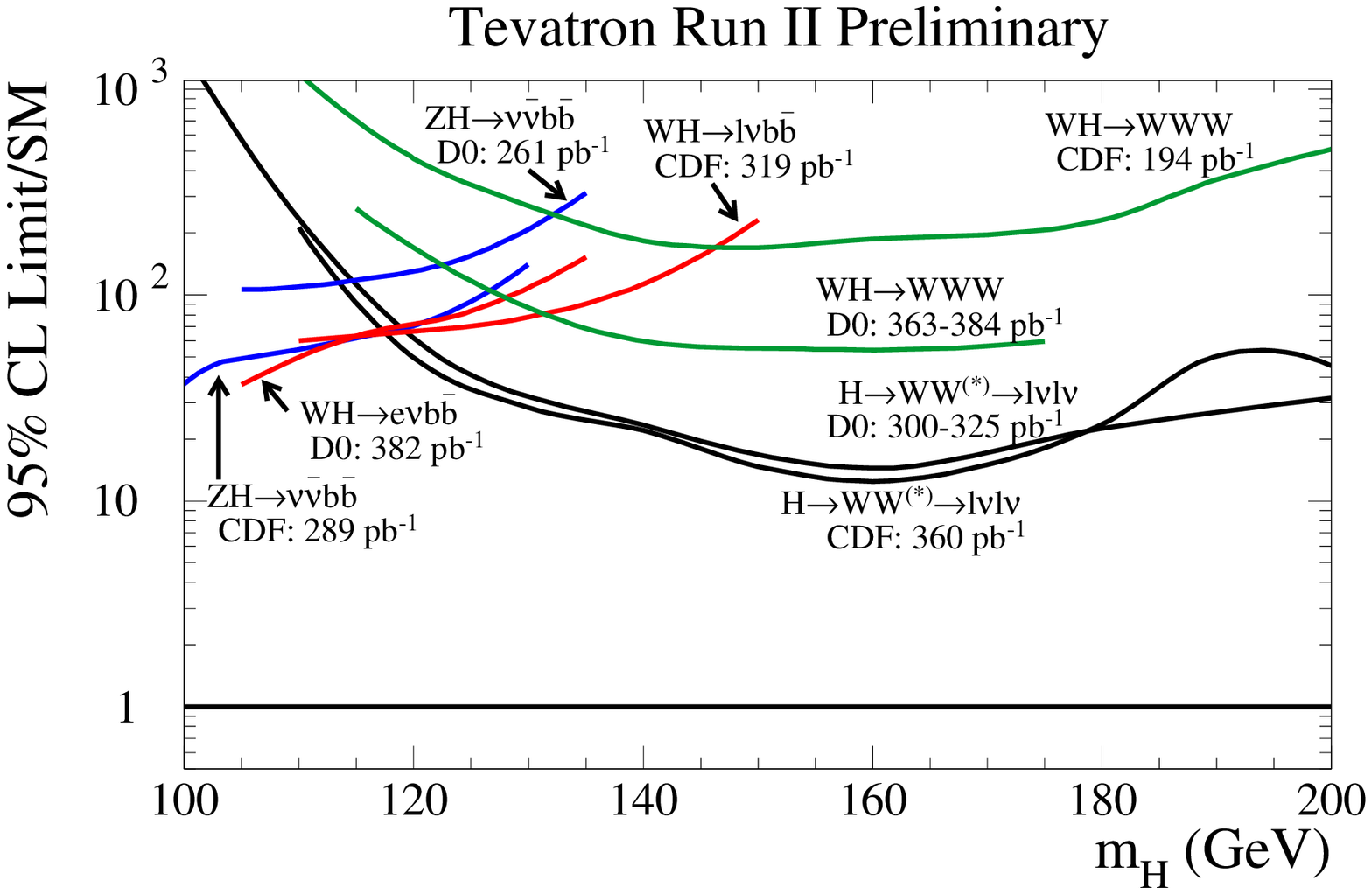}
\includegraphics[width=0.49\textwidth,height=6cm]{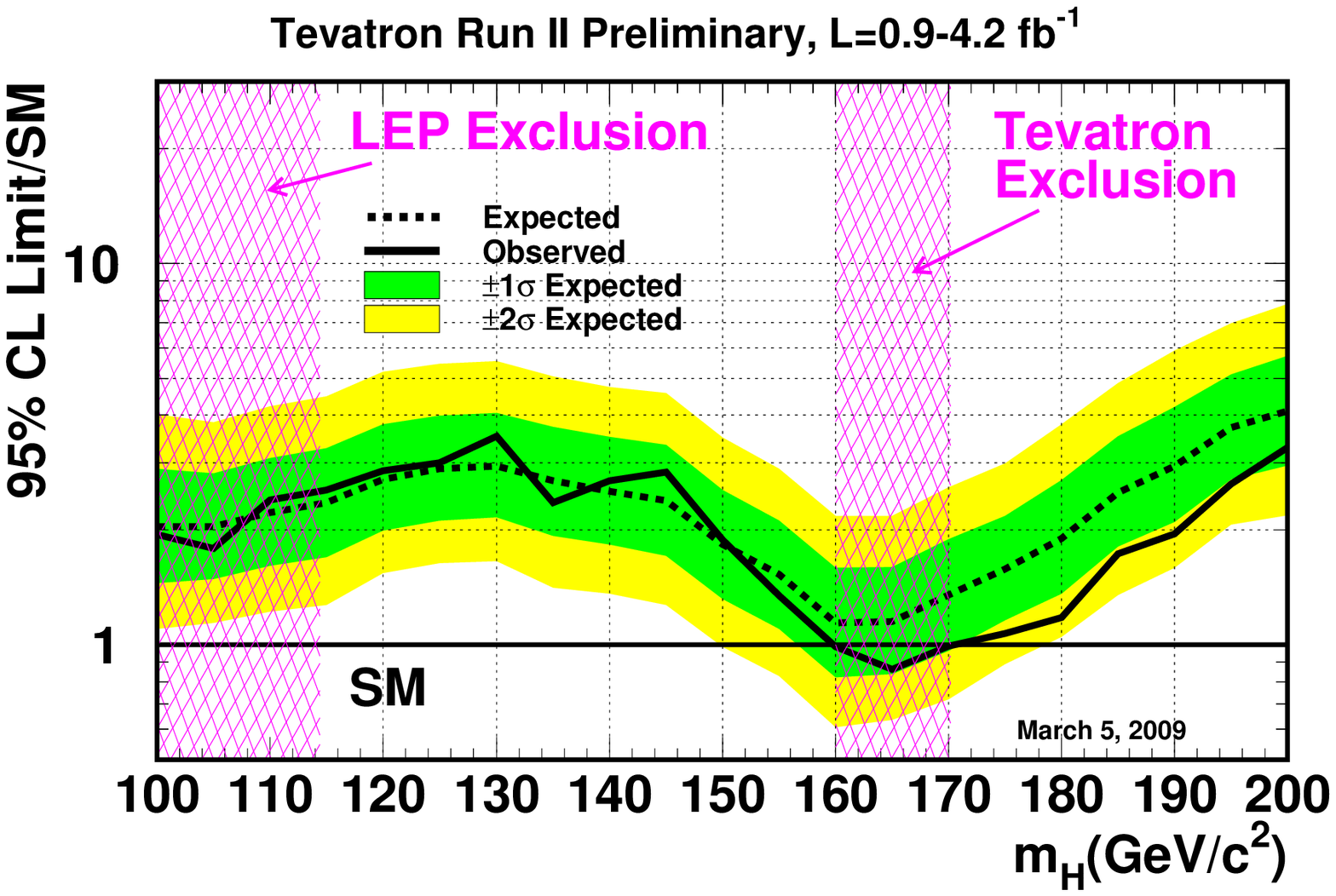}
\vspace*{-0.7cm}
\end{center}
\caption{Comparison of progress between summer 2005 and winter 2008/9.
Left: ratio of observed cross-section limit and expected SM cross-section, status summer 2005.
Right: current combined CDF and D\O\ limits at 95\% CL. 
Status winter 2008/9: Note that a region between 160 and 170 GeV mass is excluded.
}
\label{fig:cdf-d0-sm}
\vspace*{0.4cm}
\end{figure}

The large progress in sensitivity increase between results 
from summer 2005 (left plot from~\cite{as06}) and recent combined 
CDF and D\O\ results (right plot from~\cite{summary-limits})
are shown in Fig.~\ref{fig:cdf-d0-sm} with up to 4.2~fb$^{-1}$.
The achieved sensitivity in the various search channels is summarized in Table~\ref{tab:limits}.

\begin{table}[htb]
\renewcommand{\arraystretch}{0.8} % enlarge line spacing
\caption{Summary of observed and expected limits (where available) as factor compated to the SM expectation at 95\% CL from CDF and D\O. The note numbers refer to CDF and D\O\ notes, respectively.
\label{tab:limits} }
{
\begin{center}
\begin{tabular}{l|c|c|c|r|r|c}
Channel              & experiment & $m_{\rm H}$ &$\cal L$ 
                                                      & \multicolumn{2}{c|}{limit factor}& Reference \\ 
                     &            & (GeV)       &(fb$^{-1}$) & ~~~obs.~~~ &  exp. & note\\ \hline
$\rm H$$\to$WW$\to$$\rm \ell\nu\ell\nu$ 
                           & CDF  &160 &3.6 &1.5  &1.5  & 9500~\cite{cdf-ggHww} \\
                           & D\O  &160 &4.2 &1.7  &1.8  & 5871~\cite{d0-ggHww} \\
$\rm WH\to l\nu\bb$        & CDF  &115 &2.7 &5.6  &4.8  & 9596~\cite{cdf-WHbb} \\
                           & D\O  &115 &2.7 &6.7  &6.4  & 5828~\cite{d0-WHbb} \\
$\rm WH\to WWW$            & CDF  &160 &2.7 &25\phantom{.0} &20\phantom{.0} & 7307~\cite{cdf-WHww} \\
                           & D\O  &160 &3.6 &10\phantom{.0}   &18\phantom{.0}   & 5873~\cite{d0-WHww} \\ 
$\rm ZH\to \ell\ell\bb$    & CDF  &115 &2.7 &7.1  &9.9  & 9665~\cite{cdf-llbb} \\
                           & D\O  &115 &4.2 &9.1  &8.0  & 5876~\cite{d0-llbb} \\
$\rm ZH\to \nn\bb$         & CDF  &115 &2.1 &6.9  &5.6  & 9642~\cite{cdf-zHbb} \\
                           & D\O  &115 &2.1 &7.5  &8.4  & 5586~\cite{d0-zHbb} \\
W/ZH,\,VBF,\,ggH           & CDF  &115 &2.0 &31\phantom{.0} &25\phantom{.0} & 9248~\cite{cdf-sm_tautau} \\
$(H\to\tau^+\tau^-)$       & D\O  &115 &1.0 &27\phantom{.0} &28\phantom{.0} & 5883~\cite{d0-sm_tautau} \\
$\rm H \to \gamma\gamma$   & CDF  &--  &3.0 &--   &--   & 9586~\cite{cdf-sm_gammagamma} \\
                           & D\O  &115 &4.2 &16\phantom{.0} &19\phantom{.0} & 5858~\cite{d0-sm_gammagamma} \\
$\rm t\bar tH$             & CDF  &--  &0.3 &--   &--   & 9508~\cite{cdf-sm_ttH} \\
                           & D\O  &115 &2.1 &64\phantom{.0} & 45\phantom{.0}& 5739~\cite{d0-sm_ttH}
\end{tabular}
\end{center}
}
\end{table}

\begin{figure}[hp]
\vspace*{-0.4cm}
\includegraphics[width=0.49\textwidth,height=6cm]{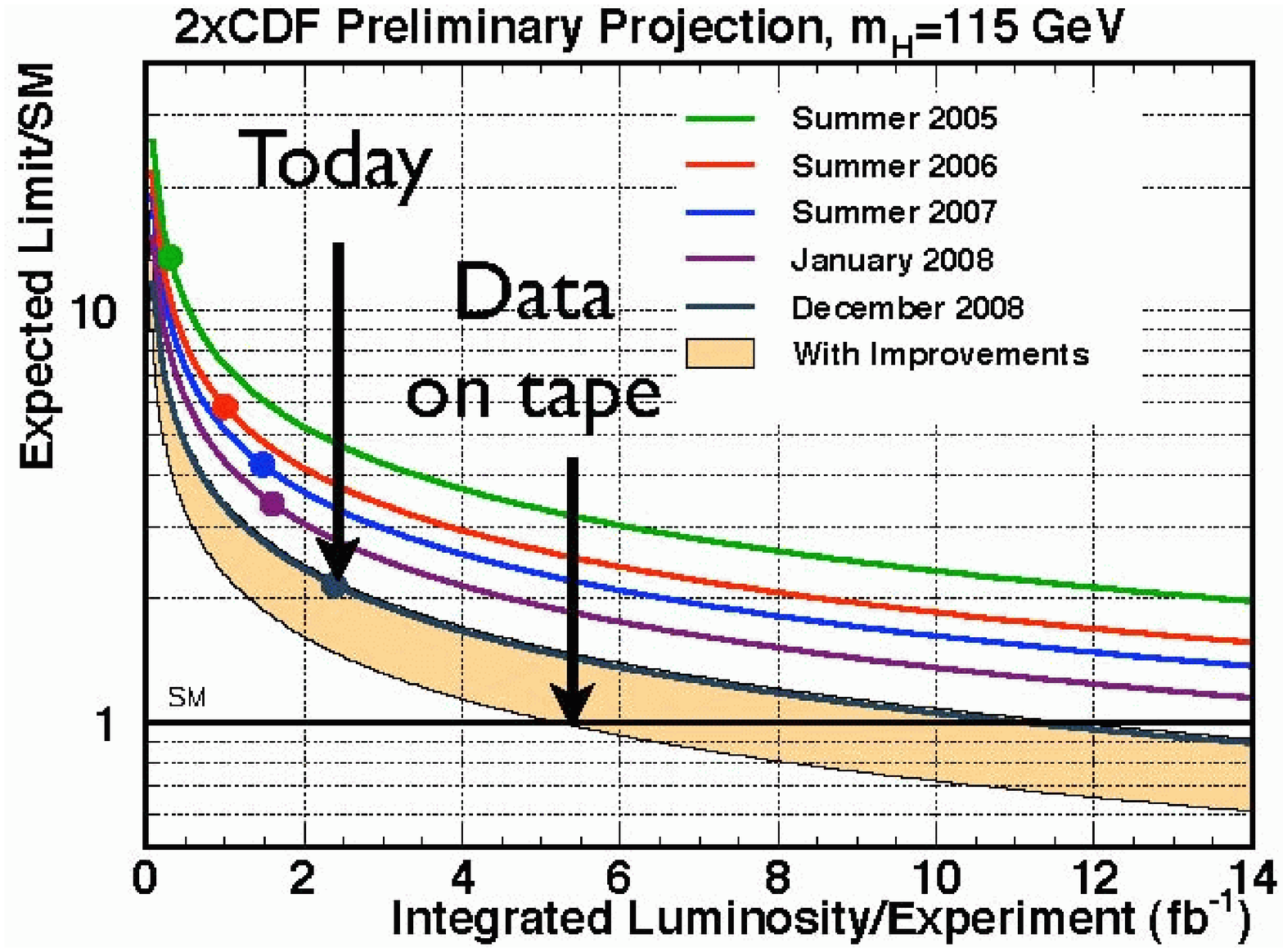} \hfill
\includegraphics[width=0.49\textwidth,height=6cm]{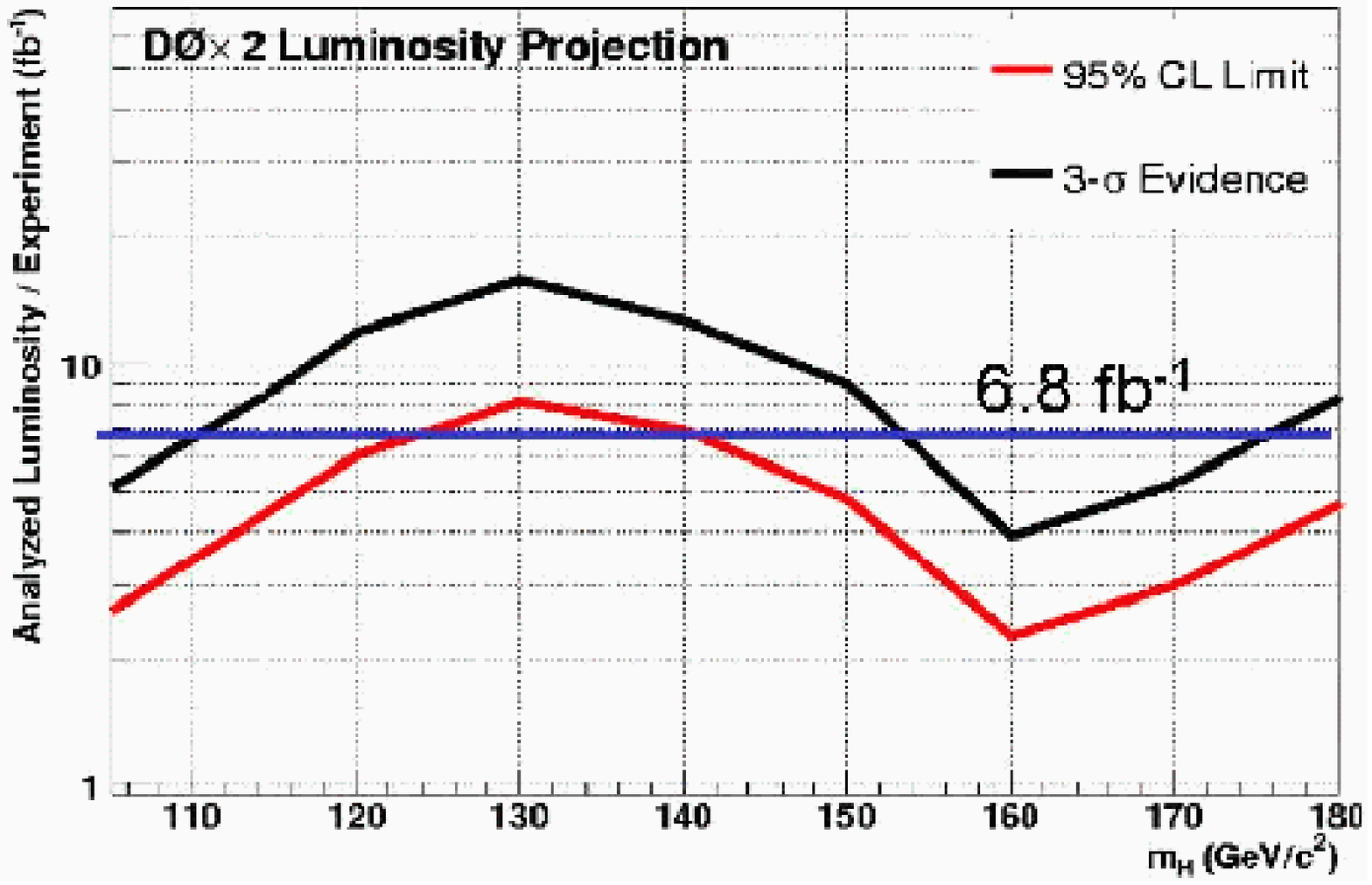}
\vspace*{-0.2cm}
\caption{Outlook.
Left: 
for a 115~GeV Higgs boson.
Right: for a mass range between 105 to 180~GeV.
}
\label{fig:sm_outlook}
\end{figure}

Improvements will continue to come from optimized b-quark tagging, 
and also from larger e/$\mu$ acceptance, 
better jet mass resolution, 
and from using advanced analysis techniques.
Higher Higgs boson sensitivities will also result from the increase in luminosity. 
Currently (winter 2008/9) about 5~fb$^{-1}$ are recorded per experiment,
and the total delivered luminosity will increase up to about 8.5~fb$^{-1}$ 
(6.8~fb$^{-1}$ recorded) as illustrated in Fig.~\ref{fig:lumi}.
Resulting sensitivity estimates are shown in Fig.~\ref{fig:sm_outlook} (from~\cite{sm_outlook}).

\clearpage
\section{Beyond the SM}

\subsection{$\rm \bb h$, $\rm \bb H $, $\rm \bb A$}

\begin{figure}[htbp]
\vspace*{-0.4cm}
\begin{minipage}{0.4\textwidth}
\includegraphics[width=\textwidth]{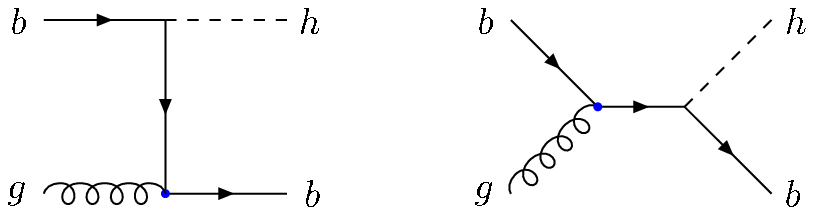}
\includegraphics[width=\textwidth]{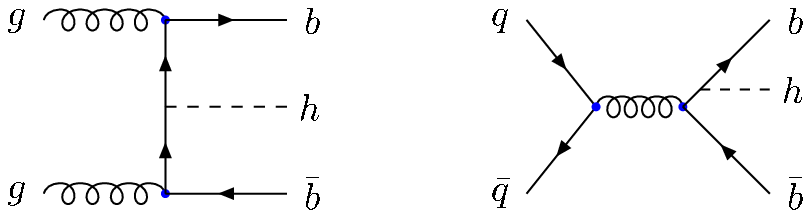}
\end{minipage} \hfill
\begin{minipage}{0.4\textwidth}
\includegraphics[width=0.9\textwidth]{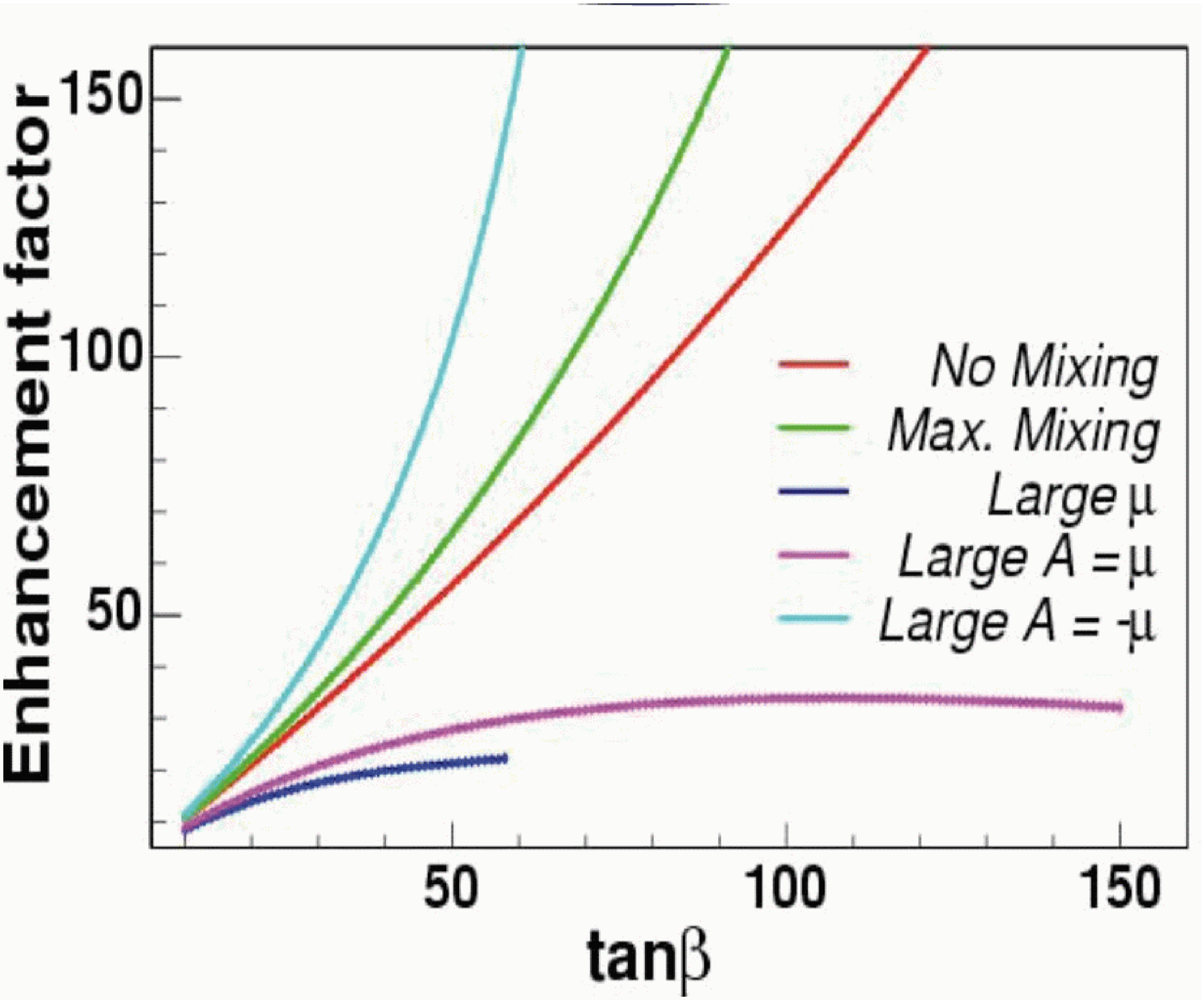}
\end{minipage}
\vspace*{-0.3cm}
\caption{D\O. Left: leading-order Feynman diagrams for 
               neutral Higgs boson production in the five-flavor scheme (top) and 
               four-flavor scheme (bottom).
Right: enhancement factor as a function of $\tan\beta$.
        } \label{fig:bba}
\end{figure}

Higgs boson production processes in association with b-quarks in $\rm p \bar p$
collisions have been calculated in two ways: in the five-flavor scheme~\cite{5fns}, 
where only one b-quark has to be present in the final state, while in the four-flavor 
scheme~\cite{4fns}, two b-quarks are explicitly required in the calculation. 
Both calculations are 
available at next-to-leading order (NLO QCD), and agree taking into account the theoretical 
uncertainties.
Figure~\ref{fig:bba} (from~\cite{d0-bba}) illustrates these processes for h production at 
leading order (LO), 
and analogous diagrams can be drawn for the H and A bosons.
The cross-section depends on $\tan^2\beta$ and on other 
Supersymmetric parameters as given by 
$
\sigma\times BR_{\rm SUSY} \approx 2 \sigma_{\rm SM} \tan^2\beta/(1+\delta_b)^2 \times 
9/(9+(1+\delta_b)^2),
$
where $\delta_b = k \tan\beta$ with $k$ depending on the SUSY parameters, 
in particular also on $A_t$, 
the mixing in the scalar top sector, the gluino mass, the $\mu$ parameter, stop and sbottom masses.
The dependence of the production cross-section enhancement factor on $\tan\beta$ is shown in 
Fig.~\ref{fig:bba} (right plot from~\cite{d0-bba}). At tree-level the production cross-section rises with 
$\tan^2\beta$.

\begin{figure}[htbp]
\vspace*{-0.2cm}
\centering
\includegraphics[width=0.32\textwidth,height=6cm]{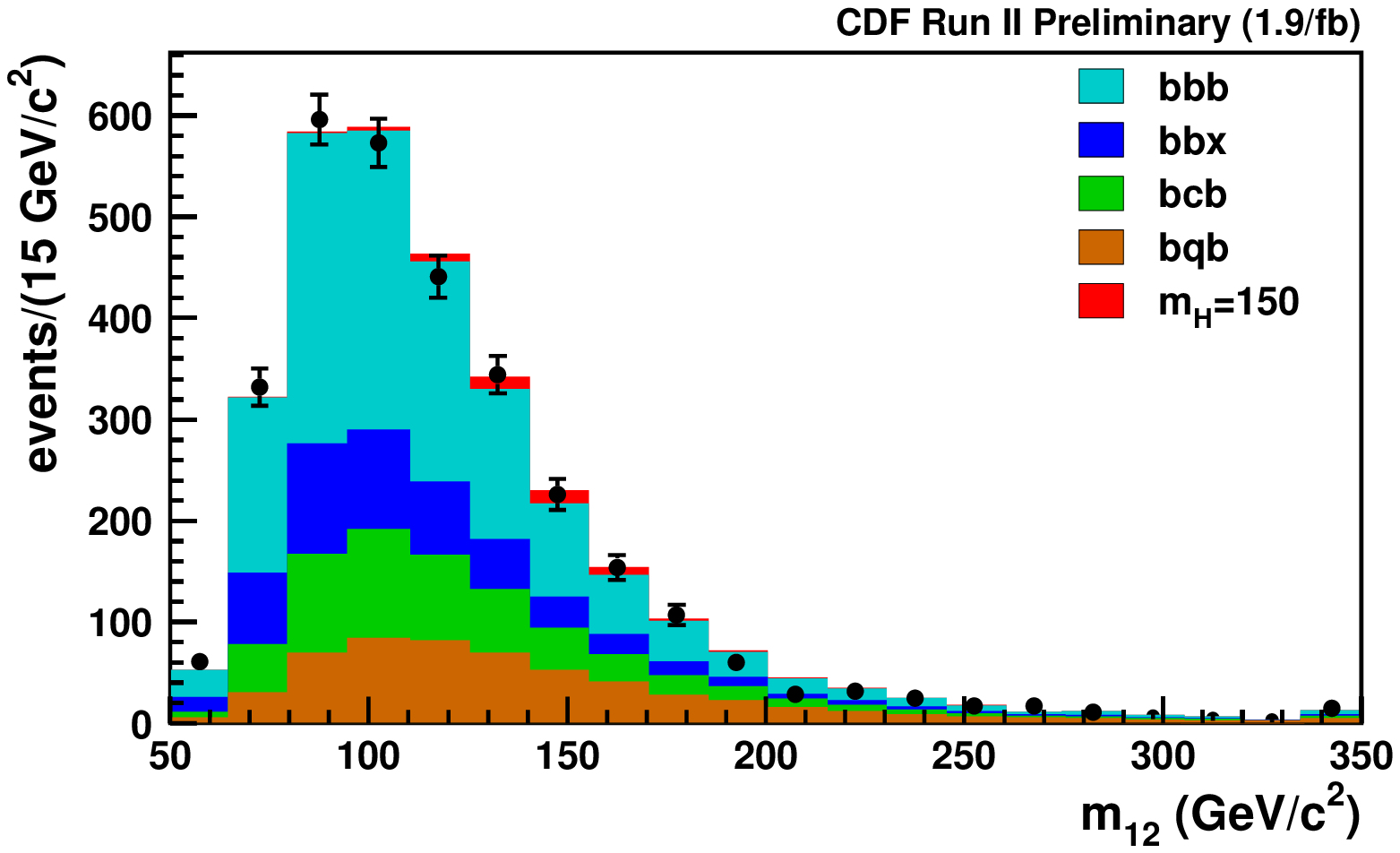} \hfill
\includegraphics[width=0.32\textwidth,height=6cm]{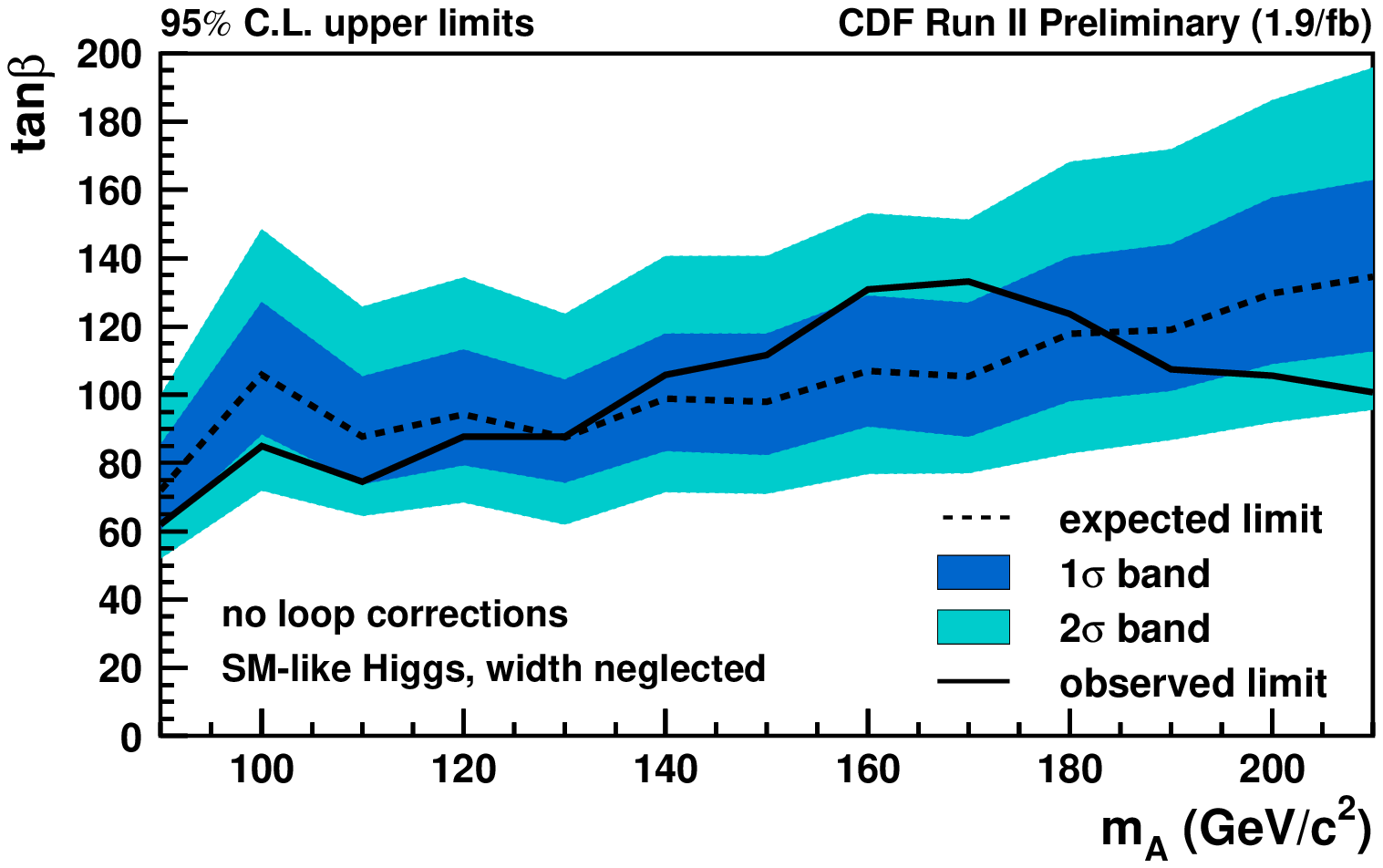} \hfill
\includegraphics[width=0.32\textwidth,height=6cm]{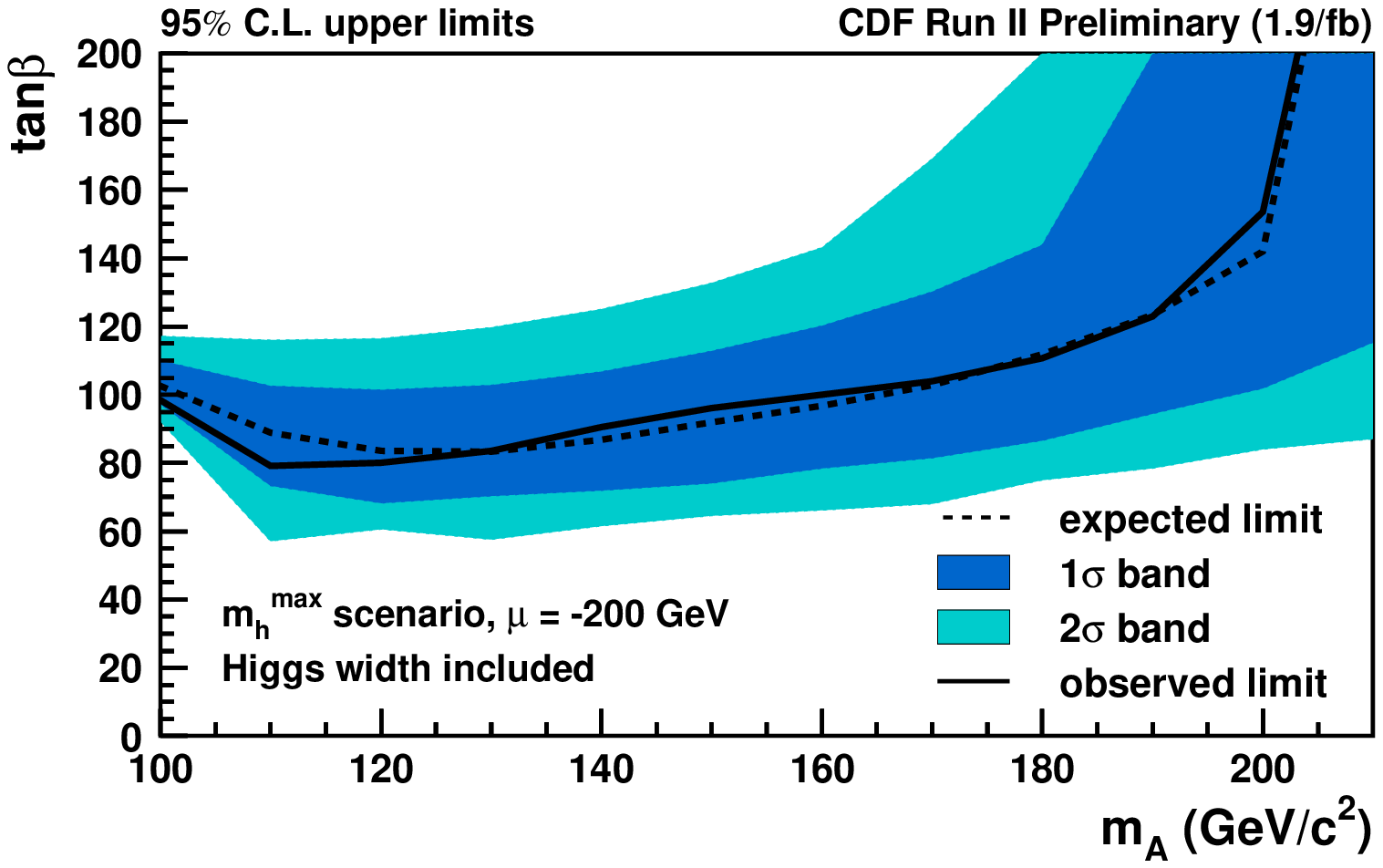}
\vspace*{-0.5cm}
\caption{
CDF. $\rm b\bar b A$($\rm A\to \bb$).
Left: invariant mass of the two most energetic jets $m_{\rm A}=150$~GeV.
Center: limits on $\tan\beta$ in the gereral Two Higgs Doublet Model (THDM).
Right: limits on $\tan\beta$ in the MSSM for the mhmax scenario.
} 
\label{fig:cdf-bba-xsec}
\vspace*{-0.6cm}
\end{figure}

\begin{figure}[htbp]
\centering
\includegraphics[width=0.32\textwidth,height=6cm]{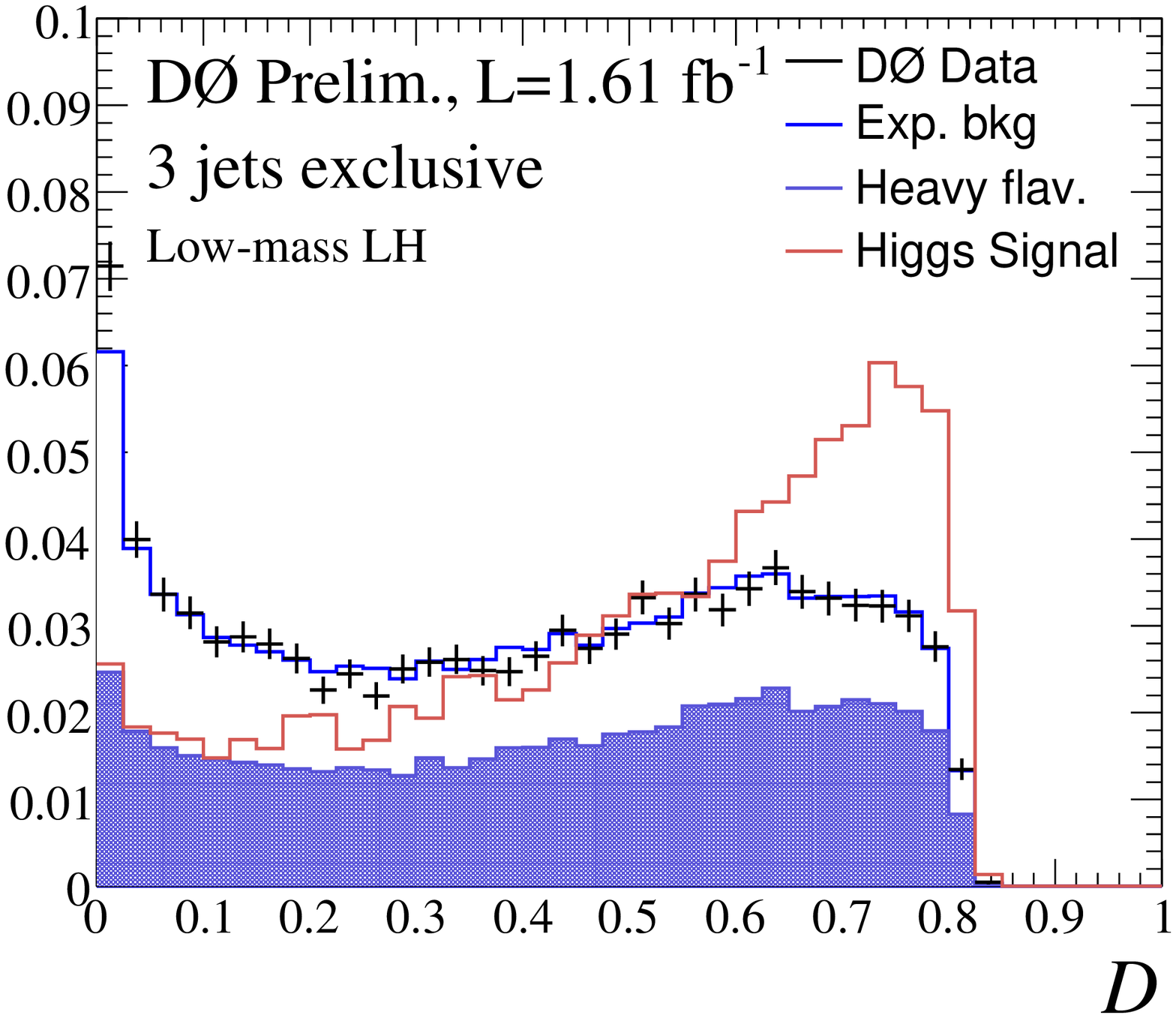} \hfill
\includegraphics[width=0.32\textwidth,height=6cm]{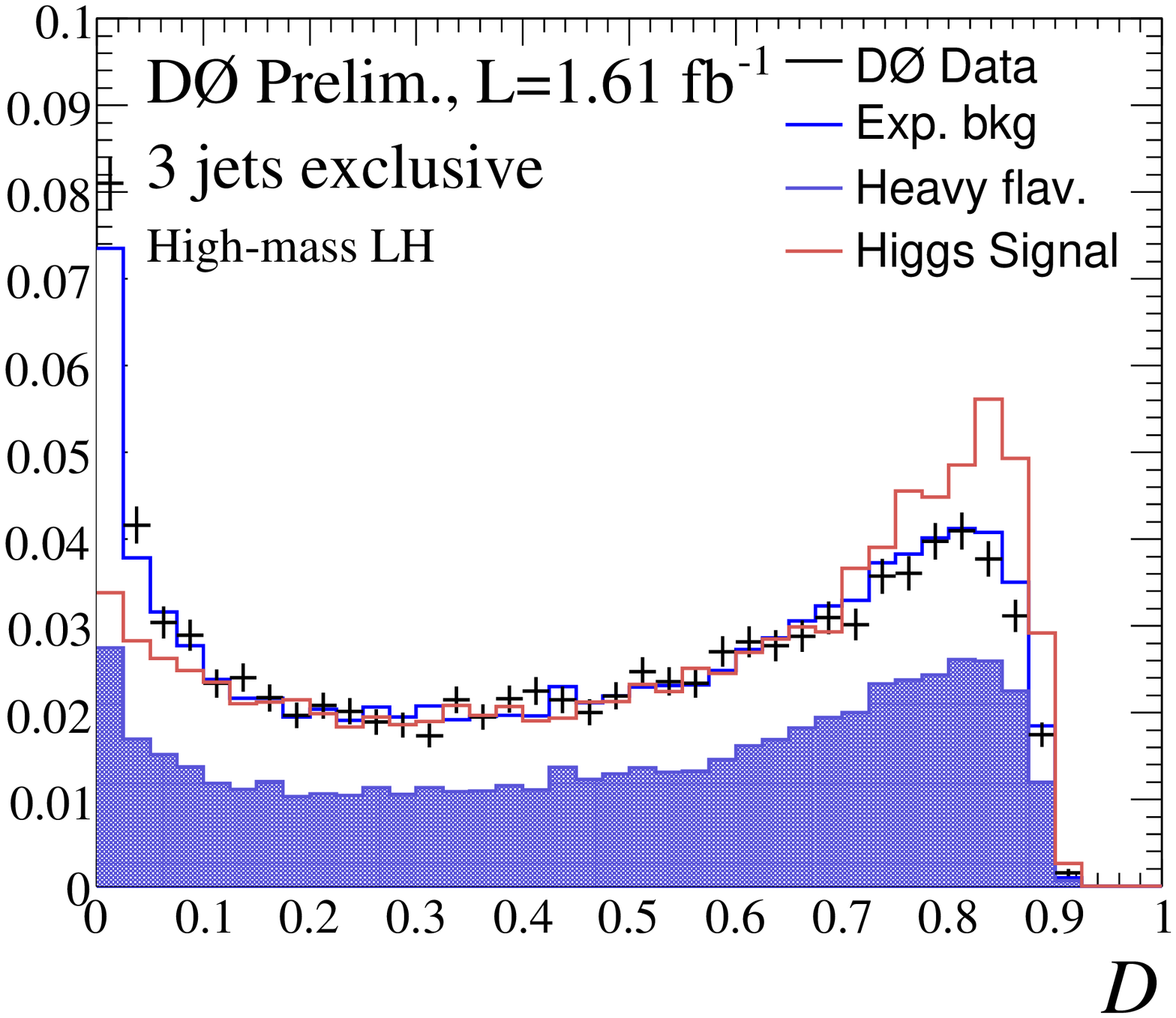} \hfill
\includegraphics[width=0.32\textwidth,height=6cm]{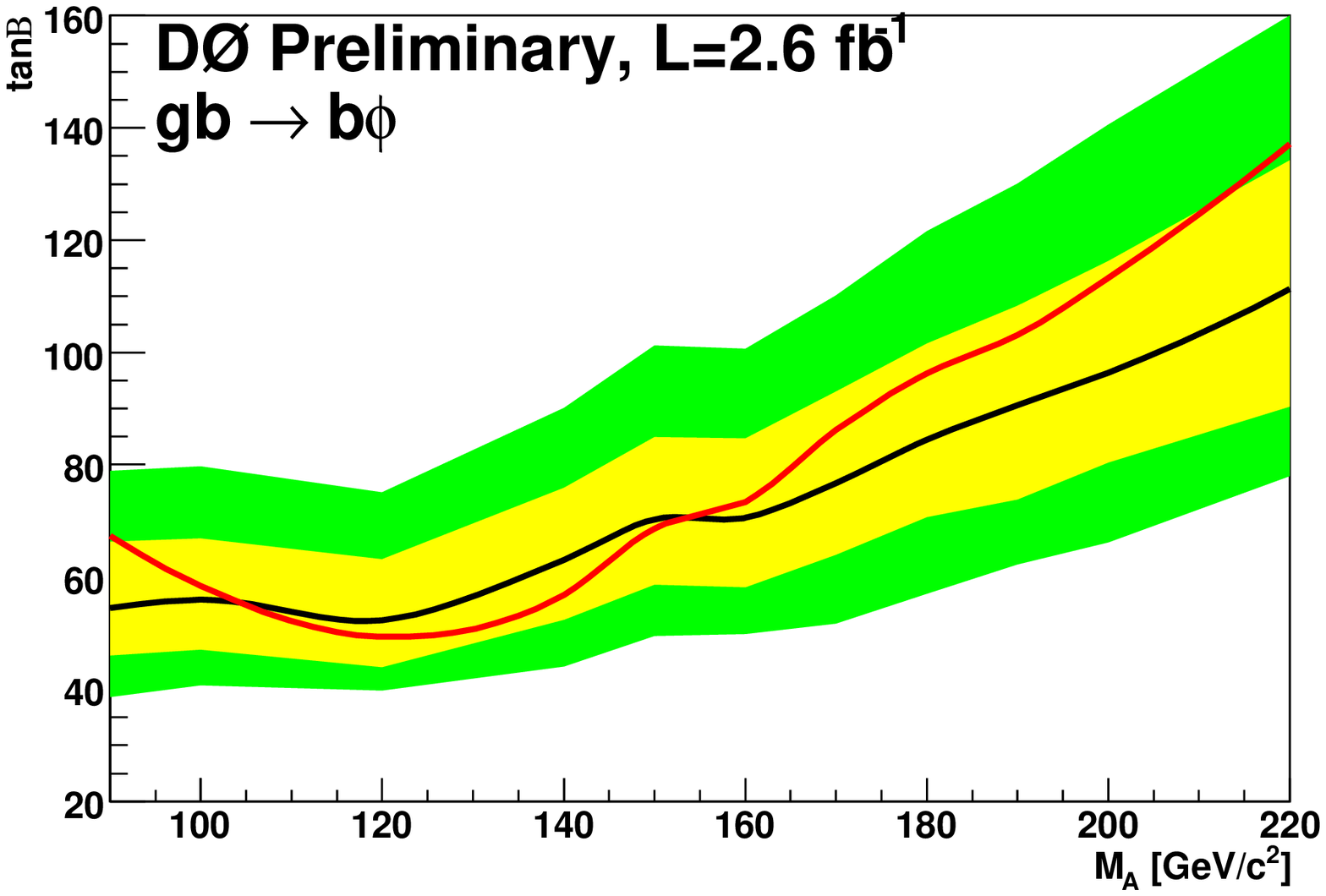}
\vspace*{-0.3cm}
\caption{
D\O. $\rm b\bar b A$($\rm A\to \bb$).
Left: discriminant variable output, low-mass.
Center: discriminant variable output, high-mass.
Right: limit at 95\% CL.
}
\label{fig:d0-bba-xsec}
\vspace*{-0.1cm}
\end{figure}

There is no indication of a $\rm \bb A$ production in the data.
Results from CDF and D\O\ are shown in 
Fig.~\ref{fig:cdf-bba-xsec} (from~\cite{cdf-bbA}) and
Fig.~\ref{fig:d0-bba-xsec} (from~\cite{d0-bba2008}). 
In the CDF results based on 1.9~fb$^{-1}$ data statistical and systematic errors 
contribute about equally, therefore, with about 8~fb$^{-1}$ data,
cross-section sensitivities could be improved by about 20\%, thus $\tan\beta$ sensitivities by about
10\%. Estimates reported in 2005~\cite{as06} were too optimistic.

\subsection{$\rm h,H,A\to \tautau$}
\begin{figure}[htbp]
\includegraphics[width=0.32\columnwidth,height=6cm]{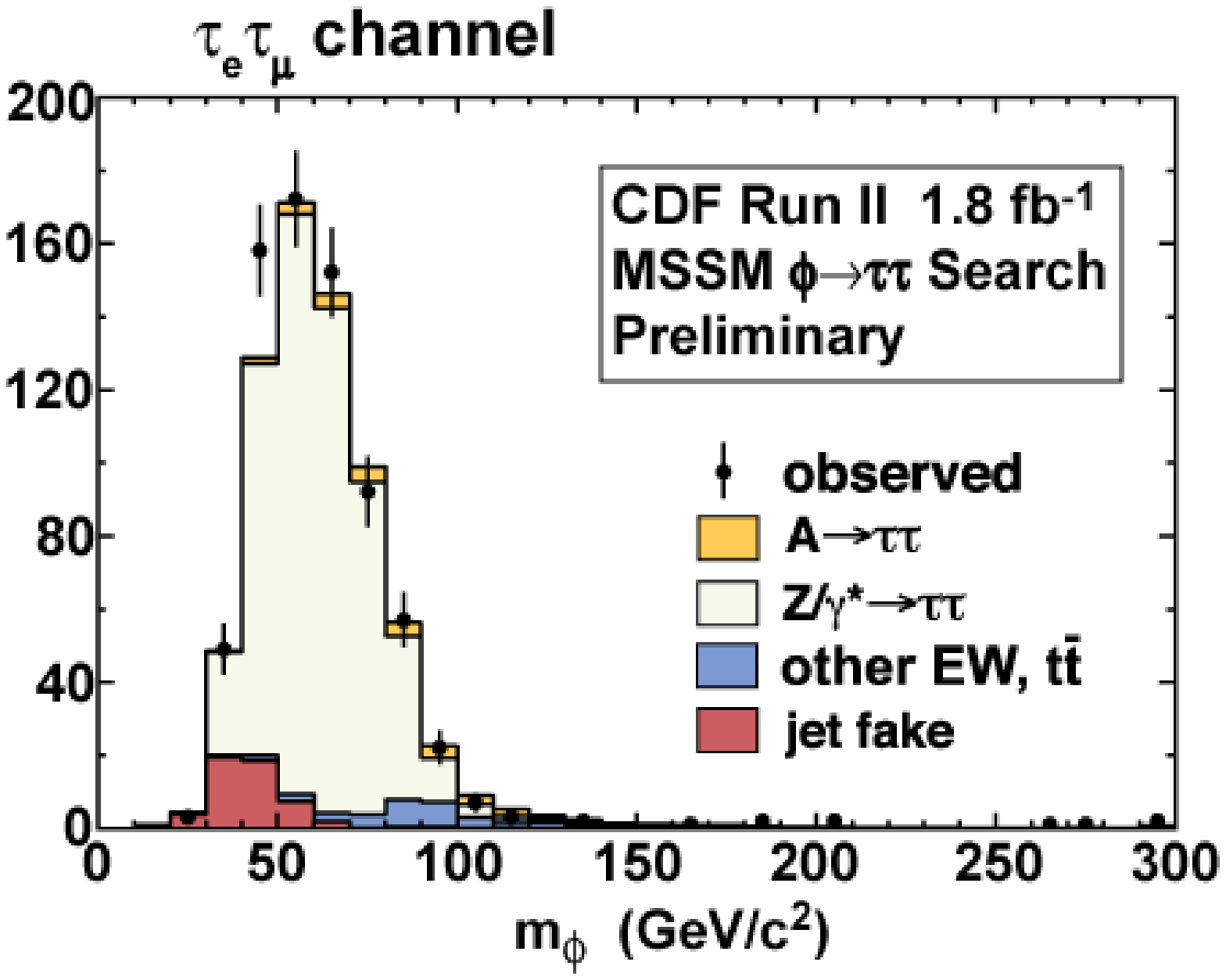} \hfill
\includegraphics[width=0.32\columnwidth,height=6cm]{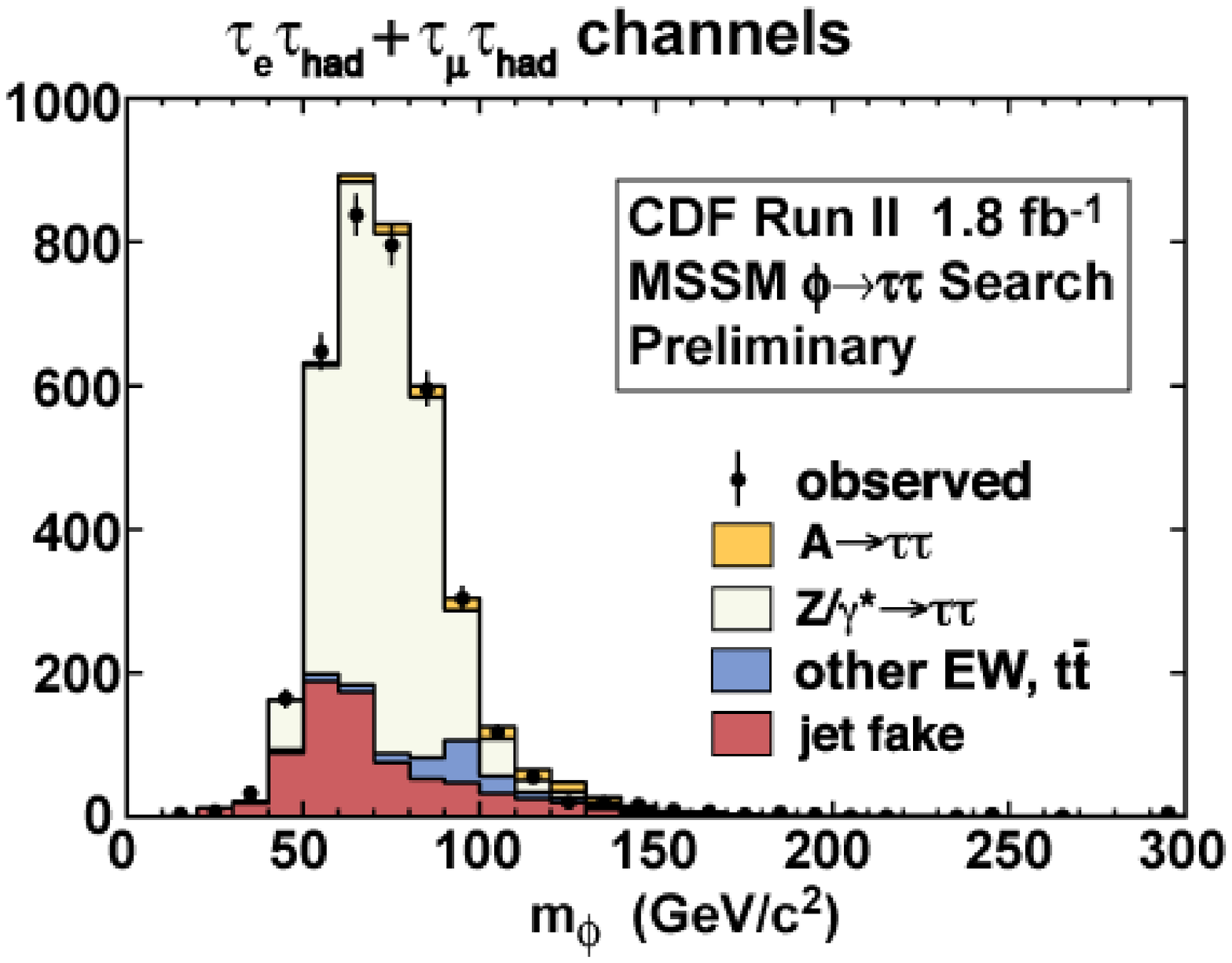} \hfill
\includegraphics[width=0.32\columnwidth,height=6cm]{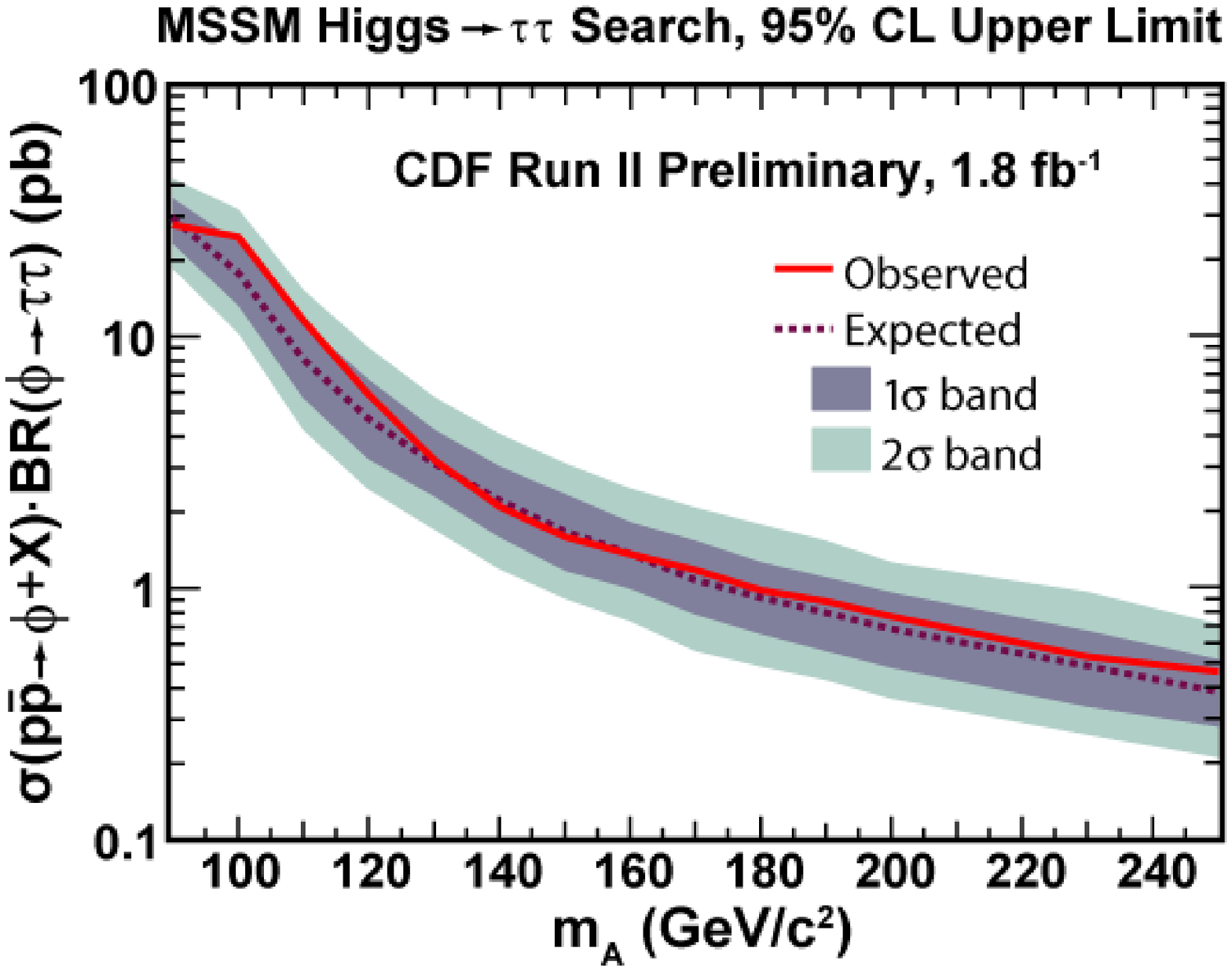}
\vspace*{-0.4cm}
\caption{
CDF. 
Left: invariant mass $\rm e\mu$ channel for $\rm \Phi=A$ with $m_{\rm A}$ = 140~GeV.
Center: invariant mass $\rm \ell\tau_h$ channel where $\ell$ represents an electron or a muon.
Right: cross-section limit at 95\% CL.
} \label{fig:cdf_tautau}
\end{figure}

\begin{figure}[bp]
\vspace*{-0.5cm}
\includegraphics[width=0.32\columnwidth,height=6cm]{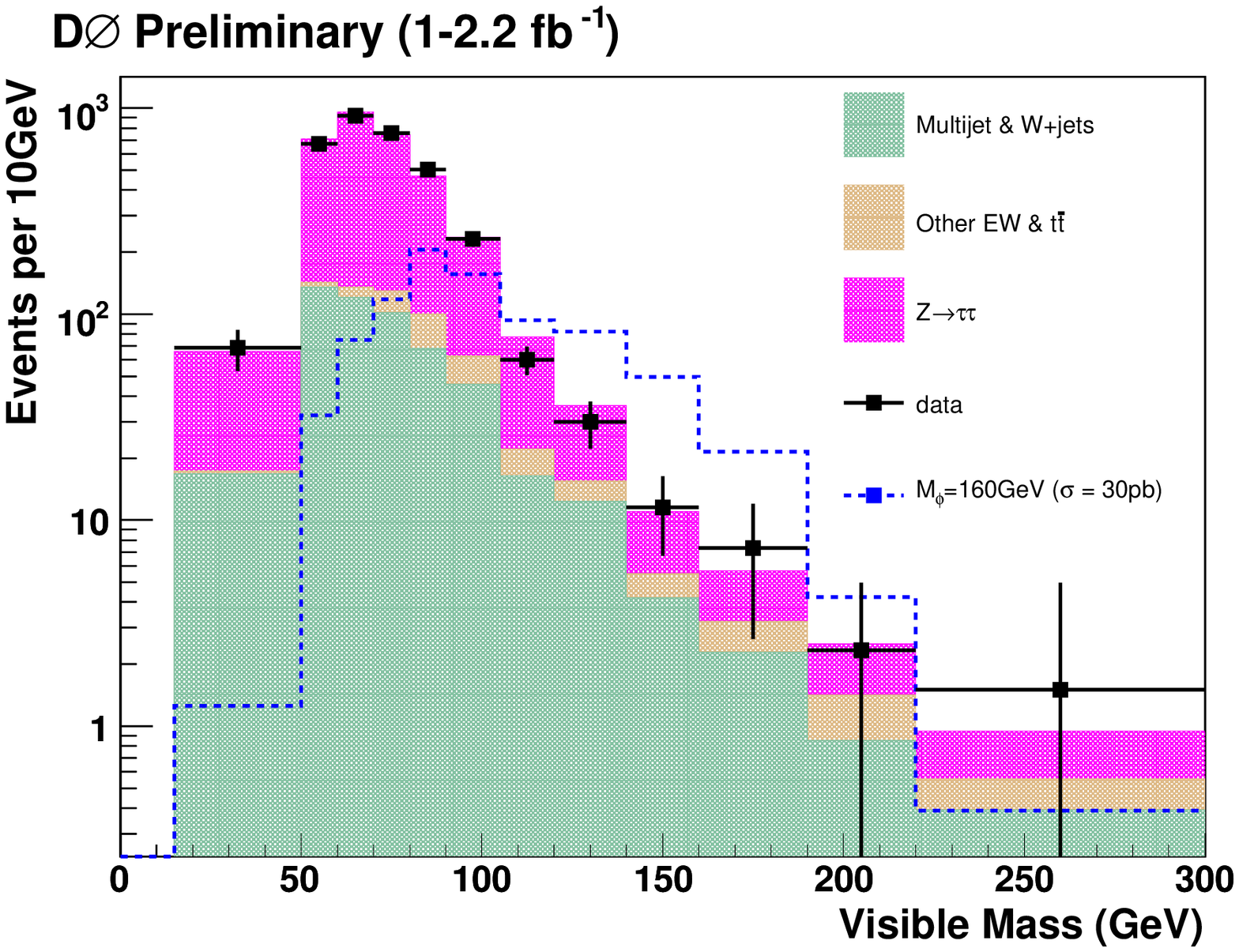} \hfill
\includegraphics[width=0.32\columnwidth,height=6cm]{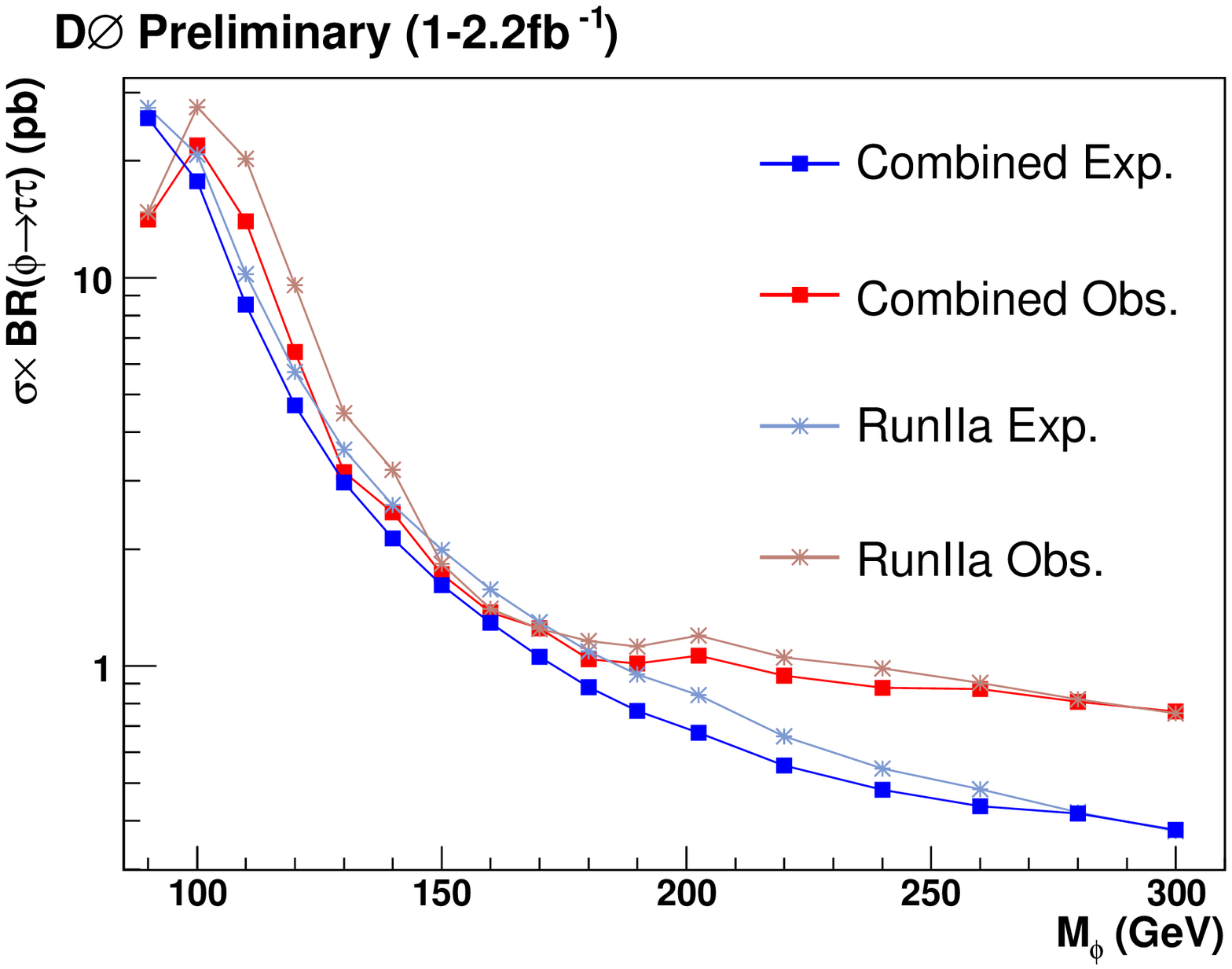} \hfill
\includegraphics[width=0.32\columnwidth,height=6cm]{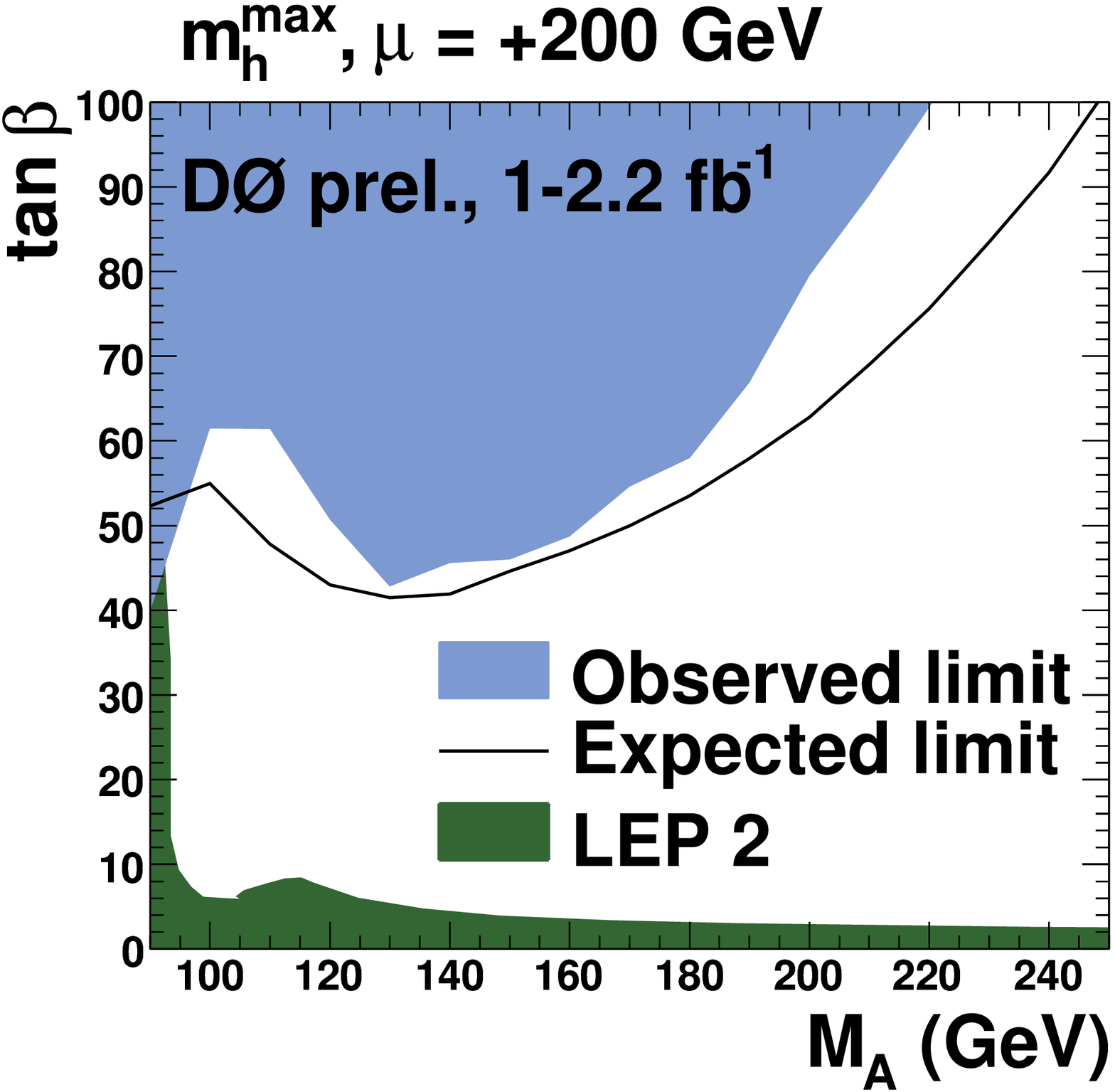}
\vspace*{-0.4cm}
\caption{
D\O. 
Left: visible mass for $\rm \Phi=A$ with $m_{\rm A}$ = 160~GeV.
Center: cross-section limit at 95\% CL.
Right: MSSM limit at 95\% CL.
} \label{fig:d0_tautau}
\end{figure}

\vspace*{-0.1cm}
The signature for $\rm h,H,A\to \tautau$ opens additional 
possibilities for a Higgs boson discovery.
Results from CDF and D\O\ are shown in
Fig.~\ref{fig:cdf_tautau} (from~\cite{cdf-tautau}) and
Fig.~\ref{fig:d0_tautau} (from~\cite{d0-tautau}).

\clearpage
\subsection{$\rm H^+$}

\begin{figure}[bp]
\vspace*{-5mm}
\begin{center}
\includegraphics[width=0.49\textwidth,height=6cm]{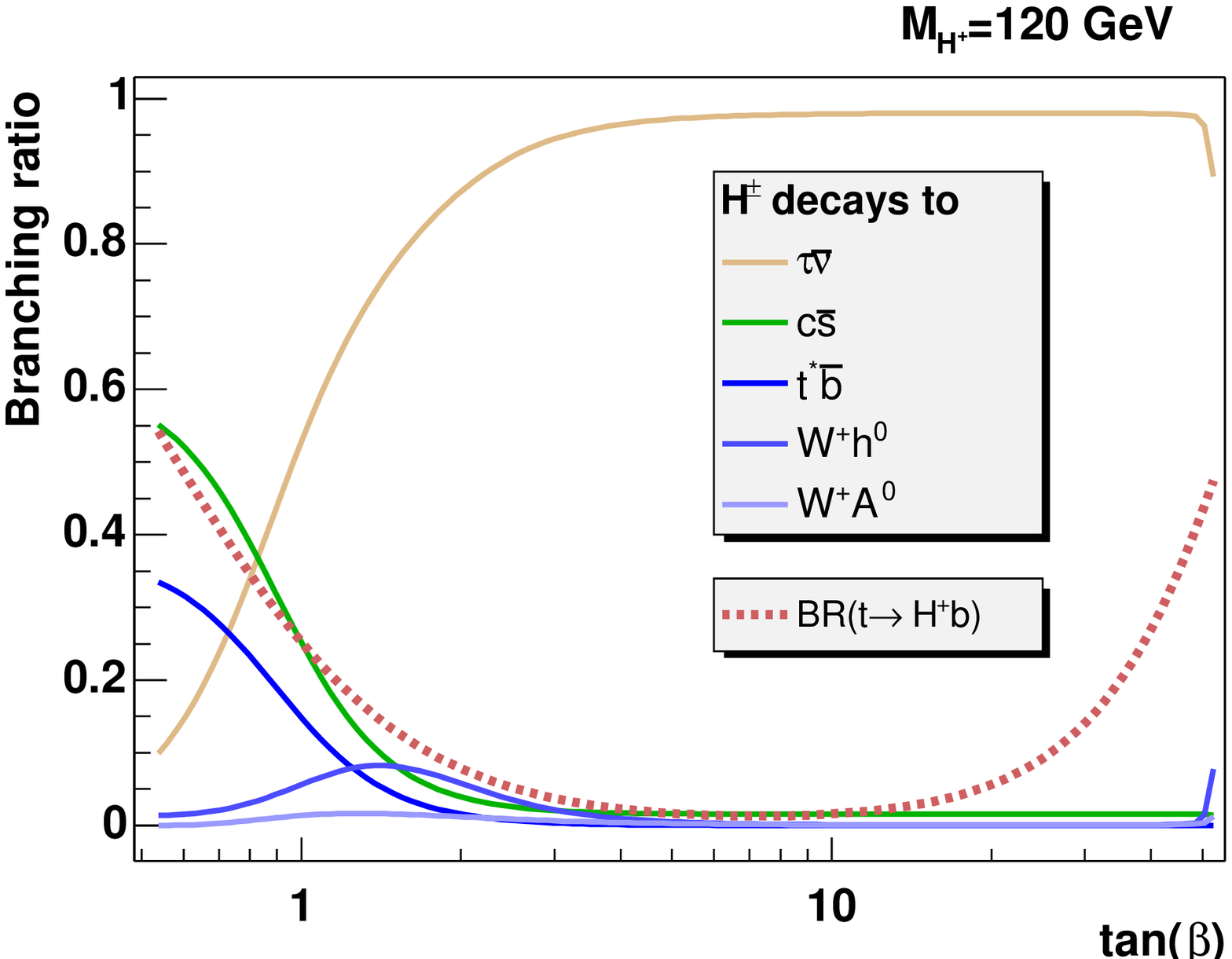} \hfill
\includegraphics[width=0.49\textwidth,height=6cm]{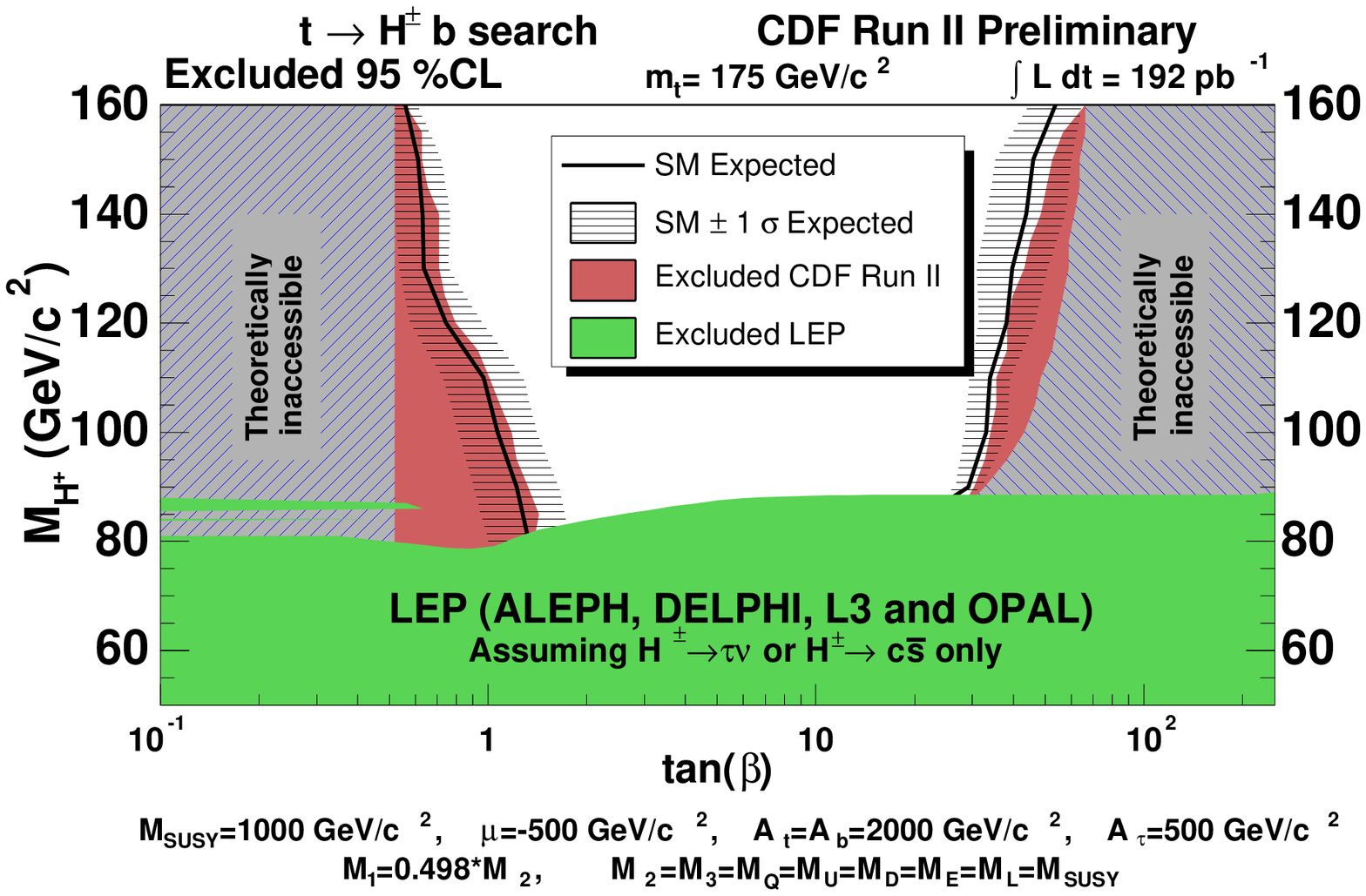}
\end{center}
\vspace*{-5mm}
\caption{
Left: branching ratios for a 120 GeV charged Higgs boson production in top decays and 
charged Higgs boson decays as a function of $\tan\beta$ in the MSSM.
Right: CDF. Limits on the charged Higgs boson mass as function of $\tan\beta$ 
for a specific set of MSSM parameters.
} \label{fig:cdf-tbh}
\end{figure}

\begin{figure}[bhp]
\begin{center}
\includegraphics[width=0.32\textwidth,height=5cm]{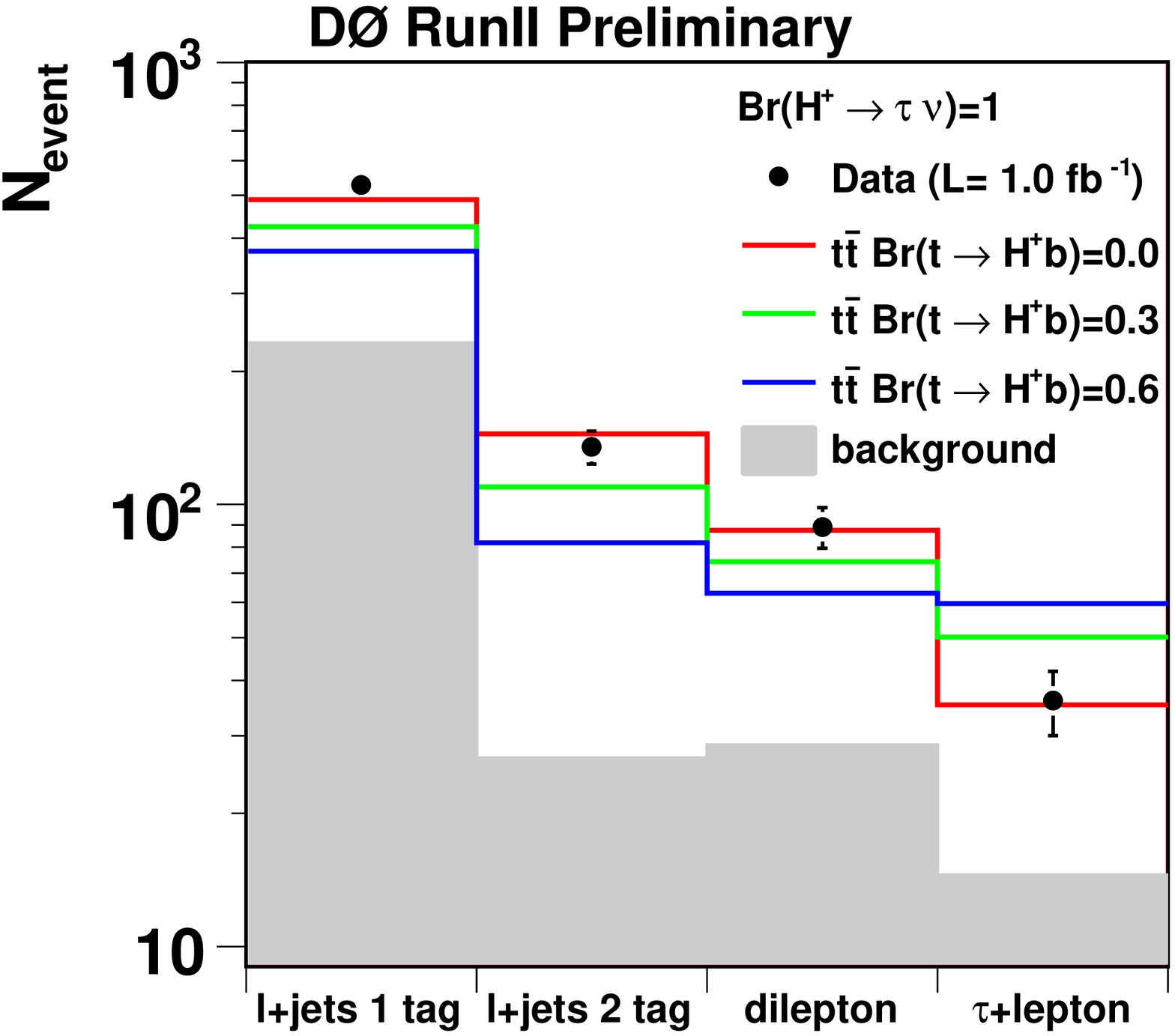}
\includegraphics[width=0.32\textwidth,height=5cm]{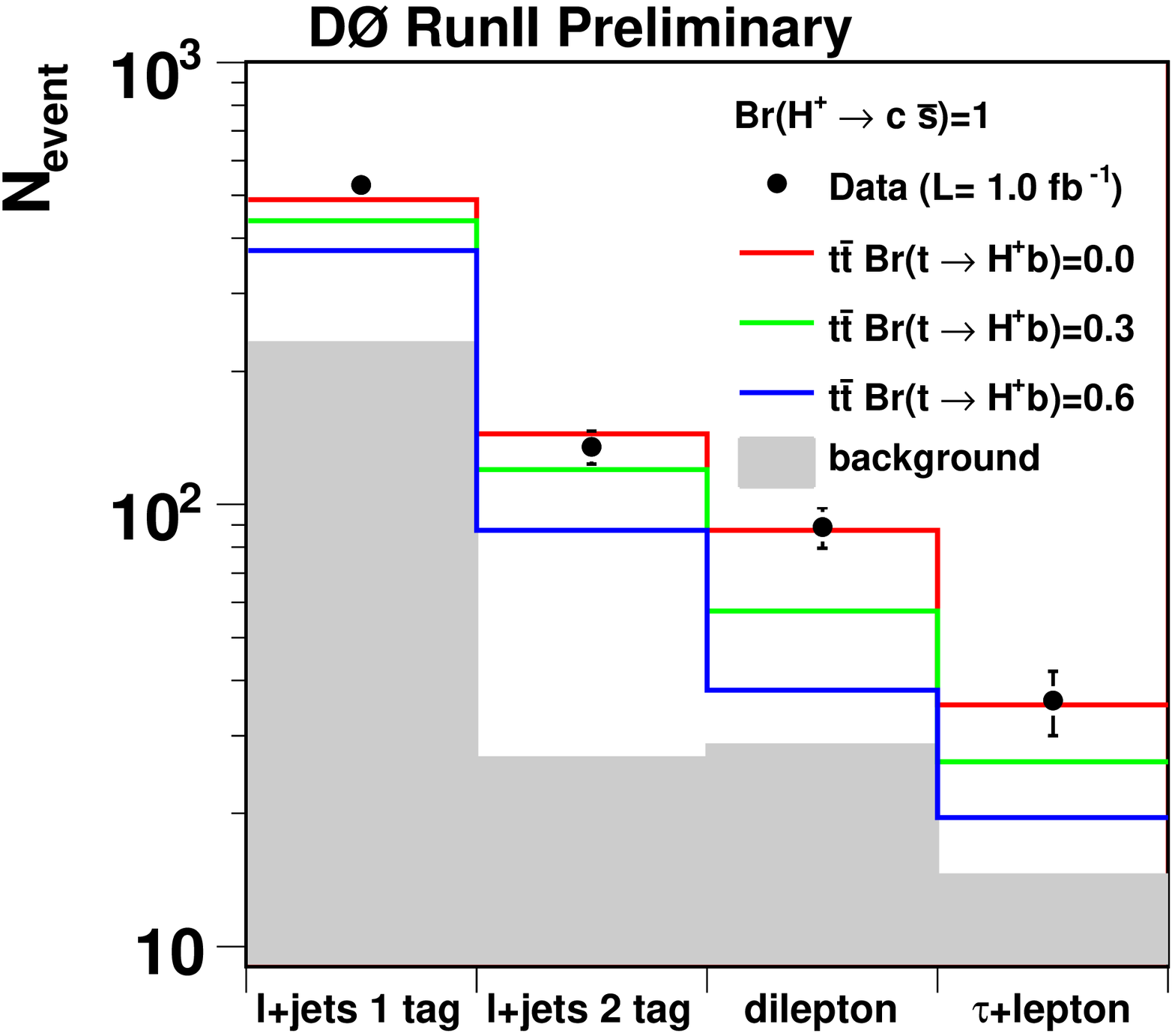}
\includegraphics[width=0.32\textwidth,height=5cm]{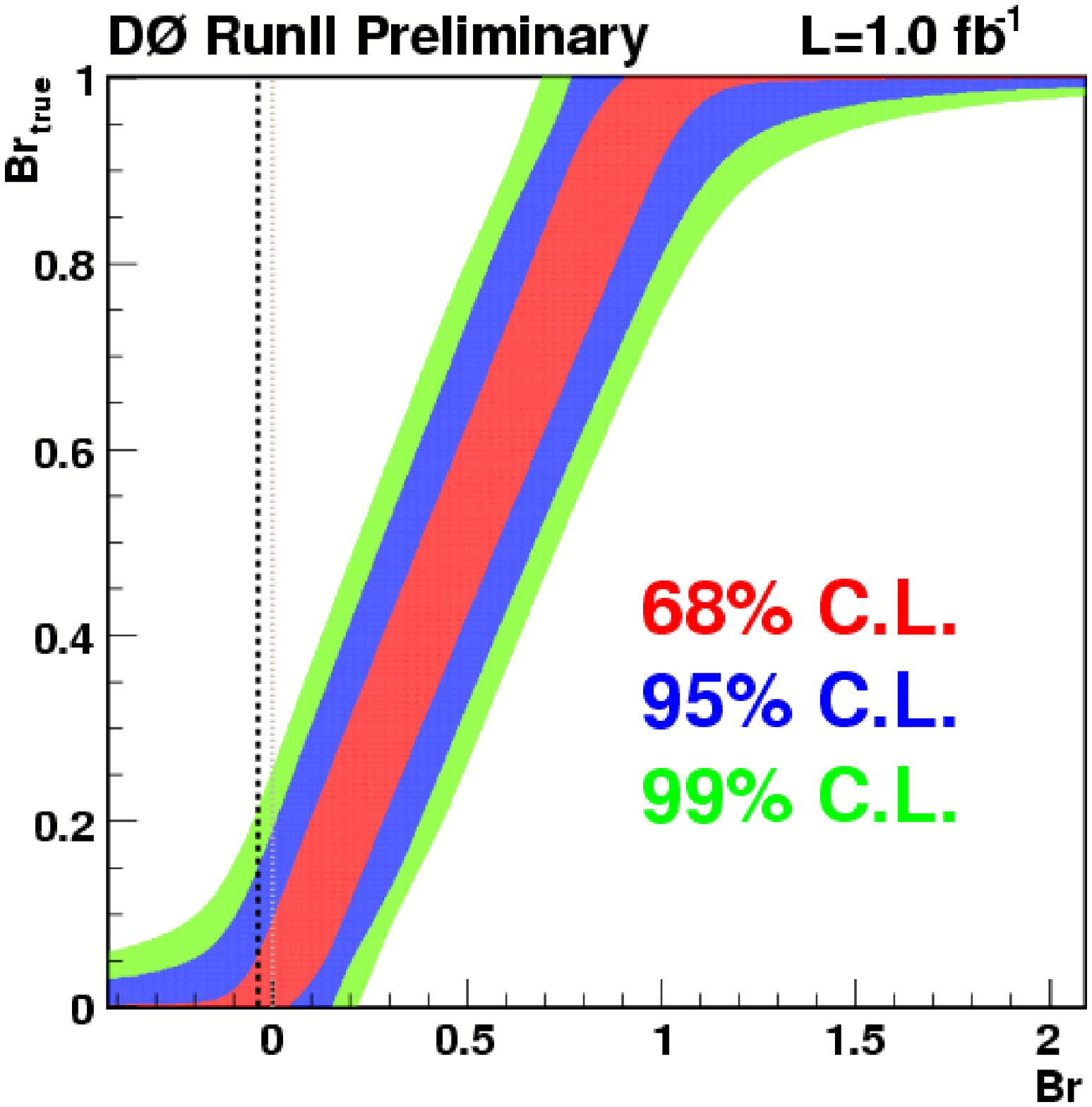}
\end{center}
\vspace*{-0.6cm}
\caption{
D\O. 
Left: variation of number of expected events for $\rm t\to H^+b$ ($\rm H^+\to \tau^+\nu$).
Center: variation of number of expected events for $\rm t\to H^+b$ ($\rm H^+\to cs$).
Right: BR($\rm t\to H^+b$) limits and type II THDM leading order calculation expectation.
} \label{fig:d0_thb_low1}
\end{figure}

\begin{figure}[thp]
\begin{center}
\includegraphics[width=0.32\textwidth,height=5cm]{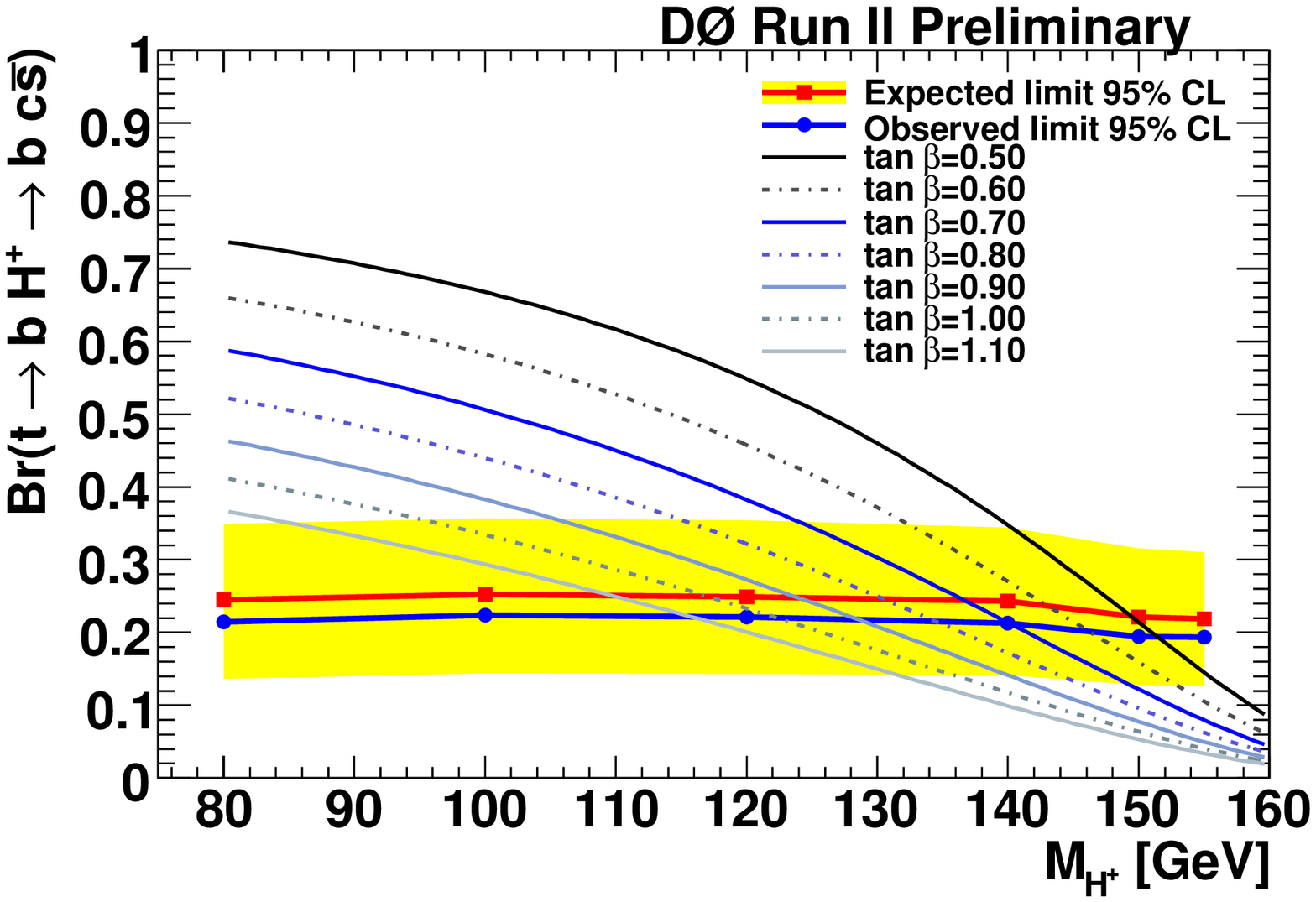}
\includegraphics[width=0.32\textwidth,height=5cm]{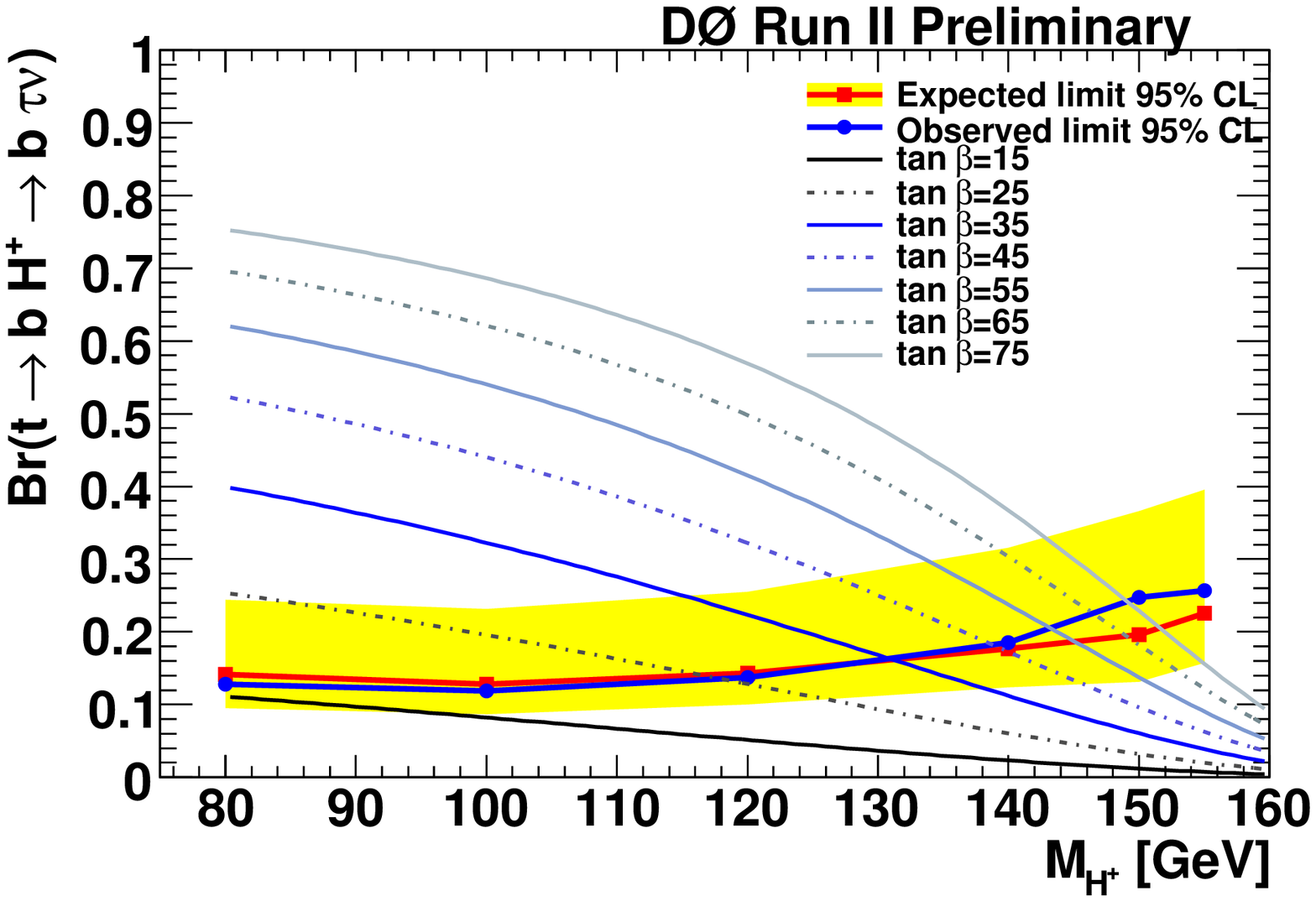}
\includegraphics[width=0.32\textwidth,height=5cm]{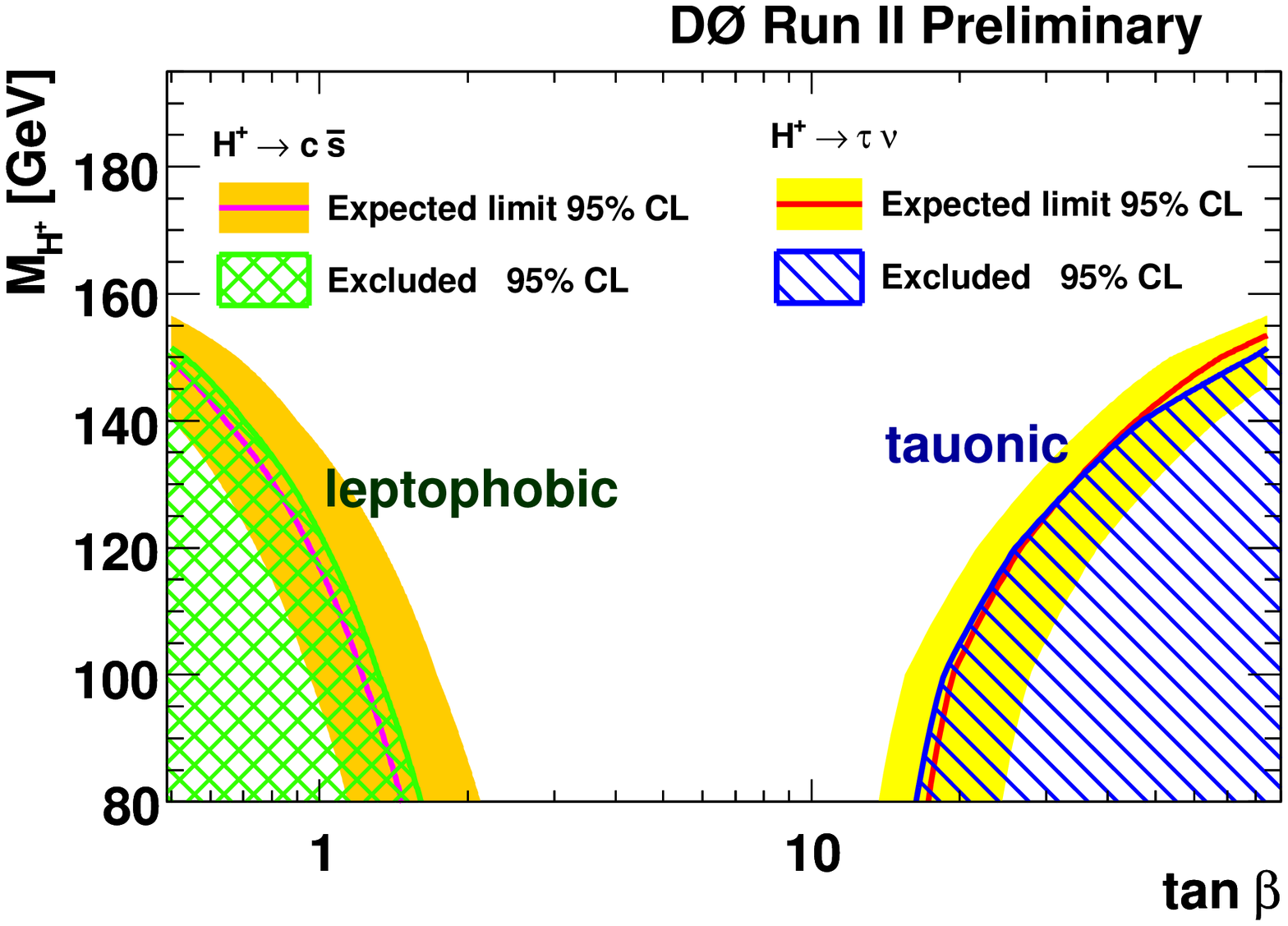}
\end{center}
\vspace*{-0.6cm}
\caption{
D\O. 
Left: BR($\rm t\to H^+b)$($\rm H^+\to cs$) limit at 95\% CL.
Center: BR($\rm t\to H^+b)$($H^+\to \tau^+\nu$) limit at 95\% CL.
Right: limits on the charged Higgs boson mass as function of $\tan\beta$ 
for a specific set of MSSM parameters.
} \label{fig:d0_thb_low2}
\end{figure}

The decay of top quarks $\rm t\to H^+b$ is possible in general Higgs boson models 
with two Higgs boson doublets. 
The expected top and charged Higgs boson branching fractions are shown in 
Fig.~\ref{fig:cdf-tbh} (left plot from~\cite{d0-tbh}) as a function of $\tan\beta$
for a specific MSSM parameter set.
The expected SM top decay rate would be modified.
No deviation from the SM top decay rates is observed.
Results from CDF for 0.192~fb$^{-1}$ are shown in Fig.~\ref{fig:cdf-tbh} 
(right plot from~\cite{cdf-tbh})
and from D\O\ for 1~fb$^{-1}$ are shown in Figs.~\ref{fig:d0_thb_low1} and~\ref{fig:d0_thb_low2}
(from~\cite{d0-tbh}).

\begin{figure}[thp]
\begin{center}
\includegraphics[width=0.32\textwidth,height=5cm]{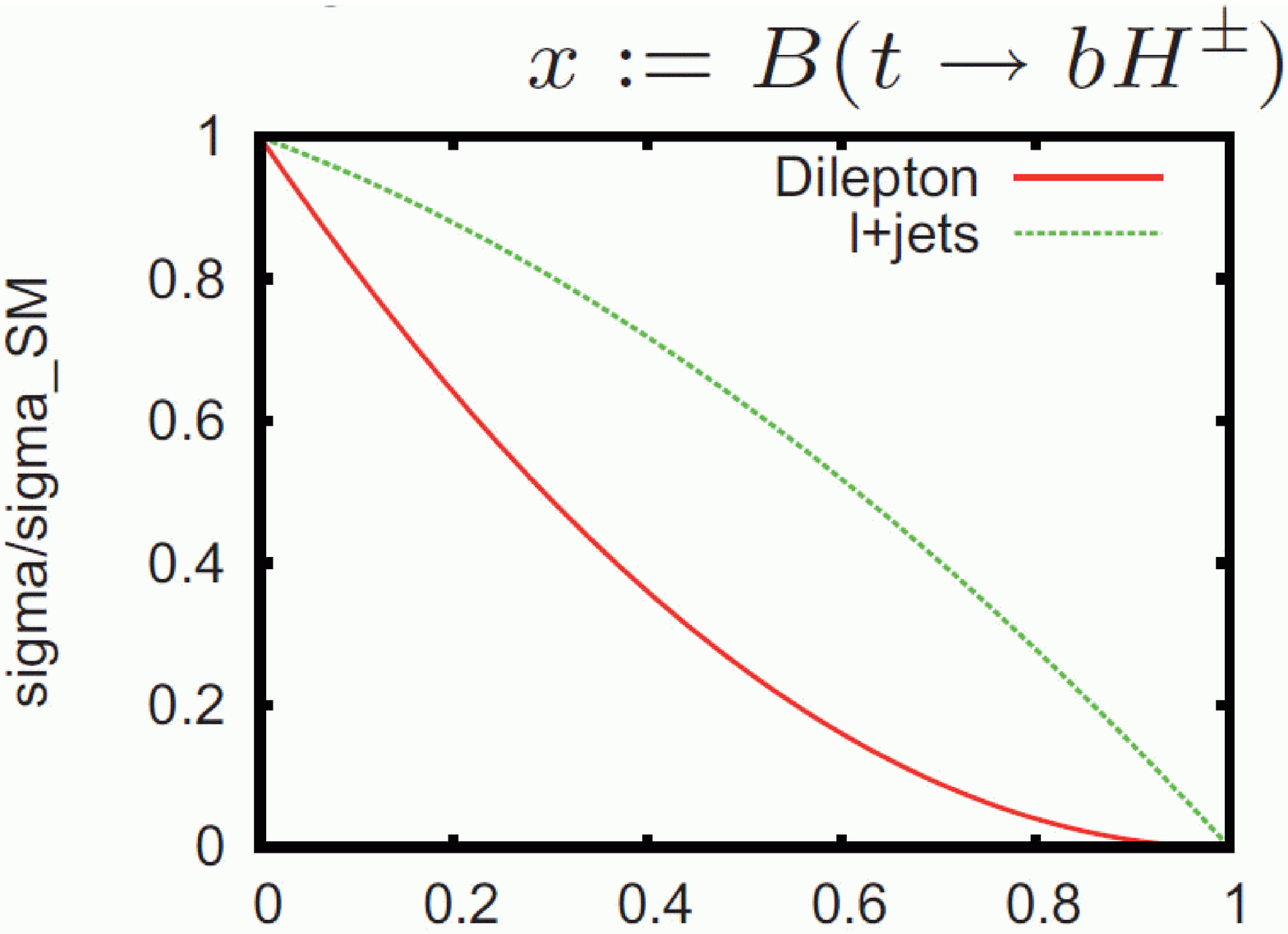}
\includegraphics[width=0.32\textwidth,height=5cm]{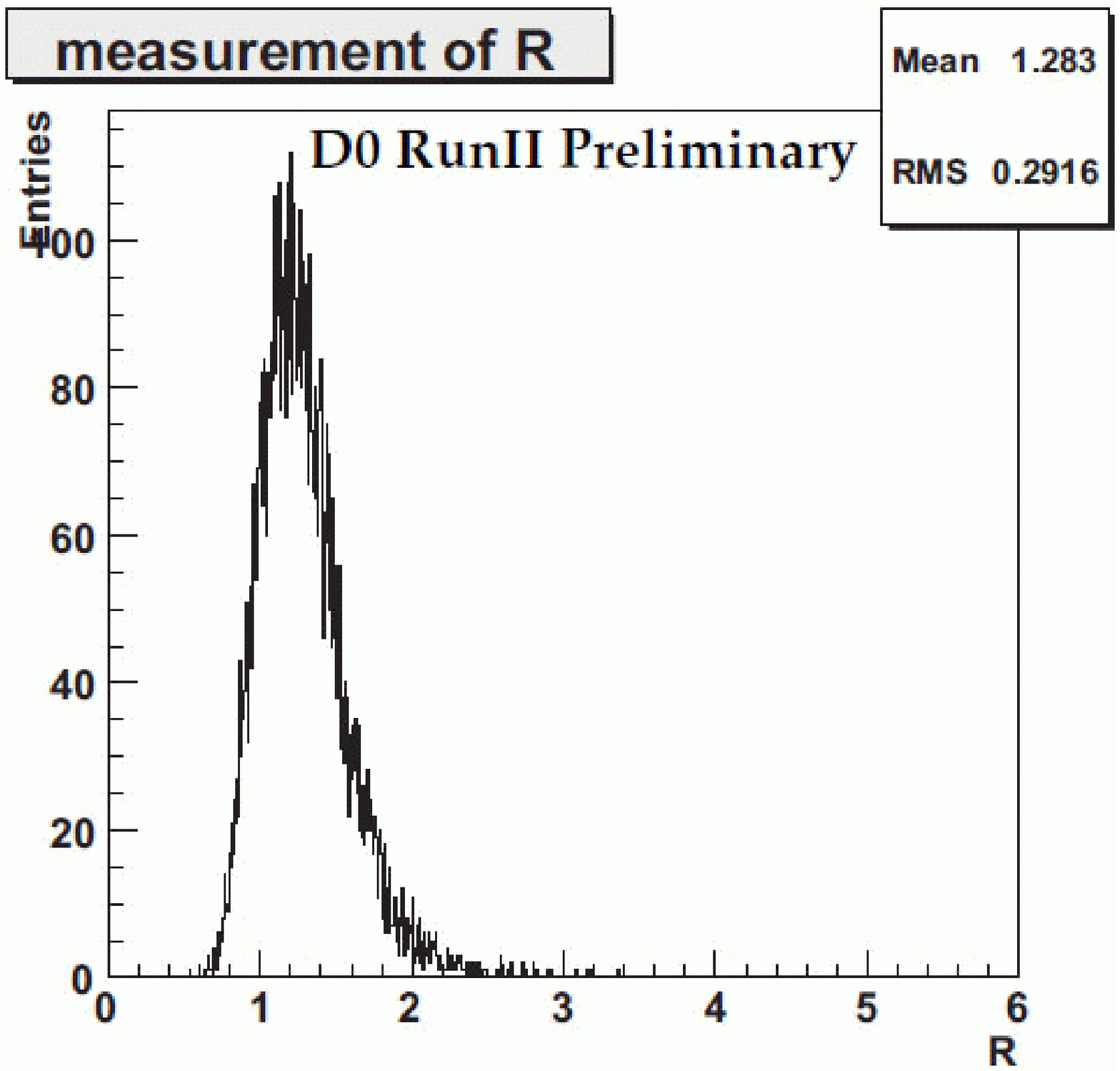}
\includegraphics[width=0.32\textwidth,height=5cm]{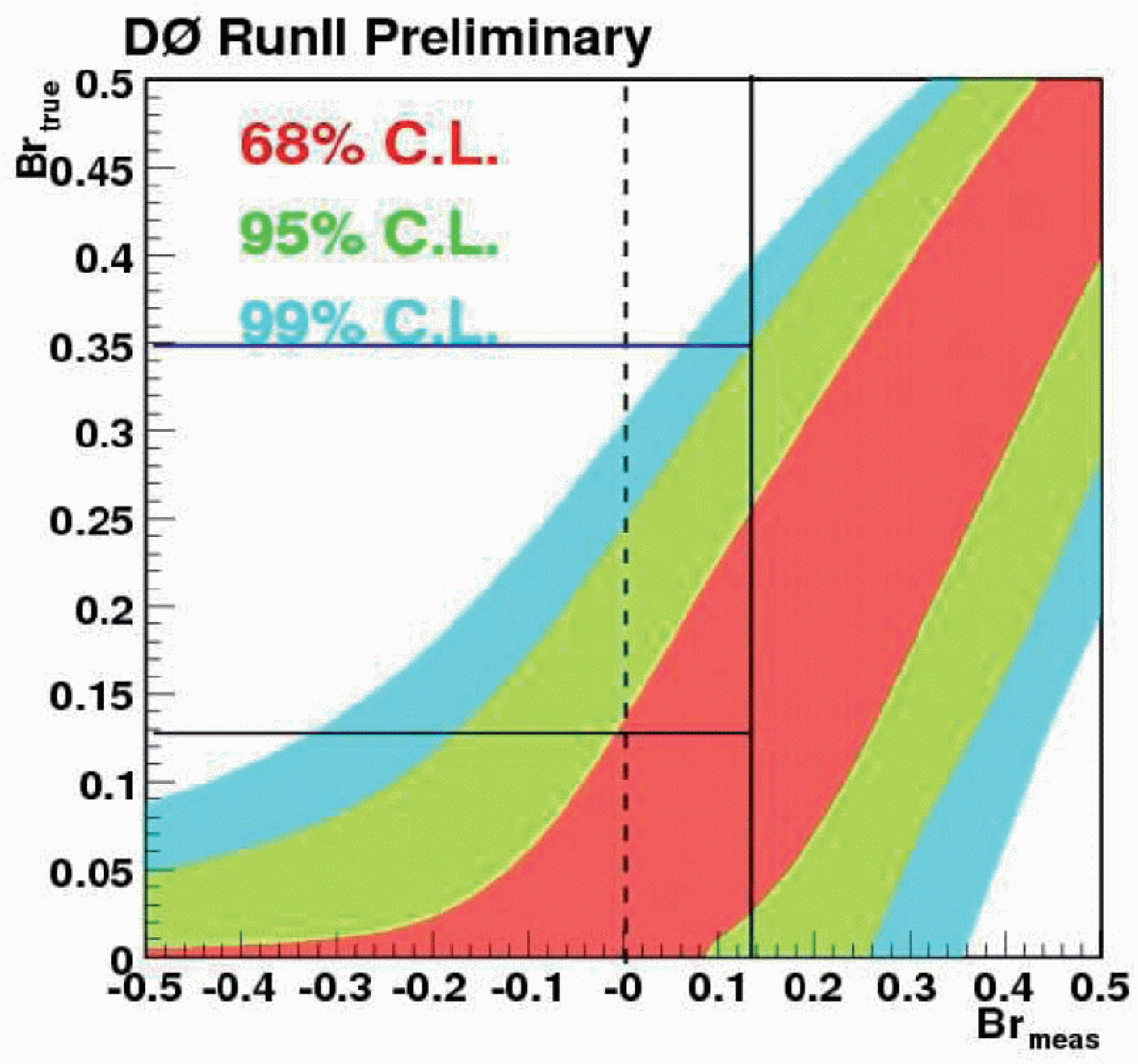}
\end{center}
\vspace*{-0.6cm}
\caption{
D\O.
Left: modified cross-sections relative to the SM cross-section as functions of 
BR$(\rm t\to H^+b$).
Center: cross-section ratio, $R$, distribution generated from the 10,000 pseudo-experiments.
Right: Feldman-Cousins confidence interval bands as functions of measured and
generated branching fraction BR$(\rm t\to H^+b$). 
For a leptophobic 80~GeV charged Higgs boson, BR($\rm H^+\to cs$)=1, BR$(\rm t\to H^+b$) 
limits  at 95\% CL are 0.35 (observed, solid line) and 0.25 (expected, dotted line).
} \label{fig:d0_tbh_ratio}
\end{figure}

An independent search has been carried out by D\O\ based on measuring the ratio 
$R = \sigma(t\bar t)_{\rm \ell+jets} / \sigma(t\bar t)_{\rm \ell^+\ell^-}$~\cite{d0-tbh_ratio}. 
In the SM $R=1$, while a decay $\rm t\to H^+b$ changes this ratio. Results are summarized in
Fig.~\ref{fig:d0_tbh_ratio} (from~\cite{d0-tbh_ratio}) for a leptofobic charged Higgs boson.
 
\begin{figure}[hbtp]
\begin{center}
\includegraphics[width=0.32\textwidth,height=5cm]{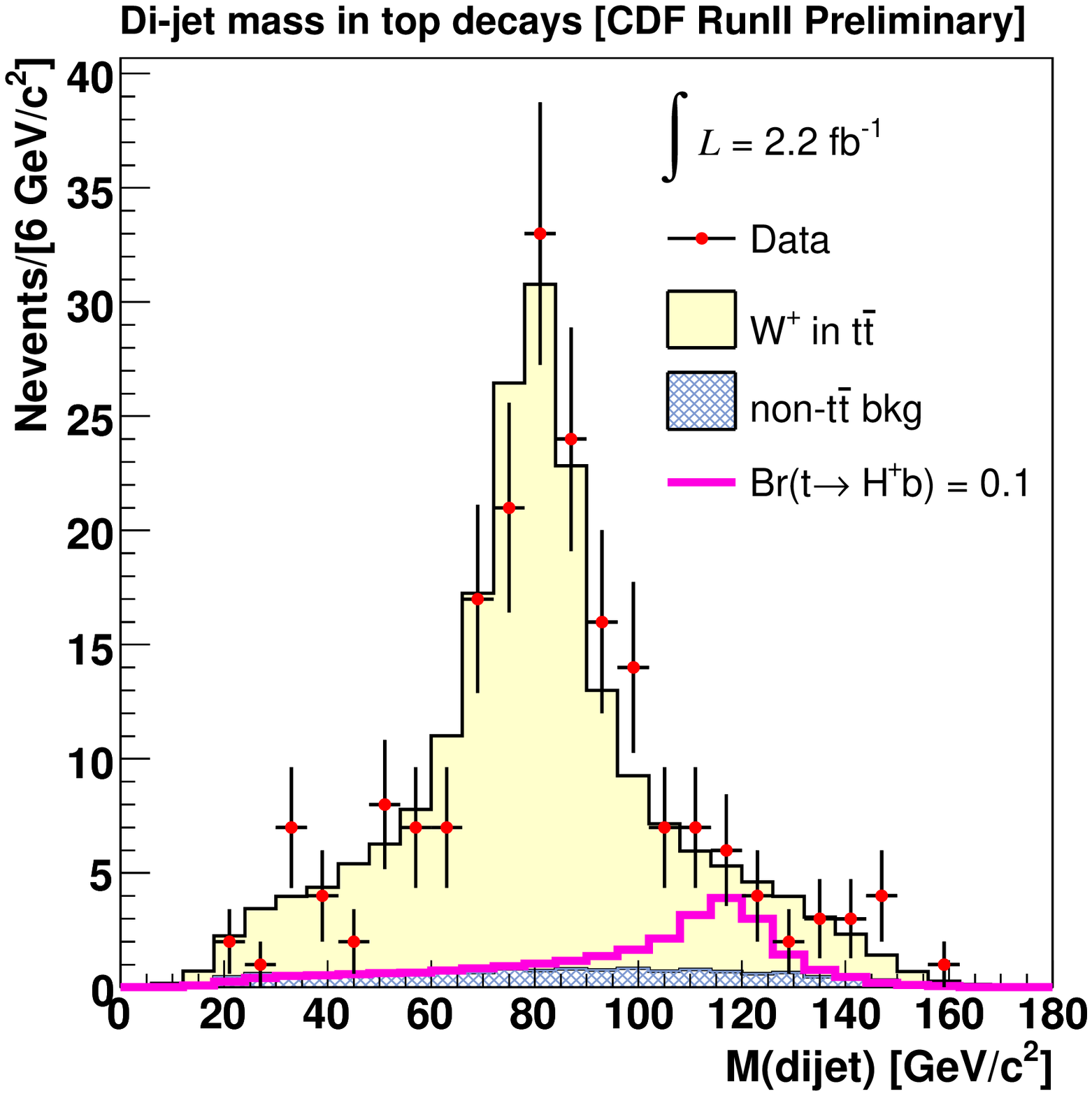}\hfill
\includegraphics[width=0.32\textwidth,height=5cm]{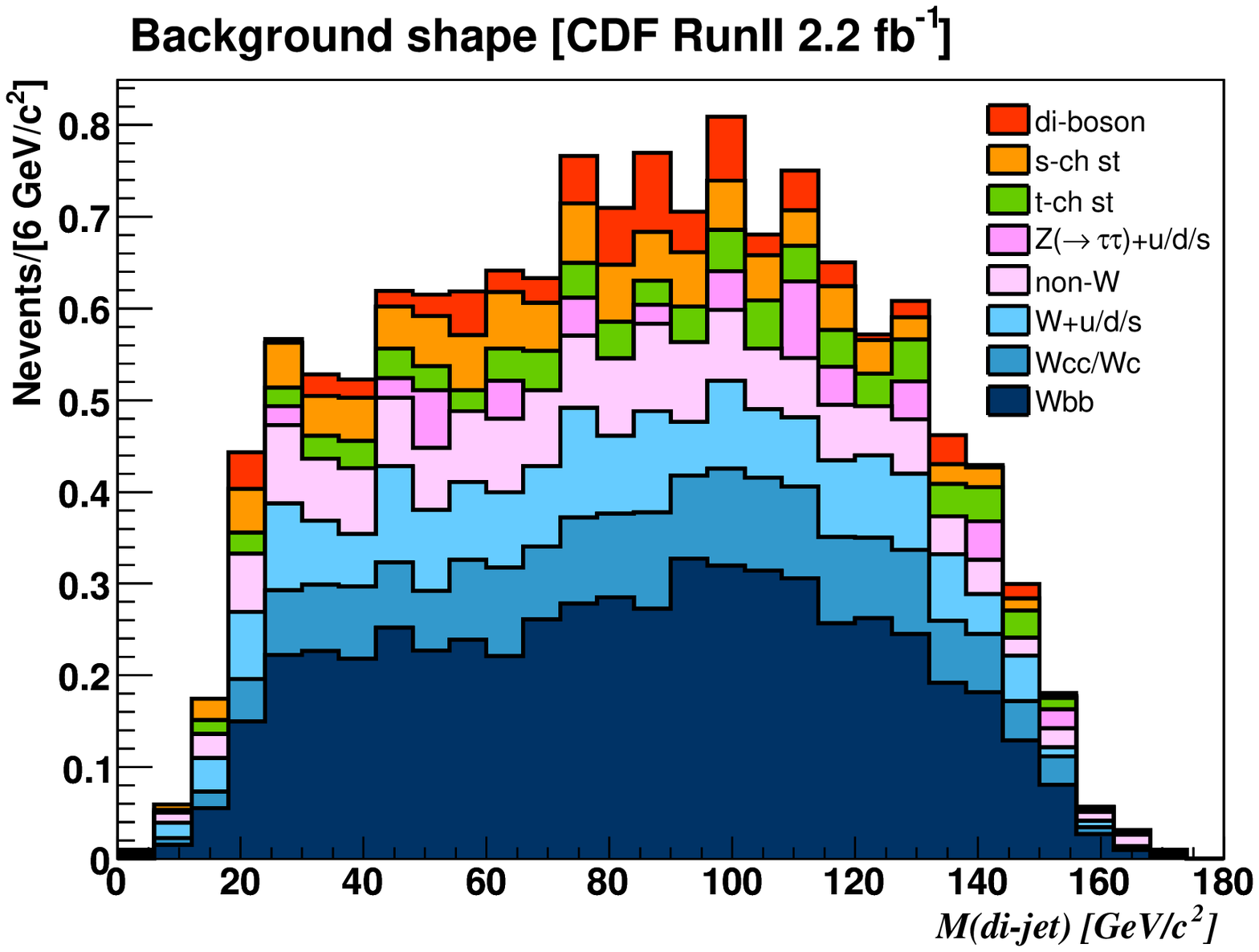}\hfill
\includegraphics[width=0.32\textwidth,height=5cm]{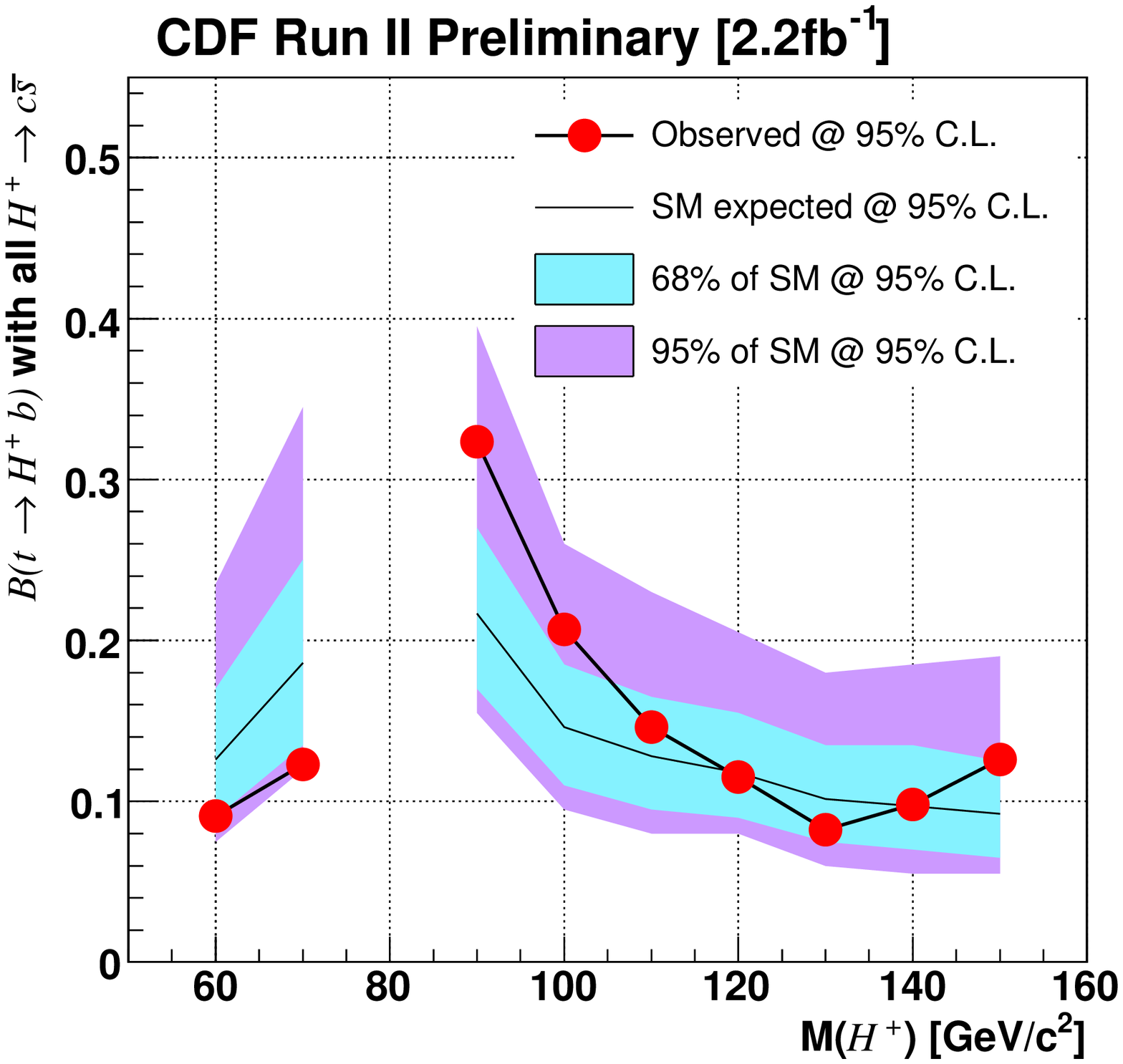}
\end{center}
\vspace*{-0.6cm}
\caption{
CDF. 
Left: di-jet mass in top decays.
Center: background contributions to the di-jet mass in top decays.
Right: model-independent BR($\rm t\to H^+b$) limit for BR($\rm H^+\to c \bar s)=1$.
In order to cover any generic anomalous charged Higgs boson, the search is extended below the W mass.
\label{fig:cdf_tbh_cs}}
\end{figure}

A recent charged Higgs boson search by CDF focuses on 
the reaction $\rm t\to H^+b~(H^+\to c\bar s)$~\cite{cdf-tbh_cs} based on 2.2~fb$^{-1}$ data.
The hadronic charged Higgs boson decay mode would allow a precise $\rm H^+$ mass reconstruction.
Results are summarized in Fig.~\ref{fig:cdf_tbh_cs} (from~\cite{cdf-tbh_cs}).

\begin{figure}[hbtp]
\begin{center}
\includegraphics[width=0.32\textwidth,height=5cm]{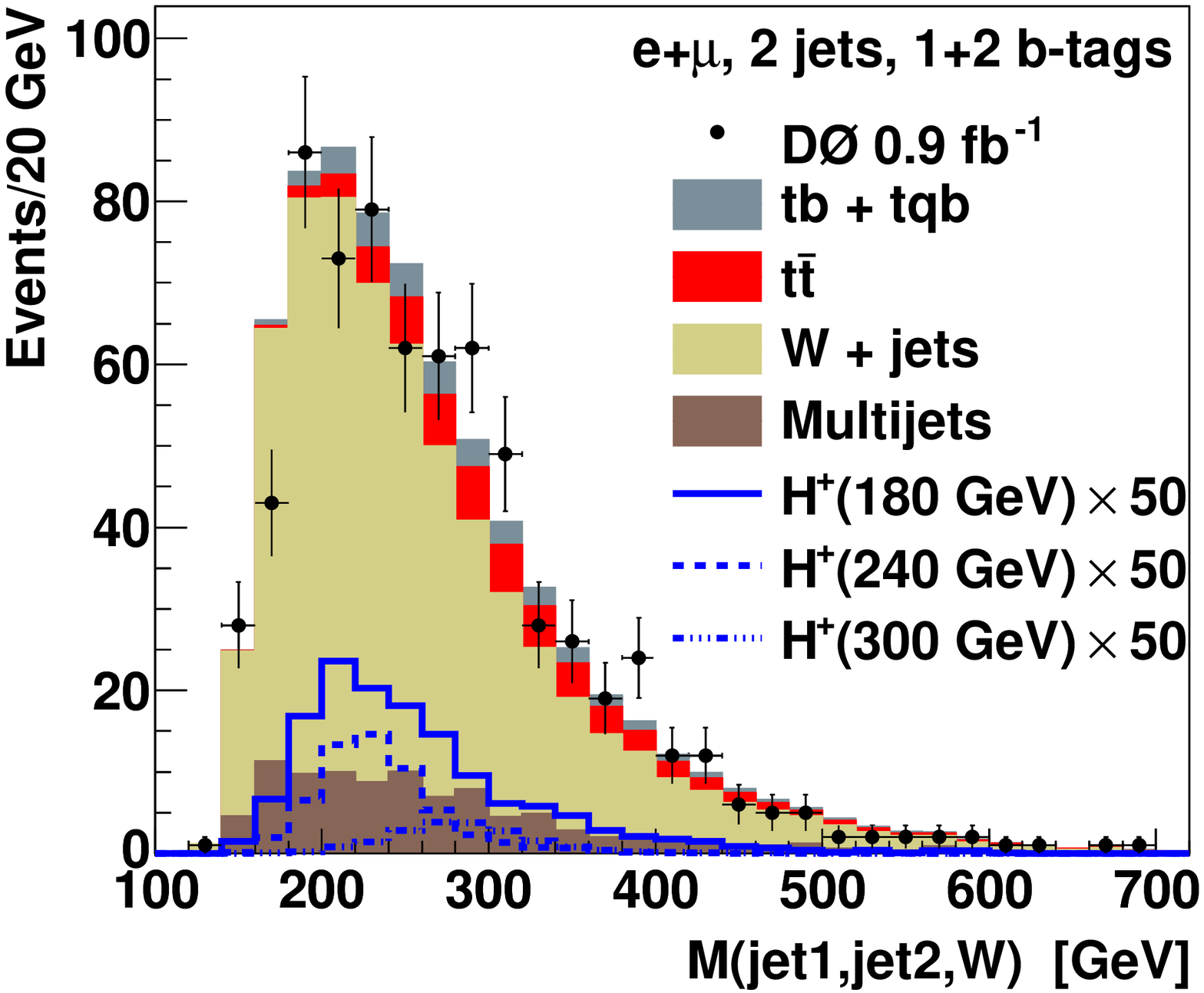}\hfill
\includegraphics[width=0.32\textwidth,height=5cm]{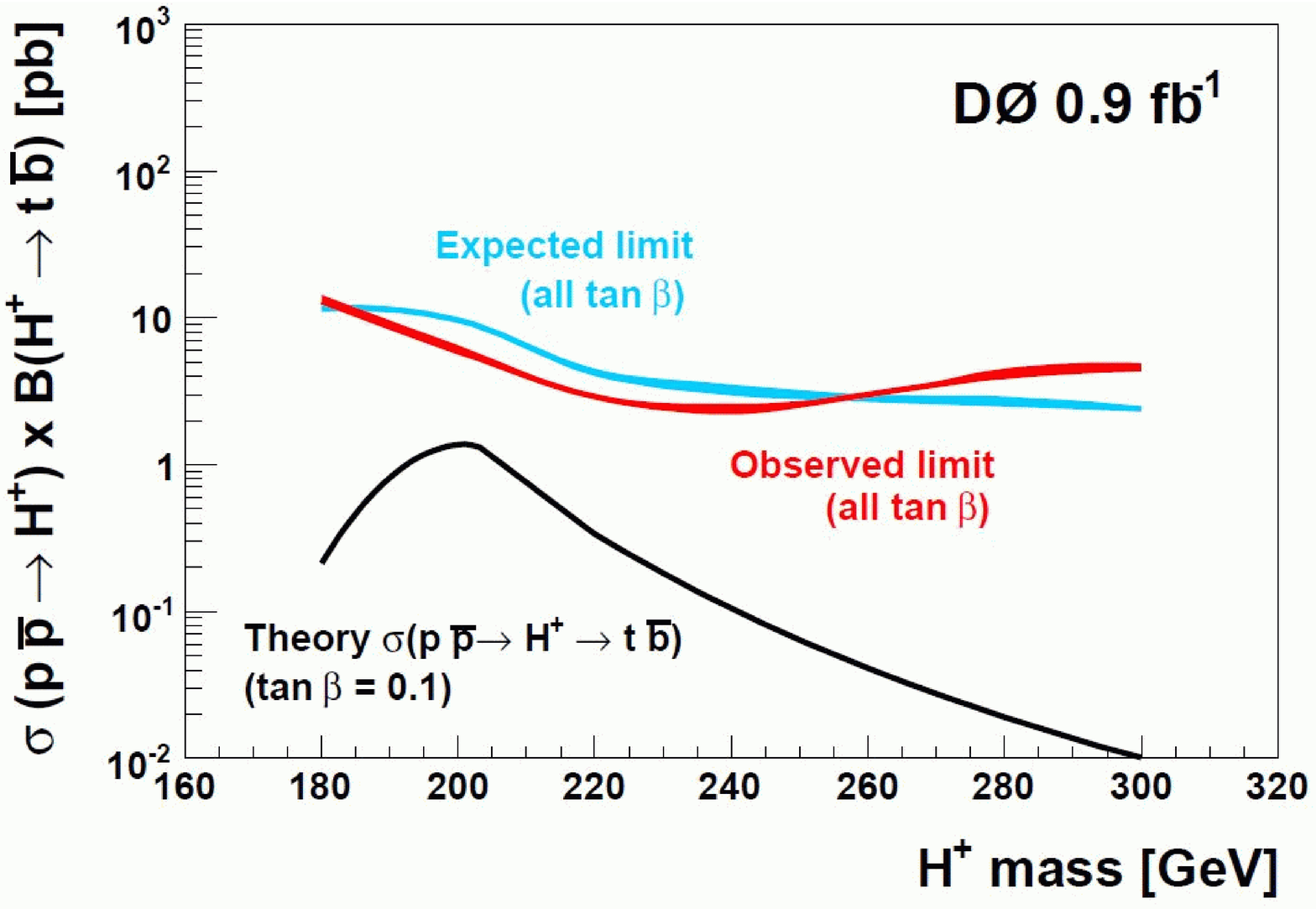}\hfill
\includegraphics[width=0.32\textwidth,height=5cm]{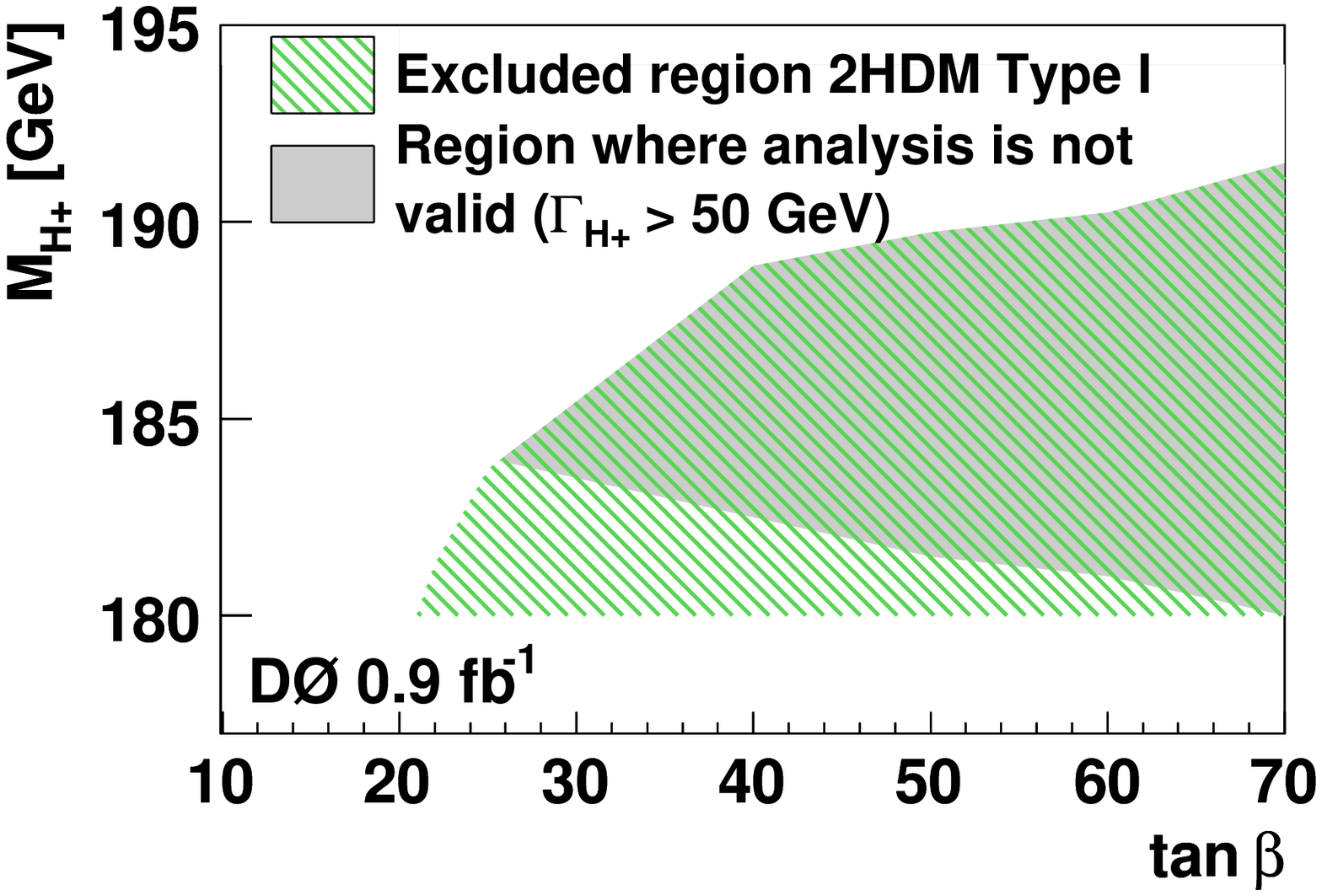}
\end{center}
\vspace*{-0.6cm}
\caption{
D\O. $p\bar p \to H^+\to t\bar b$.
Left: invariant charged Higgs boson mass (type III model).
Center: cross-section limit in THDM (type II).
Right: THDM excluded regions for model type I.
} \label{fig:d0_hp_high}
\end{figure}

In the high mass regime $(m_{\rm H^+} > m_{\rm t})$ the search for charged Higgs bosons 
has been performed similar as for the single top s-channel analysis with ${\cal L}=0.9$~fb$^{-1}$.
The reaction is $\rm q\bar q'\to H^+\to t\bar b\to W^+b\bar b\to \ell^+\nu\bb$~\cite{d0_hp_high_mass},
where $\ell$ represents an electron or a muon.
Results are summarized in Fig.~\ref{fig:d0_hp_high} (from~\cite{d0_hp_high_mass}).

\subsection{$\rm H\to \gamma\gamma$}

\begin{figure}[hbtp]
\includegraphics[width=0.32\textwidth,height=5cm]{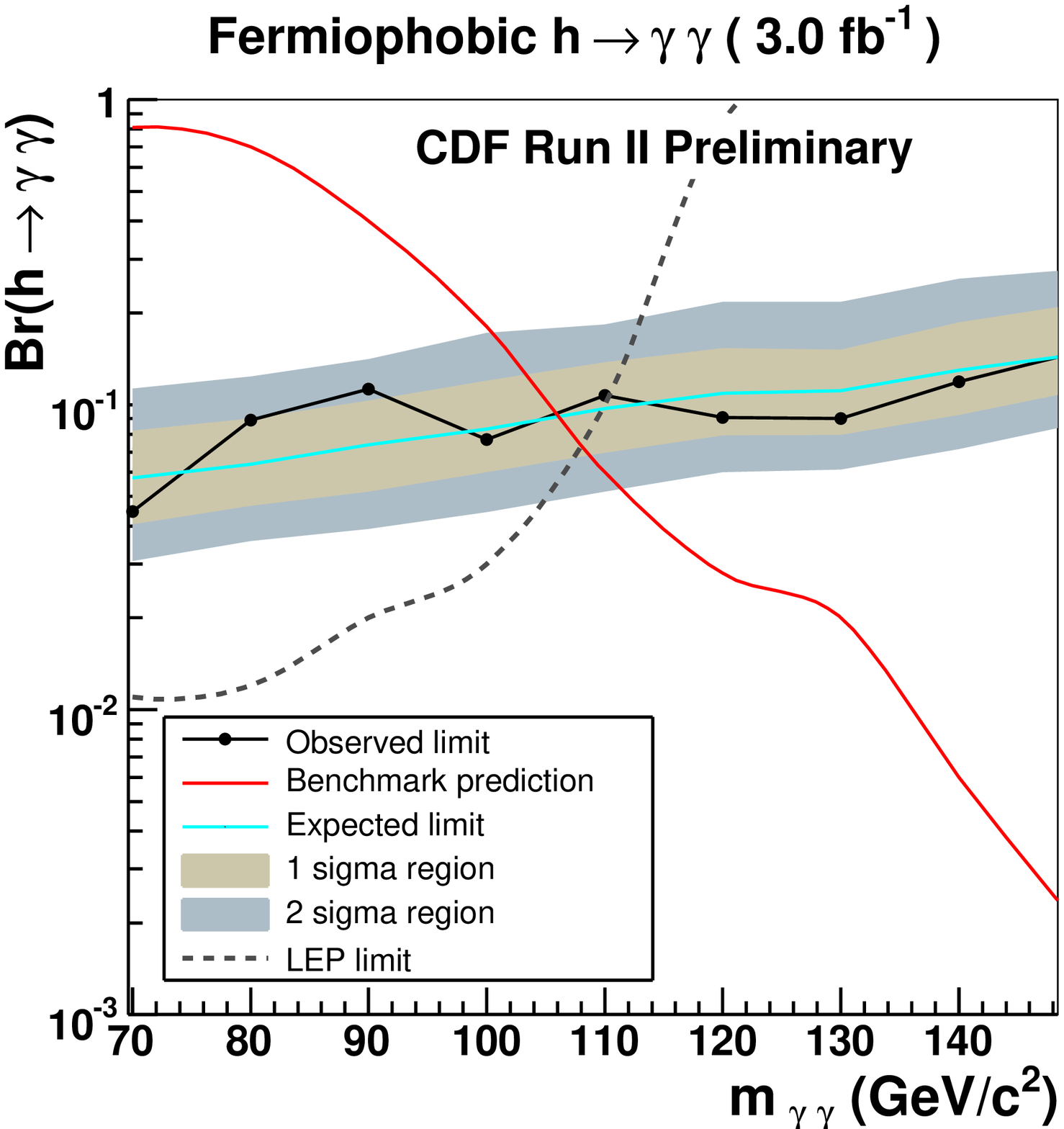}\hfill
\includegraphics[width=0.32\textwidth,height=5cm]{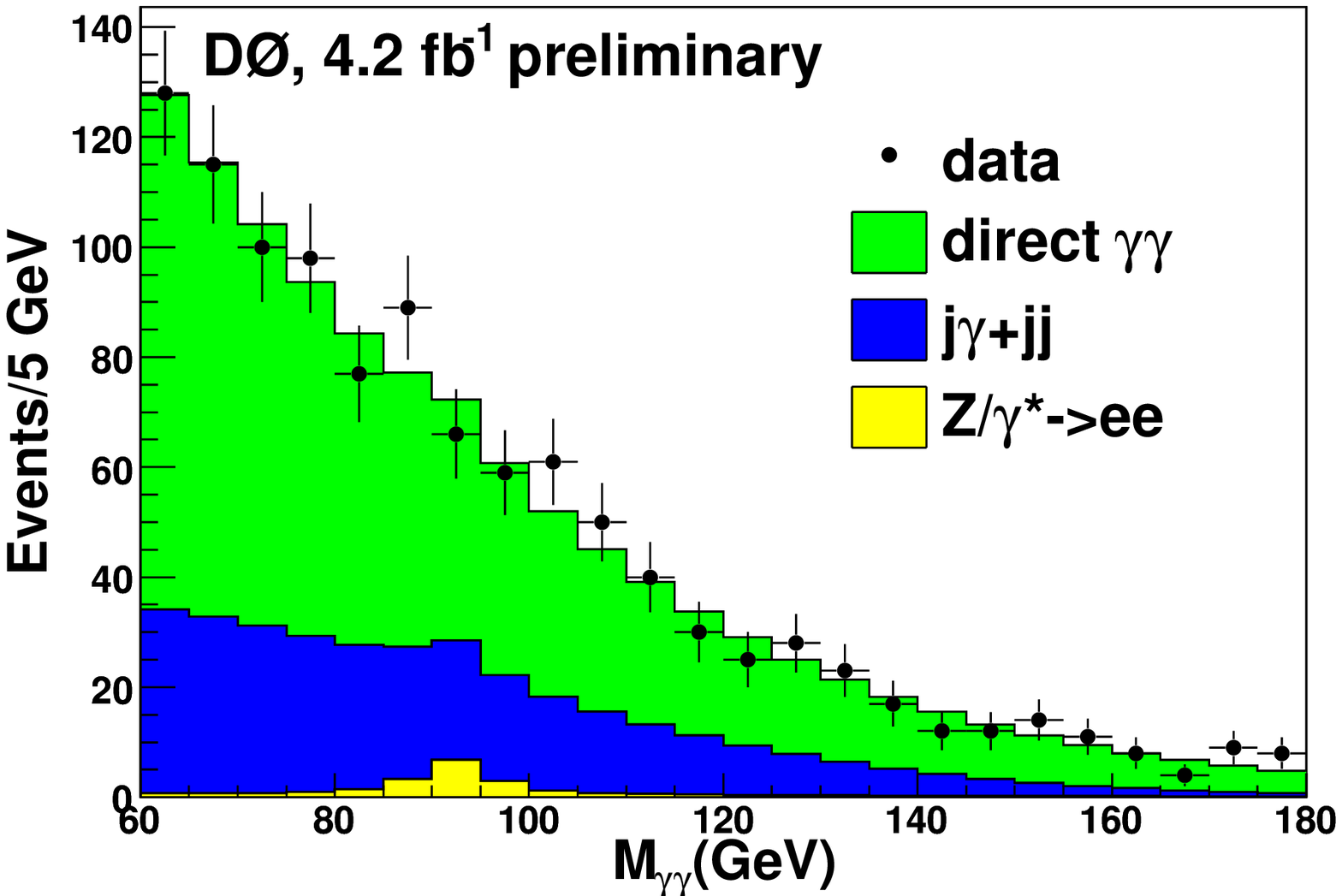}\hfill
\includegraphics[width=0.32\textwidth,height=5cm]{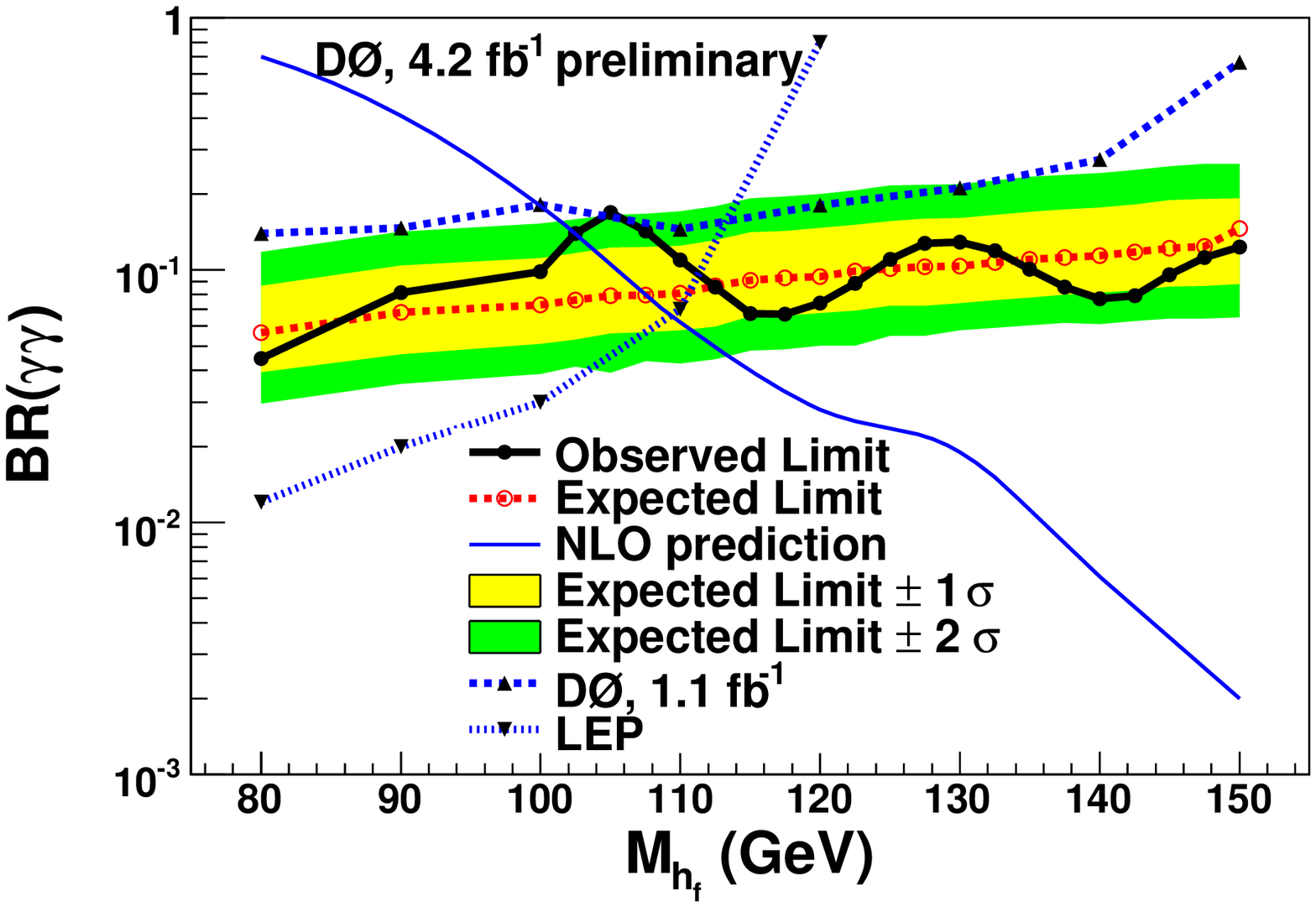}
\vspace*{-0.1cm}
\caption{Fermiophobic Higgs.
$\rm H\to \gamma\gamma$.
Left: CDF limits at 95\% CL with ${\cal L}=3.0$~fb$^{-1}$.
Center: D\O\ invariant $\gamma\gamma$  mass distribution with ${\cal L}=4.2$~fb$^{-1}$. 
Right: D\O\ limits at 95\% CL with ${\cal L}=4.2$~fb$^{-1}$.
} \label{fig:d0-h-gamma}
\end{figure}

In fermiophobic Higgs boson models, the dominant decay mode could be 
$\rm H\to \gamma\gamma$. The Higgs boson could be produced in the associated production
with a vector boson and vector boson fusion (VBF) production mechanisms.
No indication of such reactions have been observed and limits are set as shown in 
Fig.~\ref{fig:d0-h-gamma} from CDF (left plot~\cite{cdf-sm_gammagamma}) 
and from D\O\ (center and right plots~\cite{d0-gammagamma}). 

\vspace*{1cm}
\subsection{$\rm H^{++}$}

\begin{figure}[hbtp]
\includegraphics[width=0.32\columnwidth,height=5cm]{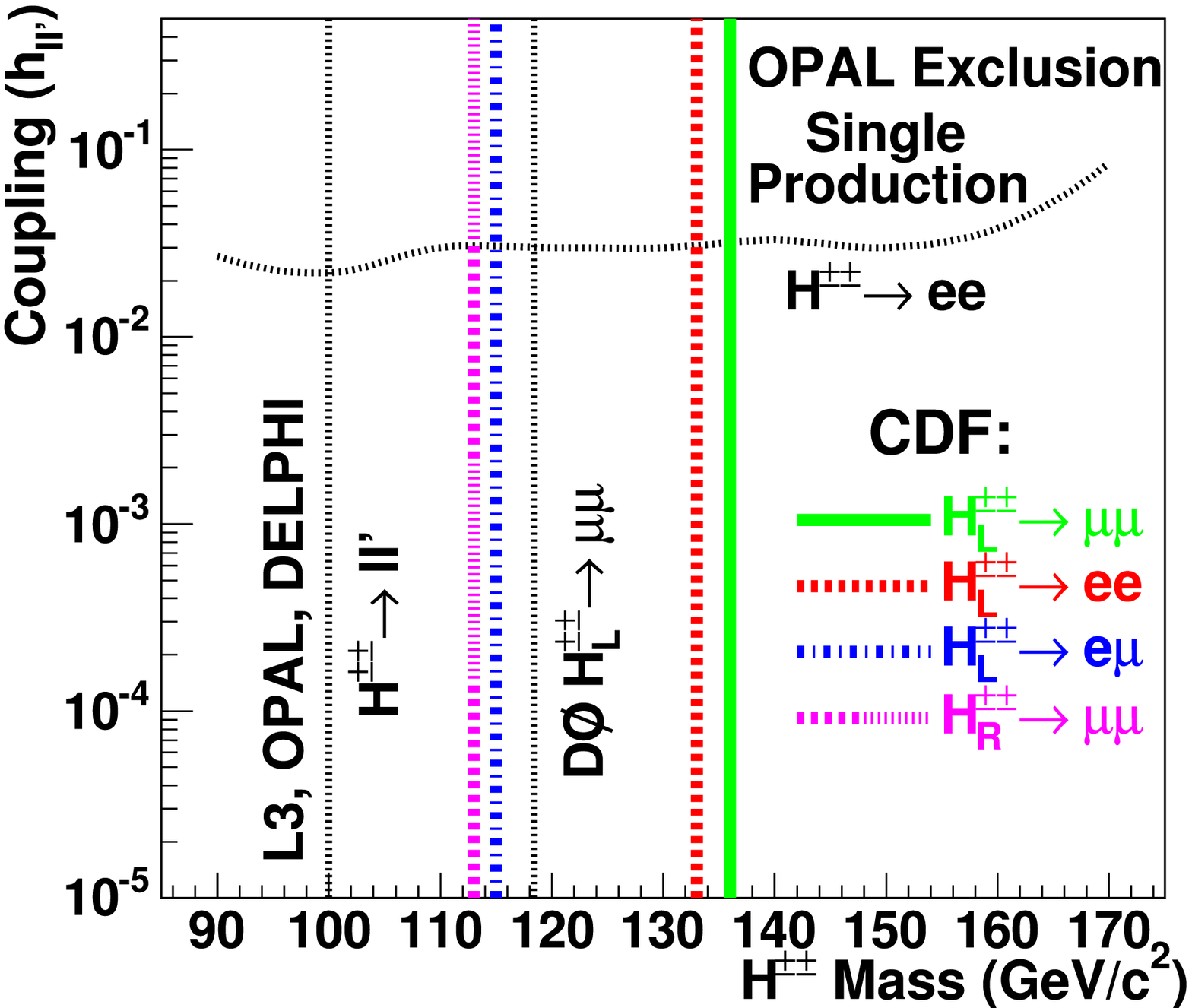} \hfill
\includegraphics[width=0.32\columnwidth,height=5cm]{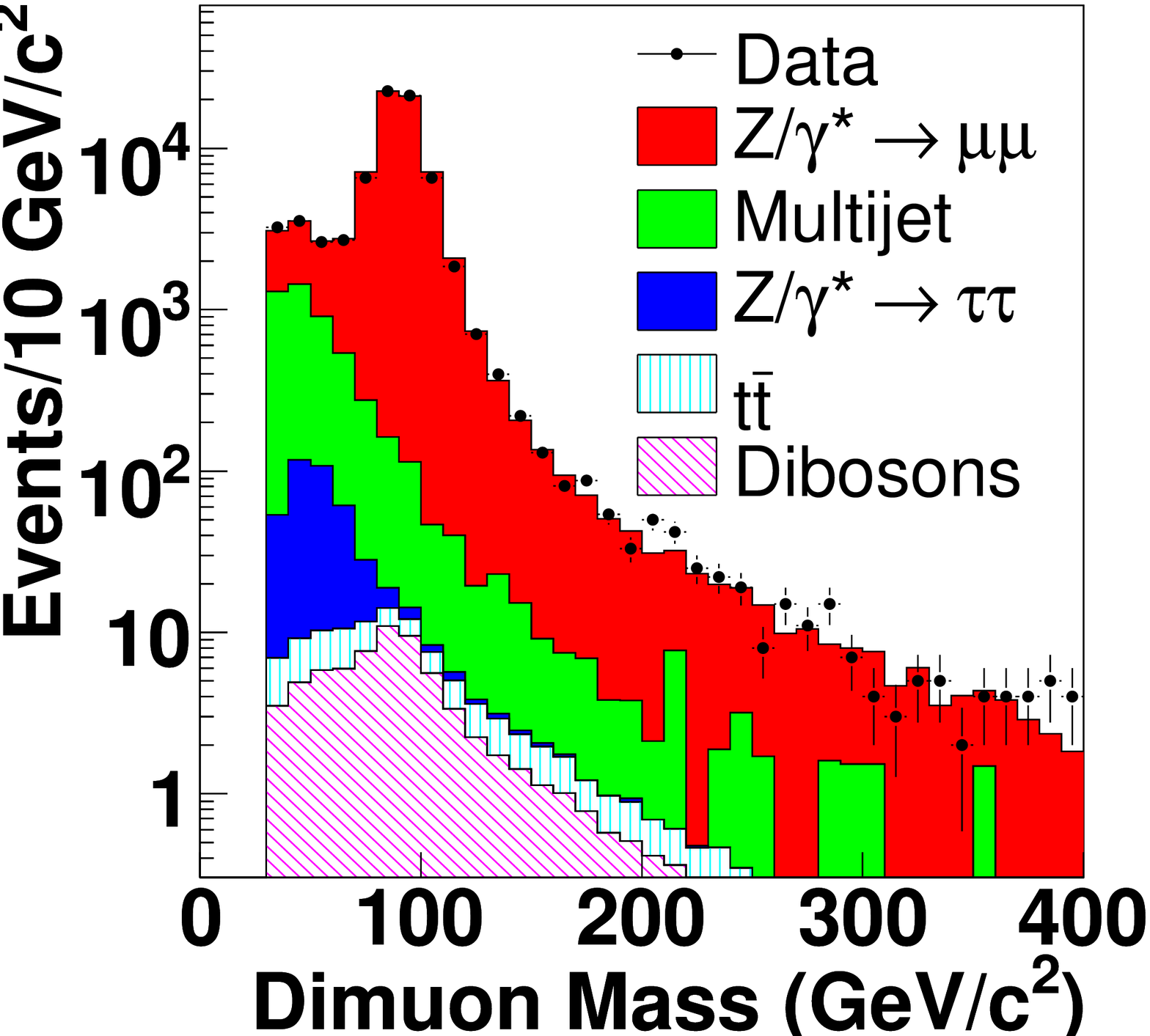} \hfill 
\includegraphics[width=0.32\columnwidth,height=5cm]{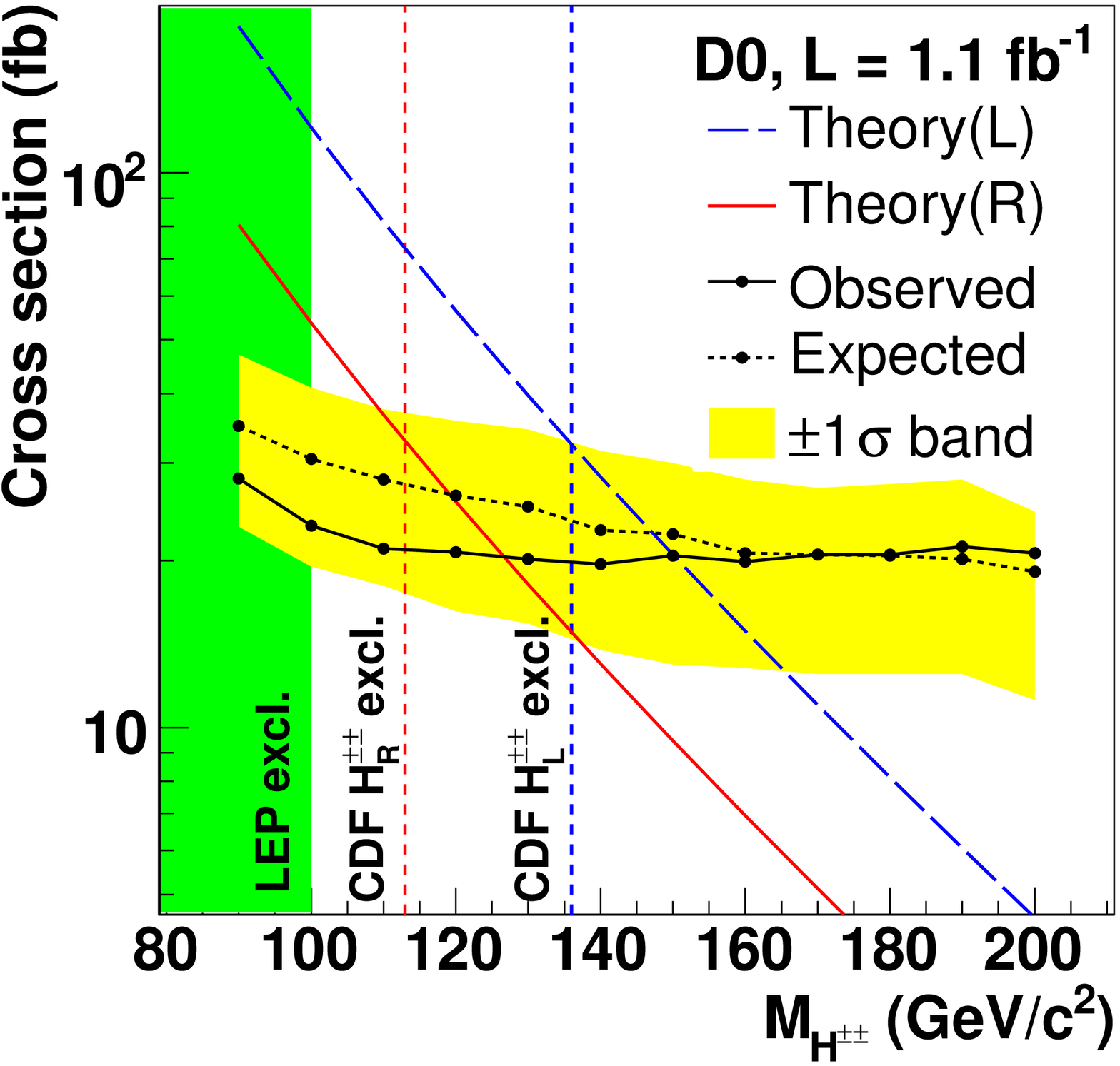} 
\vspace*{-0.1cm}
\caption{
Left: CDF doubly-charged Higgs boson mass limits at 95\% CL.
Center: D\O\ di-muon mass spectrum.
Right: D\O\ doubly-charged Higgs boson mass limits at 95\% CL.
} \label{fig:d0-hh}
\end{figure}

The possibility of doubly-charged Higgs boson exists in models with Higgs boson triplets.
Pairs of like-sign charged leptons are expected from the decay of the doubly-charged
Higgs bosons. No indication has been observed in the data. 
The di-muon mass spectrum and limits on the doubly-charged Higgs boson mass
are shown in Fig.~\ref{fig:d0-hh} 
from CDF (left plot~\cite{cdf-hpp}) 
and from D\O\ (center and right plots)~\cite{d0-hpp}).

\section{Conclusions}
Much has been learned from the searches for Higgs bosons at LEP.
The Tevatron Run-II searches for Higgs bosons are well under way and already have set 
several limits exceeding some previous LEP limits.
For the SM Higgs boson, searches for gluon fusion with WW decays, associated production
WH with $\rm \bb$ and WW decays, and $\rm ZH\to \nu\nu\bb$ decays have been performed
previously.
Updates of these searches have been reported and 
compared to a previous report~\cite{as06} with results from summer 2005.
More recently, searches for the SM reactions $\rm ZH\to \ell\ell\bb$, $\rm H\to \gamma\gamma$ and
$\rm t\bar t\to t\bar tH$ have been performed in addition.
Beyond the Standard Model, the searches at the Tevatron for
$\rm b\bar b A$, H$^+$, H$^{++}$, $\rm h\rightarrow\gamma\gamma~and~\tau^+\tau^-$ have led to
new limits on couplings and masses. 
The close collaboration of phenomenologists and experimentalists is crucial 
to fully exploit the potential of the collected data.
The sensitivity of the SM Higgs searches is evolving rapidly, 
significantly faster than the increase in sensitivity from improved statistics alone.
The first direct SM exclusion beyond the LEP results 
was achieved at high mass, around 170~GeV, in summer 2008. 
Incorporating the ongoing improvements, by the end of 2010 running, 
exclusion over virtually the full mass range favoured by the electroweak fits 
is achievable. In addition three sigma evidence will be possible over much of the 
same range.

\section*{Acknowledgments}
I would like to thank 
my colleagues from the CDF and D\O\ Higgs working groups for discussions
and comments on the manuscript, in particular 
Peter Bussey, Craig Group, Matthew Herndon, Thomas Wright and Un-ki Yang from the CDF Collaboration,
Gavin Davies, Sebastien Greder, Phillip Gutierrez and Alex Melnitchouk from the D\O\ Collaboration, and
Oscar St\aa l from the phenomenology group at Uppsala University.

\end{document}